\documentclass[epsfig,12pt]{article}
\usepackage{amsfonts}
\usepackage{amssymb}
\usepackage{graphicx}
\usepackage{amsmath}

\input epsf.sty
\newtheorem{theorem}{Theorem}
\newtheorem{acknowledgement}[theorem]{Acknowledgement}

\input{tcilatex}
\makeatletter \@addtoreset{equation}{section}

\begin{document}

\title{%
\rightline{\mbox{\normalsize
{Lab/UFR-HEP0607/GNPHE/0607/VACBT/0607}}} \textbf{Topological string in
harmonic space and correlation functions in }$S^{3}$ \textbf{stringy}
\textbf{cosmology}}
\author{ El Hassan Saidi$^{1,3,4}$\thanks{%
h-saidi@fsr.ac.ma} \ and Moulay Brahim Sedra$^{1,2,3,4}$\thanks{%
sedra@ictp.it} \\
{\small 1\textit{. Lab/UFR-Physique des Hautes Energies,}} {\small \textit{%
Facult\'{e} des Sciences de Rabat, Morocco.}}\\
{\small 2\textit{. Lab Physique de la Mati\`{e}re et Rayonnement, Facult\'{e}
des Sciences Kenitra, Morocco }}\\
{\small 3. \textit{GNPHE, Groupement National de Physique des Hautes
Energies, Si\`{e}ge focal: FS Rabat, Morocco.}}\\
{\small 4\textit{. Virtual African Centre for Basic Science and Technology,
VACBT, }}\\
{\small \textit{Focal point, LabUFR-PHE, Rabat, Morocco.}}}
\maketitle

\begin{abstract}
We develop the harmonic space method for conifold and use it to study local
complex deformations of $T^{\ast }S^{3}$ preserving manifestly $SL\left(
2,C\right) $ isometry. We derive the perturbative manifestly $SL\left(
2,C\right) $ invariant partition function $\mathcal{Z}_{top}$ of topological
string B model on locally deformed conifold. Generic $n$ momentum and
winding modes of $2D$ $c=1$ non critical theory are described by highest $%
\upsilon _{\left( n,0\right) }$ and lowest components $\upsilon _{\left(
0,n\right) }$ of $SL\left( 2,C\right) $ spin $s=\frac{n}{2}$ multiplets $%
\left( \upsilon _{\left( n-k,k\right) }\right) $, $0\leq k\leq n$ and are
shown to be naturally captured by harmonic monomials. Isodoublets ($n=1$)
describe uncoupled units of momentum and winding modes and are exactly
realized as the $SL\left( 2,C\right) $ harmonic variables $U_{\alpha }^{+}$
and $V_{\alpha }^{-}$. We also derive a dictionary giving the passage from
Laurent (Fourier) analysis on $T^{\ast }S^{1}$ ($S^{1}$) to the harmonic
method on $T^{\ast }S^{3}$ ($S^{3}$). The manifestly $SU\left( 2,C\right) $
covariant correlation functions of the $S^{3}$ quantum cosmology model of
Gukov-Saraikin-Vafa are also studied.\newline
\textbf{Keywords: }\newline
\textit{harmonic analysis on conifold, topological string B model on }$%
T^{\ast }S^{3}$\textit{, ground ring of 2D }$\mathit{c=1}$\textit{\ string,
Hartle-Hawking wave function and }$\mathbb{S}^{3}$\textit{\ quantum cosmology%
}.
\end{abstract}




\section{Introduction}

\qquad In the few last years, there has been a growing interest in the study
of the interface between 10D superstring theory and cosmology $\cite{1}$-$%
\cite{5}$. This particular interest has been propelled more by recent
developments in stringy cosmology; in particular by: (\textbf{1}) the
remarkable observation on the link between Hartle-Hawking (HH) wave $\Psi
=\Psi \left( \mathrm{t},\overline{\mathrm{t}}\right) $\ and the partition
function $\mathcal{Z}_{top}=\mathcal{Z}_{top}\left( \mathrm{t}\right) $ of
the topological string B model on the conifold $T^{\ast }S^{3}$ $\cite{6}$-$%
\cite{14}$. (\textbf{2}) The intimate relation of $\mathcal{Z}_{top}$ to the
partition function $\mathcal{Z}_{c=1}$ of the $\ c=1$ non critical bosonic
string propagating on a circle at the self-dual radius $\cite{15}$-$\cite{18}
$. (\textbf{3}) the interpretation of $\mathcal{Z}_{top}$ as an exact "wave
function of the universe" in the mini-superspace sector of superstring
theory $\cite{19}$-$\cite{20}$. These correspondences have allowed moreover
to shed light on the link between the entropy of black holes and topological
string B-model partition function $\cite{21}$-$\cite{29}$. They have also
been used in $\cite{30}$ for the set up of Gukov-Saraikin-Vafa (GSV) $S^{3}$
stringy cosmology model.

\qquad The aim of this paper is to complete partial results on $\mathcal{Z}%
_{top}\left( \mathrm{t}\right) $ and $\Psi \left( \mathrm{t},\overline{%
\mathrm{t}}\right) $ by using the harmonic space method. Our contribution
involves the following three: First we develop a manifestly covariant $%
SL\left( 2,C\right) $ harmonic analysis to approach conifold geometry and it
submanifolds with isometries $SL\left( 2,C\right) /C^{\ast }$, $SU\left(
2,C\right) $ and $SU\left( 2,C\right) /U\left( 1\right) $. Second we use
this covariant formalism to study the $SL\left( 2,C\right) $ invariant
partition function of B model topological string theory on conifold. Third
we construct the manifestly $SL\left( 2,C\right) $ covariant correlation
functions of the scaling field of stringy wave function for $S^{3}$ quantum
cosmology of Gukov-Saraikin-Vafa.

\qquad To make a general idea on the content of this paper, recall that,
roughly, $\mathcal{Z}_{top}$ is a function of the moduli space of conifold
local complex deformations $\left\{ \mathrm{t}_{+n},\widetilde{\mathrm{t}}%
_{-n}\text{, ...}\right\} $, $n\geq 0$. The latters are generated by an
analytic function $\mathrm{t=t}\left( x,y,z,x\right) $ living on $T^{\ast
}S^{3}$ $\cite{31,131}$; and so $\mathcal{Z}_{top}$ can be thought of as a
functional living on complex three dimensional conifold; $\mathcal{Z}_{top}=%
\mathcal{Z}_{top}\left[ T^{\ast }S^{3}\right] $. Moreover like any function $%
F$ on this manifold , $\mathcal{Z}_{top}$ should sit in\ a reducible
representation of the $SL\left( 2,C\right) $ isometry of $T^{\ast }S^{3}$.
However, though the $SL\left( 2,C\right) $ representation group structure of
$\mathcal{Z}_{top}\left( T^{\ast }S^{3}\right) $ is quite known, its
explicit expression with manifest $SL\left( 2,C\right) $ symmetry is still
missing.\ Notice by the way that explicit relations with manifest $SL\left(
2,C\right) $ isometry of $\mathcal{Z}_{top}\left( T^{\ast }S^{3}\right) $
are important in the study of higher order corrections in complex moduli $%
\left\{ \mathrm{t}_{+n},\widetilde{\mathrm{t}}_{+n},\text{ ...}\right\} $;
they are equally useful for the study of the couplings of states involving
both momentum and winding modes. Recall also that when looking for explicit
results, the set of local complex deformations $\mathrm{t=t}\left(
x,y,z,w\right) $ of the conifold $xy-zw=\mu $ is in general restricted to:
\newline
(\textbf{a}) the particular subset of local complex deformations $\mathrm{%
t|=t}\left( x,y\right) \mathrm{|}=\mathrm{t}\left( x,y,0,0\right) $
generating a special class of conifold local complex deformations. More
precisely, perturbation by $\mathrm{t}\left( x,y\right) \mathrm{|}$ deals
with the local deformations of the complex one dimension submanifold $\left(
xy-zw=\mu \right) |_{z=w=0}=\mu $. Going beyond this technical restriction,
would be then an interesting task.\newline
(\textbf{b}) using an approximation in which local complex deformations are
split as $\mathrm{t}\left( x,y\right) \mathrm{|}=\tau \left( x\right) +%
\widetilde{\tau }\left( y\right) $ with Laurent expansion
\begin{equation}
\tau \left( x\right) =\sum_{n\geq 1}\mathrm{t}_{n}x^{n},\qquad \widetilde{%
\tau }\left( y\right) =\sum_{m\geq 1}\mathrm{t}_{-m}y^{m}.
\end{equation}%
In this approximation, couplings type $\tau \left( x\right) \times
\widetilde{\tau }\left( y\right) $ and higher orders in $\tau \left(
x\right) $ and $\widetilde{\tau }\left( y\right) $ are treated as small
fluctuations and so ignored. This procedure is a priori valid provided that
the conditions $\tau \left( x\right) <<\mu $ and $\widetilde{\tau }\left(
y\right) <<\mu $ are fulfilled. Under this approximation, couplings type
\begin{equation}
t_{n}t_{-m}x^{n}y^{m}
\end{equation}%
involving both $x$ and $y$ mode variables are then treated as higher order
corrections ($\tau \left( x\right) \widetilde{\tau }\left( y\right) \simeq
\mathcal{O}\left( 2\right) $). Though this method allows useful
simplifications; there is however a price to pay; it is a selective method
since only a part of the excitation modes associated with the deformation
monomials $x^{n}y^{m},$ \ $x^{n}z^{m},\ \ x^{n}w^{m},$ \ $y^{n}z^{m},$ \ $%
y^{n}w^{m}$ and\ $z^{n}w^{m}$ with $k=m+n$, that is
\begin{eqnarray*}
&&\sum_{m+n=k}t_{n}t_{-m}x^{n}y^{m},\text{\quad }%
\sum_{m+n=k}t_{n}s_{m}x^{n}z^{m},\quad \sum_{m+n=k}t_{n}s_{-m}x^{n}w^{m},%
\text{ \ } \\
&&\sum_{m+n=k}t_{-n}s_{-m}y^{n}z^{m},\quad
\sum_{m+n=k}t_{-n}s_{-m}y^{n}w^{m},\text{\quad }%
\sum_{m+n=k}s_{n}s_{-m}z^{n}w^{m},
\end{eqnarray*}%
is considered. From 2D $c=1$ string view, this approximation disregards not
only winding modes $\left( s_{n}z^{n}+s_{-m}w^{m}\right) $ and their
interactions $s_{n}s_{-m}z^{n}w^{m}$; but ignores as well the contributions
coming from couplings between the $t_{\pm n}$-$s_{\pm m}$ mode components of
the $SL\left( 2,C\right) $ spin $j=\frac{k}{2}$ multiplet. Notice that the
omitted zero mode $n=m=0$ in the expansion of $\tau \left( x\right) $ and $%
\widetilde{\tau }\left( y\right) $ corresponds just to the usual global
deformation parameter $\mu $ ($t_{0}=\mu $); while positivity of integers $n$
and $m$ is required by their interpretations as highest weights of $SL\left(
2,C\right) $ representations.

\qquad Motivated by these partial results, we address the question of
building a manifestly $SL\left( 2,C\right) $ harmonic formalism for the B
model topological string partition function $\mathcal{Z}_{top}\left[ T^{\ast
}S^{3}\right] $ and the HH wave function $\Psi \left[ S^{3}\right] $ of GSK
model. With this harmonic formalism at hand, one is able to study local
complex deformations of conifold and complete the literature partial
results. We show in particular that conifold local complex deformations and
manifestly $SL\left( 2,C\right) $ topological string computations, involving
both momenta and winding modes as well as the $SU\left( 2,C\right) $
covariant analysis of the correlation functions of GSK quantum cosmology
model get a natural formulation in harmonic space language.

\qquad Among our results, we mention the three following: In the
perturbative complex deformation approach, the locally deformed conifold is
given by
\begin{equation*}
u^{+\alpha }v_{\alpha }^{-}=\mu +\xi \left( u^{+},v^{-}\right) .
\end{equation*}%
The local complex deformations,\ captured by the harmonic function $\xi
\left( u^{+},v^{-}\right) $, can be expressed in terms of the irreducible
modes $\zeta ^{\pm n}$ living on $T^{\ast }S^{2}$ with harmonic expansions,%
\begin{equation*}
\zeta ^{+n}=\zeta _{\left( \alpha _{1}...\alpha _{n}\right) }u^{+\alpha
_{1}}...u^{+\alpha _{n}},\qquad \widetilde{\zeta }^{-n}=\widetilde{\zeta }%
^{\left( \beta _{1}...\beta _{n}\right) }v_{\beta _{1}}^{-}...v_{\beta
_{n}}^{-}\quad .
\end{equation*}%
They generate special infinitesimal local complex deformations using the
harmonic variables $u^{+\alpha }$ and $v_{\alpha }^{-}$ and the $SL\left(
2\right) $ tensors $\zeta _{\left( \alpha _{1}...\alpha _{n}\right) }$ and $%
\widetilde{\zeta }^{\left( \beta _{1}...\beta _{n}\right) }$. The use of the
harmonic sections $\zeta ^{\pm n}$ allows to avoid\ the technical
difficulties one encounters in the standard complex analysis and \ permits
to go beyond the $T^{\ast }S^{1}$ restriction currently used in literature.

\qquad For the B model topological string on conifold, the harmonic space
partition function preserving manifestly $SL\left( 2,C\right) $ symmetry
factorizes as usual as $\mathcal{Z}_{top}\left( \zeta ^{n},\widetilde{\zeta }%
^{m}\right) =\exp \left[ \mathcal{F}\left( \zeta ^{n},\widetilde{\zeta }%
^{m}\right) \right] $ with free energy
\begin{equation*}
\mathcal{F}=\sum_{g=0}^{\infty }\left( \frac{g_{s}}{\mu }\right) ^{2g-2}%
\mathcal{F}_{g}
\end{equation*}%
but instead of Laurent modes, we have now the harmonic sections $\zeta ^{\pm
n}$. For instance, we find that the leading terms of the genus zero free
energy read as,%
\begin{eqnarray}
\mathcal{F}_{0} &=&-\frac{1}{g_{s}^{2}}\sum_{n>0}\frac{\mu ^{n-2}}{n}%
\int_{T^{\ast }S^{2}}\left( \zeta ^{n}\widetilde{\zeta }^{-n}\right)   \notag
\\
&&+\frac{1}{g_{s}^{2}}\sum_{n_{1}+n_{2}+n_{3}=0}\mu ^{\frac{\left\vert
n_{1}\right\vert +\left\vert n_{2}\right\vert +\left\vert n_{3}\right\vert -2%
}{2}-2}\left( \int_{T^{\ast }S^{2}}\zeta ^{n_{1}}\zeta ^{n_{2}}\zeta
^{n_{3}}\right) +...\quad .  \label{i}
\end{eqnarray}%
In this relation, complex 3D conifold is defined as $u^{+\alpha }v_{\alpha
}^{-}=\mu $, with global complex parameter $\mu $. Restriction to $S^{3}$ is
recovered by setting $v_{\alpha }^{-}=u_{\alpha }^{-}$\ and $p=\func{Re}\mu $%
.

For GSK quantum cosmology model on $S^{3}$, we find that N points
correlation functions $G_{N}=$ $<\Phi \left( U_{1}^{\pm }\right) ...\Phi
\left( U_{N}^{\pm }\right) >$ of the conformal field factor $\Phi $ on the
3-sphere read as,%
\begin{equation}
G_{N}=\mathcal{N}\int D\mathrm{\zeta }D\overline{\mathrm{\zeta }}\left(
\dprod\limits_{i=1}^{N}\Phi _{i}\left( U_{i}^{\pm }\right) \right) \exp
\left( -\frac{1}{g_{s}^{2}}\mathcal{S}\left[ p,\mathrm{\zeta },\overline{%
\mathrm{\zeta }}\right] \right) \quad ,
\end{equation}%
with%
\begin{eqnarray}
\mathcal{S}\left[ p,\mathrm{\zeta },\overline{\mathrm{\zeta }}\right]
&=&\int_{S^{3}}\mathrm{\zeta }\frac{1}{D^{--}D^{++}+D^{++}D^{--}}\overline{%
\mathrm{\zeta }}  \notag \\
&&-\frac{1}{6p}\int_{S^{3}}\left( \mathrm{\zeta }^{2}\overline{\mathrm{\zeta
}}+\mathrm{\zeta }\overline{\mathrm{\zeta }}^{2}\right) +\mathcal{O}\left(
\frac{1}{p^{2}}\left( \mathrm{\zeta }\overline{\mathrm{\zeta }}\right)
^{2},...\right) \quad ,
\end{eqnarray}%
where $D^{--}=u^{-}\frac{\partial }{\partial u^{+}}$ and $D^{++}=u^{+}\frac{%
\partial }{\partial u^{-}}$ are harmonic derivatives to be given later. As
we see, our harmonic analysis permits to go beyond the standard computations
restricted to the $T^{\ast }S^{1}$\ conifold subvariety and also beyond the
usual analysis restricted to the large circle $x\overline{x}=\func{Re}\mu $
(with $z=0$) of $S^{3}$ parameterized by $x\overline{x}+z\overline{z}=\func{%
Re}\mu $.

\qquad The presentation of this work is as follows: In section 2, we make a
preliminary discussion by anticipating on some of the results of this paper;
this may help to make an idea on the method we will be using and the way we
want to do things. In section 3, we recall useful results on the ground ring
of the $c=1$ string theory and conformal deformations. In section 4, we
introduce harmonic analysis to study complex deformations of the conifold by
putting in front group theoretic properties of its $SL\left( 2,C\right) $
isometry group. The properties of the real slice of $T^{\ast }S^{3}$ are
recovered by imposing unimodularity condition. In section 5, we give the
classification of its complex deformations by using the harmonic frame work
and establish a dictionary giving correspondence between Fourier method on
the circle and harmonic analysis on 3-sphere. In section 6, we study the
partition function $\mathcal{Z}_{top}$ of B-model topological string on $%
T^{\ast }S^{3}$ by using harmonic frame work and derives its manifestly $%
SL\left( 2,C\right) $ expression. In section 7, we consider the $S^{3}$
quantum cosmology model within the harmonic coordinate set up. We study the
manifestly $SU\left( 2,C\right) $ invariant expression of Hartle-Hawking
wave function in the $S^{3}$ cosmology model of GSV. Then, we compute the
correlation functions of the conformal field on the full sphere $S^{3}$.
Section 8 is devoted to conclusion and in section 9 we give an appendix on
technical details regarding harmonic analysis on conifold and its
submanifolds $T^{\ast }S^{2}$ , $S^{3}$\ and $S^{2}$.

\section{Preliminaries and overview}

\qquad Before going into technical details, we think it is instructive to
anticipate this study by summarizing briefly the main results obtained in
this paper. This allows to fix the ideas on the harmonic variables method
and the way we will handle $\mathcal{Z}_{top}$ of B model topological string
on conifold and Hartlee-Hawking wave function on 3-sphere.

\qquad First of all, recall that harmonic analysis on the 2-sphere $S^{2}$
has been used successfully in the past, in particular in the study of the
manifestly off shell formulation of $4D$ $\mathcal{N}=2$ extended
supersymmetry $\cite{32}$-$\cite{36}$; $4D$ $\mathcal{N}=2$ supergravity $%
\cite{37}$-$\cite{40}$ and in the building of $4D$ Euclidean Yang Mills and
gravitational instantons $\cite{41}$-$\cite{47}$. Here, we shall go beyond
this formalism since we shall deal with harmonic analysis on conifold $%
T^{\ast }S^{3}$ and its connection with the quantum ground ring
\begin{equation}
\mathcal{V}=\mathcal{R}\otimes \overline{\mathcal{R}}
\end{equation}%
of 2D $c=1$ string theory $\cite{16,15}$; see also $\cite{48}$-$\cite{52}$.
The above mentioned Galperin \textit{et al} harmonic analysis on $S^{2}$
appears in our construction as a particular case. The point is that here we
are considering a large space namely $T^{\ast }S^{3}$; its compact real
slice $S^{3}$ is related to the 2-sphere through the usual fibration $%
S^{1}\times S^{2}$. Let us give below an overview on these tools; more
rigorous details will be given in forthcoming sections.

\subsection{Harmonic variables and 2D $c=1$\ string}

\subsubsection{Harmonic variables}

\qquad To approach $\mathcal{Z}_{top}\left( T^{\ast }S^{3}\right) $ and,
upon imposing reality condition, the Hartlee-Hawking wave function $\Psi
\left( S^{3}\right) $ we shall use the following two pairs of harmonic
variables
\begin{equation}
\left( U_{\alpha }^{+},U_{\alpha }^{-}\right) ,\qquad \alpha =1,2\quad ,
\label{b1}
\end{equation}%
and
\begin{equation}
\left( V_{\beta }^{+},V_{\beta }^{-}\right) ,\qquad \beta =1,2\quad .
\end{equation}%
Each pair transforms as an isodoublet under $SU\left( 2\right) $ and
parameterizes two separate complex spaces $C^{2}\sim R^{4}$. To fix the
ideas, we shall sometimes supplement the quantities living the u-space by an
extra sub-index and the same for the v-space quantities. For instance $%
C_{u}^{2}$ refers to the complex space parameterized by eq(\ref{b1}) and $%
SU_{u}\left( 2\right) $ the isometry group $SU\left( 2\right) $ rotating the
$U_{\alpha }^{\pm }$ variables. \newline
The $U_{\alpha }^{+}=\left( U_{1}^{+},U_{2}^{+}\right) $ and $V_{\beta
}^{-}=\left( V_{1}^{-},V_{2}^{-}\right) $ are four complex holomorphic
variables parameterizing $C^{4}$; the $U_{\alpha }^{-}=\overline{U^{+\alpha }%
}$ and $V^{+\beta }=\overline{V_{\beta }^{-}}$ are their complex conjugate.
Viewed collectively, these variables are subject to the homogeneous
constraint eq%
\begin{eqnarray}
U_{\alpha }^{+}\qquad &\longrightarrow &\qquad \lambda U_{\alpha }^{+}\quad ,
\notag \\
V_{\beta }^{-}\qquad &\longrightarrow &\qquad \frac{1}{\lambda }V_{\beta
}^{-},\qquad \lambda \in C^{\ast }\quad ,
\end{eqnarray}%
so that functions generated by monomials type
\begin{equation}
\dprod\limits_{i=0}^{n}U_{\alpha _{i}}^{+}V_{\beta _{i}}^{-}
\end{equation}%
are invariant under $C^{\ast }$ transformations. A simple example is given
by the projective complex three dimension surface%
\begin{equation}
\varepsilon ^{\alpha \beta }U_{\alpha }^{+}V_{\beta }^{-}=\mu ,
\end{equation}%
where $\mu $ is complex constant. Viewed separately, the $U_{\alpha }^{\pm }$
variable are restricted as%
\begin{eqnarray}
U_{\alpha }^{+}\qquad &\longrightarrow &\qquad e^{i\theta }U_{\alpha
}^{+}\quad ,  \notag \\
U_{\alpha }^{-}\qquad &\longrightarrow &\qquad e^{-i\theta }U_{\alpha
}^{-},\qquad \theta \in R\quad ,
\end{eqnarray}%
and the same thing for $V_{\alpha }^{\pm }$. In this case invariant
functions under the above $U\left( 1\right) $ symmetry are generated by
\begin{equation}
\dprod\limits_{i}U_{\alpha _{i}}^{+}U_{\beta _{i}}^{-}\quad .
\end{equation}%
Simple examples are given by the two following copies of $S^{3}$ spheres,%
\begin{eqnarray}
U^{+\alpha }U_{\alpha }^{-} &=&\left\vert U^{+1}\right\vert ^{2}+\left\vert
U^{+2}\right\vert ^{2}=p\quad ,  \notag \\
U^{\pm \alpha }U_{\alpha }^{\pm } &=&0\quad ,\qquad p=r_{1}^{2}\quad ,
\end{eqnarray}%
embedded in $C_{u}^{2}$ and
\begin{eqnarray}
V_{\beta }^{-}V^{+\beta } &=&\left\vert V_{1}^{-}\right\vert ^{2}+\left\vert
V_{2}^{-}\right\vert ^{2}=q\quad ,  \notag \\
V^{\pm \beta }V_{\beta }^{\pm } &=&0\quad ,\qquad q=r_{2}^{2}\quad ,
\end{eqnarray}%
embedded in $C_{v}^{2}$. The relations $U^{\pm \alpha }U_{\alpha }^{\pm }=0$
and $V^{\pm \beta }V_{\beta }^{\pm }=0$ reflect that $U_{\alpha }^{\pm }$
and $V_{\alpha }^{\pm }$ are commuting isodoublets. \newline
In this harmonic frame work, the usual conifold algebraic geometry equation $%
xy-zw=\mu $, with $x,y,z,w\in C$, takes the\ following $SL\left( 2,C\right) $
covariant form,%
\begin{equation}
U^{+\alpha }V_{\alpha }^{-}=\varepsilon ^{\alpha \beta }U_{\beta
}^{+}V_{\alpha }^{-}=\mu \quad ,\qquad \varepsilon ^{12}=-\varepsilon
^{21}=1\quad ,  \label{co}
\end{equation}%
where there is no place to the complex conjugate variables $V^{+\alpha }$
and $U_{\alpha }^{-}$ and where now the pair of holomorphic complex
variables $U^{+\alpha }$ and $V_{\alpha }^{-}$ form a $SL\left( 2,C\right) $
doublet,
\begin{equation}
\left(
\begin{array}{c}
U^{+} \\
V^{-}%
\end{array}%
\right) \quad .
\end{equation}%
The generators $\nabla ^{++}$, $\nabla ^{--}$ and $\nabla ^{0}=\left[ \nabla
^{++},\nabla ^{--}\right] $ of the $SL\left( 2,C\right) $ rotations of above
harmonic variables are given by%
\begin{eqnarray}
\nabla ^{++} &=&U^{+\alpha }\frac{\partial }{\partial V^{-\alpha }}\quad ,
\notag \\
\nabla ^{--} &=&V^{-\alpha }\frac{\partial }{\partial U^{+\alpha }}\quad , \\
\nabla ^{0} &=&\left( U^{+\alpha }\frac{\partial }{\partial U^{+\alpha }}%
-V^{-\alpha }\frac{\partial }{\partial V^{-\alpha }}\right) \quad .  \notag
\end{eqnarray}%
We also have the following typical relations,%
\begin{equation}
U^{+}=\nabla ^{++}\left( V^{-}\right) \quad ,\qquad V^{-}=\nabla ^{--}\left(
U^{+}\right) \quad .
\end{equation}%
A direct comparison between the usual conifold expression $xy-zw=\mu $ and
the harmonic one $U^{+\alpha }V_{\alpha }^{-}=\mu $ shows that
\begin{equation}
U^{+\alpha }=\left( x,z\right) \quad ,
\end{equation}%
and
\begin{equation}
V_{\alpha }^{-}=\left( y,w\right) \quad .
\end{equation}%
Moreover by setting $V_{\alpha }^{-}=\frac{p}{\mu }U_{\alpha }^{-}$, the
equation $U^{+\alpha }V_{\alpha }^{-}=\mu $ reduces to the 3-sphere $%
U^{+\alpha }U_{\alpha }^{-}=p$ while imposing $U^{+\alpha }=\frac{q}{\mu }%
V^{+\alpha }$ one falls on the second 3-sphere $V^{+\alpha }V_{\alpha
}^{-}=q $.

\subsubsection{Link to 2D $c=1$ string}

\qquad Beside the manifest $SL\left( 2\right) $ covariance, the harmonic
variables $U^{+\alpha }$ and $V_{\alpha }^{-}$ get a remarkable
interpretation in the $c=1$ string ground ring analysis of Witten $\cite{16}$%
. They coincide exactly with the basic conformal spin zero and ghost number
zero vertex operators $O_{\frac{1}{2},\frac{\pm 1}{2}}\times \overline{O}_{%
\frac{1}{2},\frac{\pm 1}{2}}$ generating the ground ring $\mathcal{V}=%
\mathcal{R}\otimes \overline{\mathcal{R}}$ of the 2D $c=1$ string theory
with non zero cosmological term. More precisely, we have the result,%
\begin{eqnarray}
U^{+\alpha } &\sim &O_{\frac{1}{2},\frac{+1}{2}}\times \overline{O}_{\frac{1%
}{2},\frac{\pm 1}{2}}\quad ,  \notag \\
V_{\alpha }^{-} &\sim &O_{\frac{1}{2},\frac{-1}{2}}\times \overline{O}_{%
\frac{1}{2},\frac{\mp 1}{2}}\quad ,
\end{eqnarray}%
where $\alpha =1,2$ ($\alpha =\frac{3}{2}\mp \frac{1}{2}$). Using this
correspondence and the $c=1$ string interpretation from $\cite{16,15}$, one
learns that the harmonic variables $U^{+\alpha }$ are associated with
positive left moving units of momenta at the $SU\left( 2\right) $ radius and
$V_{\alpha }^{-}$ with negative left moving ones. From the complex three
dimension conifold view, the upper components $U^{+1}$ and $V_{1}^{-}$ are
respectively associated with positive and negative momentum units while the
down components $U^{+2}$ and $V_{2}^{-}$\ are associated with positive and
negative winding mode units. This property shows that momentum and winding
units of same sign form $SU\left( 2,C\right) $\ doublets while momentum and
winding units are rotated under full $SL\left( 2,C\right) $\ symmetry.

\subsection{Conifold in harmonic framework}

\qquad One of the power of $SL\left( 2,C\right) $ harmonic analysis we are
developing here is that isometries of the 3-sphere and\ the deformed
conifold get simple realizations. For the conifold $U^{+\alpha }V_{\alpha
}^{-}=\mu $, the transformations\ generating isometries are given by the
product of two $SU\left( 2\right) $ symmetries, that is a full isometry
group given by $SU_{u}\left( 2\right) \times SU_{v}\left( 2\right) \sim
SL\left( 2,C\right) $.

\subsubsection{$SU_{u}\left( 2\right) $ isometry subgroup}

The first $SU\left( 2\right) $ isometry factor leaving $U^{+\alpha
}V_{\alpha }^{-}=\mu $\ invariant has a group parameter $\Lambda
^{++}=\Lambda ^{++}\left( U^{+},V^{-}\right) $ with the leading harmonic
expansion term,%
\begin{eqnarray}
\Lambda ^{++} &=&U_{(\alpha }^{+}U_{\beta )}^{+}\text{ }\Lambda ^{\left(
\alpha \beta \right) }+...  \notag \\
U_{(\alpha }^{+}U_{\beta )}^{+} &=&\frac{1}{2}\left( U_{\alpha }^{+}U_{\beta
}^{+}+U_{\beta }^{+}U_{\alpha }^{+}\right) \quad ,
\end{eqnarray}%
and acts on the harmonic $U^{+}$ and $V^{-}$ variables as
\begin{equation}
U^{+\alpha }\rightarrow U^{+\alpha \prime }=V^{-\alpha }\Lambda ^{++}\quad
,\qquad V_{\alpha }^{-}\rightarrow V_{\alpha }^{-\prime }=V_{\alpha
}^{-}\quad .  \label{ut}
\end{equation}%
As we see, this transformation leaves $V_{\alpha }^{-}$ invariant; and
because of the property $V^{-\alpha }V_{\alpha }^{-}=0$, it also leaves
invariant the conifold relation $U^{+\alpha }V_{\alpha }^{-}=\mu $. We also
have the relations%
\begin{equation}
\Lambda ^{++}=U^{+\alpha }U_{\alpha }^{+\prime }\quad ,\qquad U^{+\alpha
\prime }U_{\alpha }^{+\prime }=0\quad ,\qquad V_{\alpha }^{-\prime
}V^{-\alpha }=0\quad .
\end{equation}

\subsubsection{$SU_{v}\left( 2\right) $ isometry}

\qquad The second $SU\left( 2\right) $ isometry factor operates in the $V^{-}
$-space leaving the variable $U^{+}$ invariant. It has a group parameter $%
\Gamma ^{--}$ and acts as
\begin{equation}
U^{+\alpha }\rightarrow U^{+\alpha \prime \prime }=U^{+\alpha }\quad ,\qquad
V_{\alpha }^{-}\rightarrow V_{\alpha }^{-\prime \prime }+U_{\alpha
}^{+}\Gamma ^{--}\quad ,  \label{vt}
\end{equation}%
leaving invariant conifold defining equation since $U^{+\alpha }U_{\alpha
}^{+}=0$. Like in the harmonic superspace formulation of extended
supersymmetry $\cite{32}$, one discovers here also that reality of the group
parameters $\Lambda ^{++}$ and $\Gamma ^{--}$ should be thought of as
\begin{equation}
\Lambda ^{++}=\widetilde{\Lambda ^{++}}\quad ,\qquad \Gamma ^{--}=\widetilde{%
\Gamma ^{--}}\quad ,
\end{equation}%
where $\left( \sim \right) $ stands for a combination of the usual complex
conjugation $\left( -\right) $\ and the conjugation $\left( \ast \right) $
of the charge of the $U_{C}\left( 1\right) $ Cartan Weyl subgroup of $%
SU\left( 2\right) $.

\subsection{Partition function\text{\text{ }}$\mathcal{Z}_{top}\left(
T^{\ast }S^{3}\right) $}

\qquad In the usual explicit expression of the partition function of the B
model topological string on conifold, the function $\mathcal{Z}_{top}\left(
T^{\ast }S^{3}\right) $, with its standard factorisation,%
\begin{equation}
\mathcal{Z}_{top}\left( t\right) =\exp \left( \sum_{g=0}^{\infty }\left(
\frac{\mu }{g_{s}}\right) ^{2-2g}\mathcal{F}_{g}\left( t\right) \right)
\quad ,
\end{equation}%
is given by sums involving monomials $\dprod_{i=1}^{k}t_{n_{i}}%
\dprod_{j=1}^{l}\widetilde{t}_{-m_{j}}$ in the modes $t_{n}$ and $\widetilde{%
t}_{-n}$, ($n=2s\in \mathbb{N}^{\ast }$) of the local complex deformation
moduli $t=t\left( x,y,z,w\right) $ of the deformed conifold
\begin{equation}
xy-zw=\mu +t\left( x,y,z,w\right) \quad .
\end{equation}%
In case where $T^{\ast }S^{3}$ is restricted to $S^{3}$, the modes $%
\widetilde{t}_{-n}$ coincides with $\overline{t_{n}}=t_{-n}$; the complex
conjugates of $t_{n}$. These $t_{n}$ modes are generally taken as,%
\begin{equation}
t_{\pm n}=\int_{S^{1}}e^{-in\theta }t\left( x,\overline{x}\right) \quad
,\qquad n=2s=1,2,...\quad ,
\end{equation}%
where we have solved eq $xy-zw=\func{Re}\mu $ by setting
\begin{equation}
z=w=0
\end{equation}%
and taking
\begin{eqnarray}
x &=&\sqrt{\func{Re}\mu }\exp i\theta \quad ,  \notag \\
y &=&\sqrt{\func{Re}\mu }\exp \left( -i\theta \right) \quad .
\end{eqnarray}%
From the $c=1$ string ground ring view of $\cite{16,30}$, these moduli are
associated with the conformal vertex operators
\begin{equation}
W_{s,\pm s}^{+}=e^{\left[ \pm is\sqrt{2}X+\left( 1+s\right) \sqrt{2}\varphi %
\right] }\quad ,  \label{op}
\end{equation}%
involving tachyons $V_{s,\pm s}\left( X\right) \sim \exp \left( \pm is\sqrt{2%
}X\right) $ with discrete momentum $p_{s}$ fixed as $p_{s}=\pm s\sqrt{2}$ ($%
s=\frac{n}{2}$) together with the vertex operators $\exp \left( \left(
1+s\right) \sqrt{2}\varphi \right) $ based on Liouville field $\varphi $.
The ground ring $\mathcal{V}=\mathcal{R}\otimes \overline{\mathcal{R}}$ of
the $c=1$ string theory deals with discrete primary fields effectively
described by the vertex operators $V_{s,\pm n}\left( X\right) $ with
discrete momenta $p_{n}$ given by,%
\begin{equation}
p_{n}=\pm n\sqrt{2}\quad ,\qquad \left\vert n\right\vert \leq s-1\quad .
\end{equation}%
As we will see later, these operators which, from conifold view, may be
interpreted as describing couplings between momenta and winding modes, have
a natural description in the harmonic analysis. The previous Fourrier modes $%
t_{+n}$ and $t_{-n}$\ get respectively replaced by the homogeneous harmonic
functions
\begin{equation}
\zeta ^{+n}=U_{(\alpha _{1}}^{+}...U_{\alpha _{n})}^{+}\zeta ^{\left( \alpha
_{1}...\alpha _{n}\right) }\quad ,  \label{dev}
\end{equation}%
and
\begin{equation}
\zeta ^{-n}=U_{(\alpha _{1}}^{-}...U_{\alpha _{n})}^{-}\overline{\zeta }%
^{\left( \alpha _{1}...\alpha _{n}\right) }\quad ,  \label{devv}
\end{equation}%
living on the 2-sphere $S^{2}\sim S^{3}/S^{1}$. The point is that
infinitesimally, the 3-sphere described by $U^{+\alpha }U_{\alpha }^{-}=p$\
is deformed as
\begin{equation}
U^{+\alpha }U_{\alpha }^{-}=p+\xi \left( U^{+},U^{-}\right) \quad .
\end{equation}%
As a real function, the local complex deformation function $\xi $ reads as
\begin{equation}
\xi \left( U^{+},U^{-}\right) \sim \zeta \left( U^{+}\right) +\overline{%
\zeta }\left( U^{-}\right) +O\left( \left[ \zeta +\overline{\zeta }\right]
^{2}\right)
\end{equation}%
and may be Fourrier expanded, at first order in $\zeta $ and $\overline{%
\zeta },$ as
\begin{equation}
\xi \left( U^{+},U^{-}\right) \sim \sum_{n\geq 0}\left( e^{in\theta }\zeta
^{+n}+e^{-in\theta }\overline{\zeta }^{-n}\right) +\mathcal{O}\left( \left[
\zeta +\overline{\zeta }\right] ^{2}\right) \quad ,
\end{equation}%
where $\zeta ^{+n}$ and $\overline{\zeta }^{-n}$ are as in eqs(\ref{dev}-\ref%
{devv}). Note in passing that eq(\ref{dev}) reads more explicitly as,
\begin{eqnarray}
\zeta ^{+n}\left( U^{+}\right) &=&\left( U_{1}^{+}\right) ^{n}\text{ }\zeta
^{\left( 1...1\right) }+\left( U_{1}^{+}\right) ^{n-1}\left(
U_{2}^{+}\right) \text{ }\zeta ^{\left( 1...12\right) }+...  \notag \\
&&+\left( U_{1}^{+}\right) \left( U_{2}^{+}\right) ^{n-1}\text{ }\zeta
^{\left( 12...2\right) }+\left( U_{2}^{+}\right) ^{n}\text{ }\zeta ^{\left(
2...2\right) }\quad .
\end{eqnarray}%
By using the identification $x=U^{+1}$ and $z=U^{+2}$, one sees that there
is a $1\rightarrow 1$ correspondence between the $t_{\pm n}$\ Fourrier modes
and $\zeta ^{\pm n}$\ harmonic functions on $S^{2}$; and a $1\rightarrow
\left( n+1\right) $ correspondence between $t_{+n}$ and the multiplets $%
\zeta ^{\left( \alpha _{1}...\alpha _{n}\right) }$. The particular mode $%
\zeta ^{\left( 2...2\right) }$\ corresponds, up to a sign, to $t_{n}$; while
the remaining $\left( n\right) $ modes $\zeta ^{\left( \alpha _{1}...\alpha
_{n}\right) }$ with all $\alpha _{i}\neq 2$ have however no analogue in the
usual formulation. The reason is that in the standard approach, winding
modes are ignored. \newline
The correspondence between Fourrier analysis on $S^{1}$ and harmonic one on $%
S^{3}$ is in fact more general. For instance the integral on periodic
functions living on $S^{1}$,
\begin{equation}
\int_{S^{1}}\left[ t\left( \theta \right) \right] ^{2}\sim 2\sum_{n\geq
1}t_{n}\overline{t}_{-n}\quad ,
\end{equation}%
has a counterpart in harmonic analysis on the 3-sphere. We have%
\begin{equation}
\int_{S^{3}}\left[ \xi \left( u^{+},u^{-}\right) \right] ^{2}\sim
2\sum_{n\geq 1}\left( \int_{S^{2}}\zeta ^{+n}\overline{\zeta }^{-n}\right)
\quad ,
\end{equation}%
which, by integration on the harmonic variables of the 2-sphere using
harmonic integration rules given in appendix, we obtain,%
\begin{equation}
\int_{S^{3}}\left[ \xi \left( u^{+},u^{-}\right) \right] ^{2}\sim
2\sum_{n\geq 1}\frac{1}{\left( n+1\right) }\zeta ^{\left( \alpha
_{1}...\alpha _{n}\right) }\overline{\zeta }_{\left( \alpha _{1}...\alpha
_{n}\right) }\quad ,
\end{equation}%
where now $\zeta ^{\left( \alpha _{1}...\alpha _{n}\right) }$ are $SU\left(
2\right) $ tensors with spin $s=\left( n+1\right) $ and $\overline{\zeta }%
_{\left( \alpha _{1}...\alpha _{n}\right) }$ their complex conjugates. Note
that the coefficient $\frac{1}{\left( n+1\right) }$ in RHS of above eq shows
that in a $SL\left( 2\right) $ manifestly covariant formulation, each
component of the trace $\zeta ^{\left( \alpha _{1}...\alpha _{n}\right) }%
\overline{\zeta }_{\left( \alpha _{1}...\alpha _{n}\right) }$ contributes
with the same weight. As such, the monomials type $\dprod_{i=1}^{k}t_{n_{i}}%
\dprod_{j=1}^{l}\widetilde{t}_{-m_{j}}$, with $%
n_{1}+...+n_{k}=m_{1}+...+m_{l}$ which are involved in the explicit
expression of free energy $\mathcal{F}=\mathcal{F}\left( t,\widetilde{t}%
\right) $ of topological string B model restricted to the 3-sphere, should
be thought of as given by
\begin{equation}
\int_{S^{3}}\left[ \zeta \left( U^{+}\right) +\overline{\zeta }\left(
U^{-}\right) \right] ^{k}=\sum_{j=0}^{k}\frac{k!}{j!\left( k-j\right) !}%
\int_{S^{3}}\left[ \zeta \left( U^{+}\right) \right] ^{k-j}\left[ \overline{%
\zeta }\left( U^{-}\right) \right] ^{j}\quad ,
\end{equation}%
as required by the manifestly $SU\left( 2\right) $ invariant harmonic
analysis. In this manner, one can immediately determine the $SL\left(
2\right) $ covariant expression of the partition function
\begin{equation}
\mathcal{Z}_{top}\left( \zeta ,\widetilde{\zeta }\right) =\exp \left(
\sum_{g=0}^{\infty }\left( \frac{\mu }{g_{s}}\right) ^{2-2g}\mathcal{F}%
_{g}\left( \zeta ,\widetilde{\zeta }\right) \right) \quad .
\end{equation}%
The rule is as follows: (\textbf{i}) start from the usual expression of $%
\mathcal{Z}_{top}=\mathcal{Z}_{top}\left( t,\widetilde{t}\right) $, (\textbf{%
ii}) make the substitution
\begin{equation}
\left( t,\widetilde{t}\right) \qquad \longrightarrow \qquad \left( \zeta ,%
\widetilde{\zeta }\right) \quad ,
\end{equation}%
and (\textbf{iii}) perform traces on $SL\left( 2\right) $ representations.
The result we get is as in eqs(\ref{i}); for details see section 6.\newline
With this rule, one can extend all results obtained for the large circle $%
S^{1}$ of the 3-sphere to the whole points of $S^{3}$; see also theorem of
subsubsection 6.2.2.

Using the correspondence between Fourrier analysis on $S^{1}$ and the
harmonic one on $S^{3}$, we reconsider in section 7 the cosmological toy
model of GSV; in particular the computation of the manifestly $SU\left(
2\right) $ covariant $N$ points Green functions
\begin{equation}
G_{N}=<\Phi _{1}...\Phi _{N}>\quad ,
\end{equation}%
of the conformal factor $\Phi _{i}$ describing the correlations of the
fluctuations of the $S^{3}$ real slice of the conifold $T^{\ast }S^{3}$. In
particular, we show that quantum correlations are generated by the
functional,
\begin{equation}
\mathcal{Z}\left[ J,\overline{J}\right] =\mathcal{N}\int_{\mathcal{M}}D\zeta
D\overline{\zeta }\text{ }\exp \left( -\frac{1}{g_{s}^{2}}\mathcal{S}\left[
p,\zeta ,\overline{\zeta }\right] +\int_{S^{3}}\left( J\xi +\overline{J}%
\overline{\zeta }\right) \right) \quad ,
\end{equation}%
where $g_{s}$ is the usual string coupling, $\sqrt{p}$ is the radius of the
3-sphere $S^{3}$ described by $U^{+\alpha }U_{\alpha }^{-}=p$ and $\mathcal{M%
}$\ the moduli space of local complex deformations of conifold. In this
relation the field variable
\begin{equation}
\xi =\xi \left( S^{3}\right) \simeq \zeta +\overline{\zeta }\quad ,
\end{equation}%
describes the infinitesimal complex deformations of the 3-sphere and $%
\mathcal{S}\left[ p,\zeta ,\overline{\zeta }\right] $ is given by the
following functional.%
\begin{equation}
\mathcal{S}\left[ p,\zeta ,\overline{\zeta }\right] =\int_{S^{3}}\overline{%
\zeta }\frac{2}{D^{++}D^{--}+D^{--}D^{++}}\zeta -\frac{1}{3p}%
\int_{S^{3}}\left( \zeta +\overline{\zeta }\right) ^{3}+O\left( \frac{1}{%
p^{2}}\xi ^{4}\right) \quad ,
\end{equation}%
where $D^{++},D^{--}$ and $\left[ D^{++},D^{--}\right] =D^{0}$ generate the $%
SU\left( 2,C\right) $ algebra and where the term $O\left( \frac{1}{p^{2}}\xi
^{4}\right) $ stands for higher fluctuations field variable interactions.
Besides the fact that $\xi $ describes small fluctuations around $p$, the
above mentioned approximation may be also justified in the $S^{3}$ cosmology
model where the volume of $S^{3}$ is supposed large ($p\rightarrow \infty $).

In the end of this presentation, we would like to add that in harmonic frame
work, special features on topological theory becomes manifest. For instance,
the explicit rederivation of topological Chern Simons gauge theory from the
gauging of conifold $C^{\ast }$ symmetry $\cite{131}$ is one of these
features. An other example concerns conifold local complex deformation
parameter $\xi $. A careful analysis shows that the right way to express the
local conifold equation is as $\xi =\nabla ^{++}\xi ^{--}+\nabla ^{--}\xi
^{++}$ where now there is no zero mode $\mu $, a property which is not
exactly fulfilled in above discussion. If we substitute above $\xi $ as $%
\nabla ^{++}\xi ^{--}+\nabla ^{--}\xi ^{++}$, one ends with the%
\begin{equation}
\int_{S^{3}}\xi ^{--}\xi ^{++}\text{\quad ,}
\end{equation}%
instead of $\int_{S^{3}}\overline{\zeta }\frac{2}{\left\{
D^{++},D^{--}\right\} }\zeta $ where apparent non localities disappear. We
leave these details and others to the developments of forthcoming sections.

\section{\textbf{c=1 string theory and Conifold geometry}}

\qquad In the first part of this section, we review briefly 2D target space $%
c=1$ string theory; in particular its quantum ground ring of the discrete
primary states at the $SU\left( 2\right) $ radius and its connection with
conifold geometry. In the second part, we study particular aspects of the
conifold $T^{\ast }S^{3}$ with special focus on its local complex
deformations by using standard complex analysis. This description is useful
to fix ideas and to make contact with the harmonic frame work introduced in
section 2 and which we shall develop further in section 4.

\subsection{Discrete primaries in 2D $c=1$ string theory}

\qquad As noted before, string theory with two dimensional target space time
is characterized by the existence of a special set of discrete primary
fields generating three types of ground rings denoted here below as $%
\mathcal{R}$ for chiral ring, $\overline{\mathcal{R}}$ for antichiral one
and $\mathcal{V}=\mathcal{R}\otimes \overline{\mathcal{R}}$ for their
combination.

\subsubsection{2D $c=1$\ string}

\qquad To get an idea on the structure of these ground rings, it is
interesting to start by recalling that the world sheet action $S\left[
X,\varphi ,h\right] =S$ describing 2D $c=1$ string theory is,%
\begin{eqnarray}
S &=&\frac{1}{4\pi \alpha ^{\prime }}\int d^{2}\sigma \sqrt{h}\left(
h^{ij}\partial _{i}X\partial _{j}X+h^{ij}\partial _{i}\varphi \partial
_{j}\varphi -2\sqrt{\alpha ^{\prime }}\varphi R^{\left( 2\right) }\right)
\notag \\
&&+\mathrm{\mu }\int d^{2}\sigma \sqrt{h}\exp \left( -\frac{2}{\sqrt{\alpha
^{\prime }}}\varphi \right) \quad ,  \label{ac1}
\end{eqnarray}%
where $X\left( \sigma \right) =X\left( \sigma ^{0},\sigma ^{1}\right) $ is
the matter scalar field, $h\left( \sigma \right) $ is the world sheet
metric, $R^{\left( 2\right) }\left( \sigma \right) $ the Ricci scalar, $%
\varphi \left( \sigma \right) $ the usual Liouville field and $\mathrm{\mu }$
the cosmological constant. The total holomorphic energy momentum tensor $%
T=T\left( X,\varphi ,c,b\right) $ of the quantum theory including $\left\{
b,c\right\} $ ghost system is,
\begin{equation}
T=T\left( X\right) +T\left( \varphi \right) +T_{ghost}\left( c,b\right)
\quad ,
\end{equation}%
with
\begin{eqnarray}
T\left( X\right)  &=&-\frac{1}{2}\left( \partial X\right) ^{2}\quad ,  \notag
\\
T\left( \varphi \right)  &=&-\frac{1}{2}\left( \partial \varphi \right)
^{2}+\varphi \sqrt{2}\quad , \\
T_{ghost}\left( c,b\right)  &=&-2b\partial c-\left( \partial b\right) c\quad
.  \notag
\end{eqnarray}%
One of the interesting things regarding this 2D quantum field theory is that
along with the standard vertex operators
\begin{equation}
V_{p}\left( X\right) =e^{ipX}\quad ,\qquad p\in R\quad ,
\end{equation}%
of conformal weight $h=p^{2}/2$, there are additional discrete primary
fields
\begin{equation}
V_{n/\sqrt{2}}\left( X\right) ,\qquad n\text{\ an }\func{integer}\text{,}
\end{equation}%
that are involved in the building of the above mentioned ground rings. These
$V_{n/\sqrt{2}}$ conformal states belong to representations of the
holomorphic (resp. antiholomorphic) $SU\left( 2\right) $ current algebra
generated by the field operators
\begin{equation}
T^{\pm }=e^{\pm iX\sqrt{2}}\quad ,\qquad T^{0}=i\sqrt{2}\partial X\quad ,
\end{equation}%
\ (resp. $\overline{T^{\pm }}$ and $\overline{T^{0}}$). As required by the $%
SU\left( 2\right) $ compactification condition $X\rightarrow X+2\pi $, the
momenta $p$ of the conformal vertex $V_{p}\left( X\right) $ is quantized as
\begin{equation}
p=\frac{n}{\sqrt{2}}\quad ,
\end{equation}%
with $n$ an integer. Viewed from the $SU\left( 2\right) $ holomorphic
current algebra, the $V_{n/\sqrt{2}}\left( X\right) $ field operators are
generally ranged into $SU\left( 2\right) $ spin $s$ representation with
highest weight state
\begin{equation}
V_{s,s}\left( X\right) =e^{is\sqrt{2}X}\quad .
\end{equation}%
This field operator satisfy the following OPE relations,
\begin{eqnarray}
T^{+}\left( z_{1}\right) V_{s,s}\left( X\left( z_{2}\right) \right)  &\sim
&0+\text{ regular terms}\quad \text{,}  \notag \\
T^{0}\left( z_{1}\right) V_{s,s}\left( X\left( z_{2}\right) \right)  &\sim &%
\frac{2s}{z_{1}-z_{2}}V_{s,s}\left( X\left( z_{2}\right) \right) +\text{
regular terms}\quad \text{,}
\end{eqnarray}%
where we have set $\left\vert n\right\vert =2s$, that is $s=0,\frac{1}{2}%
,1,...$ and, for simplicity, we re-named the vertex operators $V_{\pm s\sqrt{%
2}}\left( X\right) $ as $V_{s,\pm s}$ with first index $s$ referring to the $%
SU\left( 2\right) $ spin and the second one, namely $\pm s$, to momentum. In
this view, momenta are also the eigenvalues of the charge operator $T^{0}$.
By repeatedly acting with lowering operators, we get a set of $2s+1$
discrete conformal fields
\begin{equation}
\left\{ V_{s,n}\left( X\right) ,\text{ }0\leq \left\vert n\right\vert \leq
s\right\} \quad ,
\end{equation}%
forming an $su\left( 2\right) $ spin $s$ multiplet having a conformal
dimension $h=s^{2}$. The special vertices with $\left\vert n\right\vert =s$
namely%
\begin{equation}
V_{s,\pm s}\left( X\right) =e^{\pm is\sqrt{2}X}\quad ,
\end{equation}%
are standard tachyon operators with momenta $p=\pm s\sqrt{2}$ while the
remaining others $V_{s,n}$ are new objects having momenta $\left\vert
n\right\vert <s$. For the particular cases $s=0$ and $s=\frac{1}{2}$, the
corresponding vertex operators are associated with the identity operator
\begin{equation}
V_{0,0}=I\quad ,
\end{equation}%
and the two component spinor vertex $V_{\frac{1}{2},\pm \frac{1}{2}}$ given
by,%
\begin{equation}
V_{\frac{1}{2},+\frac{1}{2}}=e^{+\frac{i}{\sqrt{2}}X}\quad ,\qquad V_{\frac{1%
}{2},-\frac{1}{2}}=e^{-\frac{i}{\sqrt{2}}X}\quad .
\end{equation}%
They have conformal weights $h_{1/2}=\frac{1}{4}$.

\subsubsection{Quantum field operators}

\qquad The new $V_{s,n}$ states when coupled to the operators $V_{\omega
}\left( \varphi \right) =e^{i\omega \varphi }$ with conformal dimension $%
h_{\varphi }=\left( \omega ^{2}/2+i\omega \sqrt{2}\right) $ allow to build
spin one primary fields $W_{s,n}^{+}=W_{s,n}^{+}\left( X,\varphi \right) $.
Using the vertex operators $e^{\left( 1\pm s\right) \varphi }$ of conformal
dimension
\begin{equation}
h_{\varphi }=2\left( 1\pm s\right) -\left( 1\pm s\right) ^{2}=1-s^{2}\quad ,
\end{equation}%
it is clear that the following holomorphic operators,%
\begin{eqnarray}
W_{s,n}^{+} &=&V_{s,n}\times e^{\left( 1+s\right) \sqrt{2}\varphi }\quad
,\qquad \omega _{+}=\frac{1+s}{i}\quad ,  \notag \\
W_{s,n}^{-} &=&V_{s,n}\times e^{\left( 1-s\right) \sqrt{2}\varphi }\quad
,\qquad \omega _{-}=\frac{1-s}{i}\quad ,
\end{eqnarray}%
have conformal weights $\left( h,\overline{h}\right) =\left( 1,0\right) $
and momentum $\left( n\sqrt{2},\frac{1+s}{i}\sqrt{2}\right) $ and $\left( n%
\sqrt{2},\frac{1-s}{i}\sqrt{2}\right) $ respectively. Note in passing that
\begin{equation}
W_{0,0}^{\pm }=e^{\sqrt{2}\varphi }\quad ,
\end{equation}%
it is just the cosmological vertex operator involved in the action (\ref{ac1}%
) and,
\begin{equation}
W_{\frac{1}{2},\frac{1}{2}}^{-}=\exp \left( \frac{i\left( X+i\varphi \right)
}{\sqrt{2}}\right) \quad ,\qquad W_{\frac{1}{2},\frac{-1}{2}}^{-}=\exp
\left( \frac{-i\left( X-i\varphi \right) }{\sqrt{2}}\right) \quad ,
\label{w}
\end{equation}%
are basic operators whose role will be exhibited later on. True quantum
field operators are obtained by paring holomorphic sector with the
antiholomorphic one. Such a pairing may be achieved however in two manners.
First by using standard combination
\begin{equation}
Z_{s,n,m}^{\pm }=W_{s,n}^{\pm }\times \overline{W}_{s,m}^{\pm }\quad ,
\end{equation}%
of the $\left( h,\overline{h}\right) =\left( 1,0\right) $ holomorphic
operators $W_{s,n}^{\pm }$ with the corresponding $\left( h,\overline{h}%
\right) =\left( 0,1\right) $ antiholomorphic ones $\overline{W}_{s,n}^{\pm }$%
. These vertex operators have conformal weights $\left( h,\overline{h}%
\right) =\left( 1,1\right) $ and are used in the study of infinitesimal
deformations of above CFT$_{2}$.

The second class concerns making true quantum field operators $J_{s,n,m}$
and $\overline{J}_{s,n,m}$ by combining holomorphic and antiholomorphic
quantities, but still with conformal weights $\left( h,\overline{h}\right)
=\left( 1,0\right) $ and $\left( h,\overline{h}\right) =\left( 0,1\right) $
respectively. More precisely, for the case of the specific $\left(
1,0\right) $ operators $J_{s,n,m}$ (resp $\left( 0,1\right) $ operators $%
\overline{J}_{s,n,m}$) generating a Lie algebra of symmetries, we have to
combine $W_{s,n}^{+}$ (resp $\overline{W}_{s,n}^{+}$) with conformal spin
zero fields $\overline{O}_{s-1,n}$ (resp.$O_{s-1,n}$) as shown below,%
\begin{eqnarray}
J_{s,n,m} &=&W_{s,n}^{+}\times \overline{O}_{s-1,m}\quad ,  \notag \\
\overline{J}_{s,n,m} &=&O_{s-1,n}\times \overline{W}_{s,m}^{+}\quad .
\end{eqnarray}%
In these relations, the operators $O_{s-1,n}$\ and their antiholomorphic
counterpart are the BRST partners of
\begin{equation}
Y_{s,n}^{+}=\mathrm{c\times }W_{s,n}^{+}\quad ,\qquad \left\vert
n\right\vert <s\quad ,
\end{equation}%
which are conformal operators with spin zero and ghost number $G=1$. Indeed
starting from the vertex operators $W_{s,n}^{\pm }$ (resp $\overline{W}%
_{s,n}^{\pm }$) and using the ghost fields $\mathrm{b=b}_{zz}$ and $\mathrm{%
c=c}^{z}$ of spins $2$ and $-1$, we can build the conformal spin zero and
ghost number $G=1$ operators $Y_{s,n}^{\pm }$ as follows%
\begin{equation}
Y_{s,n}^{\pm }=\mathrm{c\times }W_{s,n}^{\pm }\quad ,\qquad \left\vert
n\right\vert <s\quad .
\end{equation}%
Cohomology of BRST operator $Q$ shows that the $Y_{s,n}^{\pm }$\ operators
should have partners $QY_{s,n}^{\pm }$ with ghost numbers $G^{\prime }=G\pm
1 $. Following $\cite{16}$, $Y_{s,n}^{+}$ field operators have partners $%
O_{u,n}$ with $u=s-1$ and momenta $\left\vert n\right\vert \leq u$ at ghost
number $G=0$ and $Y_{s,n}^{-}$ have partners at $G=2$. For the leading value
$s=1$ ($u=0$), the partner $O_{0,0}$ of $Y_{1,0}^{+}$\ is just the identity
operator $I$. For the next value $s=\frac{3}{2}$, we have two discrete
primaries $Y_{\frac{3}{2},\frac{1}{2}}^{+}$ and $Y_{\frac{3}{2},-\frac{1}{2}%
}^{+}$ of spin zero and ghost number $G=1$. The corresponding spin zero and
ghost number $G=0$ operators $O_{\frac{1}{2},\frac{1}{2}}$ and $O_{\frac{1}{2%
},\frac{-1}{2}}$ are given by,
\begin{eqnarray}
O_{\frac{1}{2},\frac{1}{2}} &=&\left( \mathrm{cb}+\frac{i}{\sqrt{2}}\left(
\partial X-i\partial \varphi \right) \right) W_{\frac{1}{2},\frac{1}{2}%
}^{-}\quad ,  \notag \\
O_{\frac{1}{2},\frac{-1}{2}} &=&\left( \mathrm{cb}-\frac{i}{\sqrt{2}}\left(
\partial X+i\partial \varphi \right) \right) W_{\frac{1}{2},\frac{-1}{2}%
}^{-}\quad ,
\end{eqnarray}%
where $W_{\frac{1}{2},\frac{1}{2}}^{-}$ and $W_{\frac{1}{2},\frac{-1}{2}%
}^{-} $ are as in eq(\ref{w}). The left moving momenta $\left( n\sqrt{2},iu%
\sqrt{2}\right) $ of these spin zero BRST invariant operators $O_{\frac{1}{2}%
,\frac{\pm 1}{2}}$ are same as for the operators $W_{\frac{1}{2},\frac{\pm 1%
}{2}}^{-}$ namely $\left( \frac{\pm \sqrt{2}}{2},\frac{i\sqrt{2}}{2}\right) $%
. Note that similar relations may be also written down for the operators $%
\overline{O}_{\frac{1}{2},\frac{\pm 1}{2}}$ involving antiholomorphic sector
objects. Note moreover that the spin zero BRST invariant operators $O_{u,n}$
and $\overline{O}_{u,n}$ generate commutative and associative ground rings.
The chiral (resp antichiral) ground ring $\mathcal{R}$ ($\overline{\mathcal{R%
}}$) is generated by the operators $O\equiv O_{u,n}$ (resp $\overline{O}=%
\overline{O}_{u,n}$) and has a multiplication law given by the short
distance product,%
\begin{equation}
O\left( z_{1}\right) O^{\prime }\left( z_{2}\right) \sim O^{\prime \prime
}\left( z_{2}\right) +\left\{ Q,F\right\} \quad ,
\end{equation}%
where, like for left hand side, the right hand side of this relation is also
BRST invariant. Along with these $\mathcal{R}$ and $\overline{\mathcal{R}}$
ground rings, we have as well the ground ring $\mathcal{V}$ combining left
and right movers. It is generated by the field operators,
\begin{equation}
V_{u,n,m}=O_{u,n}\times \overline{O}_{u,m}\quad ,
\end{equation}%
with a similar multiplication law given by multiplying the left and right
moving parts separately. Because of the specific properties of the $O_{\frac{%
1}{2},\frac{\pm 1}{2}}$ and $\overline{O}_{\frac{1}{2},\frac{\pm 1}{2}}$,
the basic generators of the above quantum ring $\mathcal{V}$ are given by
the following fundamental objects,%
\begin{eqnarray}
x &=&O_{\frac{1}{2},\frac{1}{2}}\times \overline{O}_{\frac{1}{2},\frac{1}{2}%
}\quad ,  \notag \\
y &=&O_{\frac{1}{2},\frac{-1}{2}}\times \overline{O}_{\frac{1}{2},-\frac{1}{2%
}}\quad ,  \notag \\
z &=&O_{\frac{1}{2},\frac{1}{2}}\times \overline{O}_{\frac{1}{2},-\frac{1}{2}%
}\quad , \\
w &=&O_{\frac{1}{2},-\frac{1}{2}}\times \overline{O}_{\frac{1}{2},\frac{1}{2}%
}\quad ,  \notag
\end{eqnarray}%
obeying the obvious relation
\begin{equation}
xy-zw=0\quad ,
\end{equation}%
with $x,y,z,w\in C$. As these variable operators turn out to play an
important role in the present study, it is interesting to have in mind the
following group theoretic property,%
\begin{equation}
\begin{tabular}{|l|l|l|l|l|}
\hline
operators & x & y & z & w \\ \hline
spin $\left( s,\overline{s}\right) $ & $\left( \frac{1}{2},\frac{1}{2}%
\right) $ & $\left( \frac{1}{2},\frac{1}{2}\right) $ & $\left( \frac{1}{2},%
\frac{1}{2}\right) $ & $\left( \frac{1}{2},\frac{1}{2}\right) $ \\ \hline
Cartan Charge $\left( n,m\right) $ & $\left( 1,1\right) $ & $\left(
-1,-1\right) $ & $\left( 1,-1\right) $ & $\left( -1,1\right) $ \\ \hline
\end{tabular}%
\quad \text{,}
\end{equation}%
where $\left( s,\overline{s}\right) $ stand for the spin representation of
the $SU\left( 2\right) \times SU\left( 2\right) $ symmetry and where $\left(
n,m\right) $ are the values of the (integer) charges of the Cartan-Weyl
operators $T^{0}$ and $\overline{T^{0}}$. Forgetting about the fact that the
pairs of complex variables $\left( x,z\right) $ and $\left( y,w\right) $ are
related by complex conjugation and considering complex deformations of the $%
c=1$ string theory by the conformal spin $\left( 1,1\right) $ operators,
\begin{equation}
Z^{+}=\sum_{s,n,m}W_{s,n}^{+}\times \overline{W}_{s,m}^{+}\quad ,
\end{equation}%
the previous three dimensional quadric $xy-zw=0$ get corrected as follows
\begin{equation}
xy-zw=f\left( x,y,z,w\right) \quad ,
\end{equation}%
where now $f=f\left( x,y,z,w\right) $ is a priori a general polynomial
transforming under $SU\left( 2\right) \times SU\left( 2\right) $ as
\begin{equation}
\dbigoplus\limits_{2s,2\overline{s}=0}^{\infty }\left( s,\overline{s}\right)
\quad .
\end{equation}%
For the particular case where the deformation parameter $f$ is a constant,
one has an interesting field theoretic interpretation. The constant $f$
corresponds just to the cosmological constant $\mu $\textrm{\ }($f=\mu $)
transforming as a $SU\left( 2\right) \times SU\left( 2\right) $ singlet.\
Therefore at $\mu \neq 0$, the quantum ground ring of the 2D $c=1$ string
theory at the $SU\left( 2\right) $ point is given by the ring of polynomials
on conifold $T^{\ast }S^{3}$%
\begin{equation}
xy-zw=\mu \quad .  \label{1}
\end{equation}%
The generic case where $f$ is an arbitrary local function is discussed below.

\subsection{Geometry of the ground ring}

\qquad From the point of view of the algebraic geometry of the three
dimension conifold, the parameter $\mu $ is a complex deformation of the
conic singularity $xy-zw=0$. It is a globally defined quantity since it is
independent of the local coordinates of $T^{\ast }S^{3}$; i.e,%
\begin{equation}
\frac{\partial \mu }{\partial x^{i}}=0,\qquad x^{i}=x,y,z,w\quad .
\end{equation}%
To exhibit more clearly the isometries of the conifold geometry, it is
interesting to think about the ambient complex four space $C^{4}$ where
lives $T^{\ast }S^{3}$ as the set of complex $2\times 2$ matrices $\mathcal{M%
}\left( 2,C\right) $. In this representation, the complex holomorphic vector
$\left( x,y,z,w\right) $ is now parameterized as
\begin{equation}
X=\left(
\begin{array}{cc}
x & w \\
z & y%
\end{array}%
\right) \quad ,  \label{aA}
\end{equation}%
and the conifold defining eq(\ref{1}) takes then the remarkable form%
\begin{equation}
\det X=\mu \quad .  \label{a0}
\end{equation}%
Clearly, this relation has a manifest $GL\left( 2,C\right) \sim GL\left(
1,C\right) \times SL\left( 2,C\right) $ automorphism symmetry acting,
through arbitrary matrices $M$ of $GL\left( 2,C\right) $, as follows,
\begin{equation}
X\rightarrow MXM^{-1}\quad ,\qquad M\in GL\left( 2,C\right) \quad ,
\label{a1}
\end{equation}%
Note that along with above eqs (\ref{1}, \ref{a0}), there are also other
remarkable representations of the conifold. One of the realizations of this
quadric is that obtained by the change of holomorphic variables $\left(
x,y,z,w\right) $ as $x=\left( x_{1}+ix_{2}\right) ,$ $y=\left(
x_{1}-ix_{2}\right) ,$ $z=i\left( x_{3}+ix_{4}\right) $ and $w=i\left(
x_{3}-ix_{4}\right) $, $x_{i}\in C$. With this holomorphic change\ the
algebraic geometry equation (\ref{1}) reads then as
\begin{equation}
x_{1}^{2}+x_{2}^{2}+x_{3}^{2}+x_{4}^{2}=\sum_{i=1}^{4}x_{i}^{2}=\mu \quad .
\label{2}
\end{equation}%
Instead of $SL\left( 2,C\right) $ invariance\footnote{%
The $GL\left( 1,\mathbb{C}\right) $ abelian subsymmetry of$\ GL\left( 2,%
\mathbb{C}\right) $ acts trivially of the complex holomorphic variables x,
y, z and w.} of eq(\ref{a0}), this equation has rather a manifest $O\left(
4,C\right) $ symmetry group rotating the holomorphic variables as $%
x_{i}^{\prime }=O_{ij}x_{i}$ with $O_{ij}\in C$ and $O_{ij}O_{jk}=\delta
_{ik}$. Recall also that like in the case of real rotations, here also the
complex holomorphic $O\left( 4,C\right) $ group splits as the product of the
holomorphic group product $O\left( 3,C\right) \times O\left( 3,C\right) $.
The usual trigonometric functions sine and cosine of the real group $O\left(
3,R\right) $ get now replaced by in the $O\left( 3,C\right) $ as shown below,%
\begin{equation}
\sin \vartheta _{j}\rightarrow \frac{1}{2i}\left( \lambda _{j}-\frac{1}{%
\lambda _{j}}\right) \quad ,\qquad \cos \vartheta _{j}\rightarrow \frac{1}{2}%
\left( \lambda _{j}+\frac{1}{\lambda _{j}}\right) \quad ,
\end{equation}%
where the $\lambda _{j}$'s are now non zero complex number. To fix the
ideas, we give here below the example of the $O\left( 2,C\right) $ symmetry
rotating the $x_{1}$ and $x_{2}$ variables of the following complex
holomorphic relation,
\begin{equation}
x_{1}^{2}+x_{2}^{2}=\nu \quad ,  \label{aa}
\end{equation}%
where $\nu $\ is some non zero complex number. In this case the $2\times 2$
matrices $O$ leaving (\ref{aa}) invariant reads as%
\begin{equation}
O=\frac{1}{2}\left(
\begin{array}{cc}
\lambda +\frac{1}{\lambda } & \frac{i}{\lambda }-i\lambda \\
i\lambda -\frac{i}{\lambda } & \lambda +\frac{1}{\lambda }%
\end{array}%
\right) \quad ,
\end{equation}%
with general features as,%
\begin{equation}
\det O=\pm 1\quad ,\qquad OO^{T}=O^{T}O=I\quad .
\end{equation}%
Here the $O\left( 2,C\right) $ group parameter $\lambda $, which may be also
split by using Euler representation of complex number as $\left\vert \lambda
\right\vert \exp i\vartheta $, is a non zero complex number. Using the
change of holomorphic variables $x_{1}=\left( x+y\right) /2,$ $x_{2}=\left(
x-y\right) /2i,$ eq(\ref{aa}) reads also as%
\begin{equation}
xy=\nu \quad .
\end{equation}%
This relation is invariant under $GL\left( 1,C\right) \sim C^{\ast }$
transformation generated by the scaling $x\rightarrow \lambda x$ and $%
y\rightarrow y/\lambda $ with non zero parameter $\lambda =\varrho
_{1}+i\varrho _{2}$. From this description, one clearly see that the
holomorphic $SO\left( 2,C\right) $ group is then just $GL\left( 1,C\right)
\sim C^{\ast }$; it should be thought of as the complex generalization of
the familiar isomorphism $U\left( 1\right) \sim SO\left( 2,R\right) $
dealing with real compact form associated with $\left\vert \lambda
\right\vert ^{2}=\varrho _{1}^{2}+\varrho _{2}^{2}=1$. For the non compact
case where the group parameters $\varrho _{1}$ and $\varrho _{2}$ are such
that $\varrho _{1}^{2}-\varrho _{2}^{2}=1$, we have the hyperbolic symmetry $%
SO\left( 1,1;R\right) $. This presentation extends naturally to higher
dimensional holomorphic orthogonal groups $O\left( n,C\right) $ and their
representations.

Moreover, using the factorisation $SO\left( 4,C\right) \simeq SL\left(
2,C\right) \times \widetilde{SL\left( 2,C\right) }$ allowing to express the
complex\footnote{%
For the real forms of$\ SO\left( 4,\mathbb{C}\right) $, we have the
homomorphisms $SO\left( 4,\mathbb{R}\right) \sim SU\left( 2,\mathbb{C}%
\right) \times SU\left( 2,\mathbb{C}\right) $, $SO\left( 1,3;\mathbb{R}%
\right) \sim SL\left( 2,\mathbb{C}\right) $ and $SO\left( 2,2;\mathbb{R}%
\right) \sim SL\left( 2,\mathbb{R}\right) \times SL\left( 2,\mathbb{R}%
\right) $.} holomorphic 4-vector $x_{i}$ as a holomorphic bispinor $%
x^{\alpha \overline{\beta }}$, the conifold eq(\ref{1}) can be also
rewritten as
\begin{equation}
\varepsilon _{\alpha \gamma }\varepsilon _{\overline{\beta }\overline{\delta
}}x^{\alpha \overline{\beta }}x^{\gamma \overline{\delta }}=x^{\alpha
\overline{\beta }}x_{\alpha \overline{\beta }}=\mu \quad ,\qquad \varepsilon
_{\alpha \gamma }=-\varepsilon _{\gamma \alpha },\qquad \varepsilon
^{12}=\varepsilon _{21}=1\quad ,  \label{ha}
\end{equation}%
where the sum over repeated indices is understood. For later use and also to
have more insight into these complex relations, we give here below two
realizations of eq(\ref{1}) using first free complex holomorphic coordinates
$\left( z_{1},z_{2},z_{3}\right) $ ($z_{i}\neq 0$) and, by help of Euler
representation of complex numbers, the corresponding real six dimensional
parameterization involving three non compact coordinates $\left(
r_{1},r_{2},r_{3}\right) $ and three compact ones $\left( \theta _{1},\theta
_{2},\theta _{3}\right) $. The free complex holomorphic coordinates
representation solving the conifold eq(\ref{1}) reads as,%
\begin{eqnarray}
x &=&\frac{\sqrt{\mu }}{2i}z_{1}^{\frac{1}{2}}\left( z_{3}^{\frac{1}{2}%
}-z_{3}^{-\frac{1}{2}}\right) \quad ,  \notag \\
y &=&\frac{\sqrt{\mu }}{2i}z_{1}^{-\frac{1}{2}}\left( z_{3}^{\frac{1}{2}%
}-z_{3}^{-\frac{1}{2}}\right) \quad ,  \notag \\
z &=&-\frac{\sqrt{\mu }}{2}z_{2}^{\frac{1}{2}}\left( z_{3}^{\frac{1}{2}%
}+z_{3}^{-\frac{1}{2}}\right) \quad , \\
w &=&\frac{\sqrt{\mu }}{2}z_{2}^{-\frac{1}{2}}\left( z_{3}^{\frac{1}{2}%
}+z_{3}^{-\frac{1}{2}}\right) \quad ,  \notag
\end{eqnarray}%
where the involved square roots $\sqrt{z_{i}}$ are dictated by boundary
conditions of spinors. Under the loop transformation $z_{i}\rightarrow
z_{i}\exp 2i\pi $, $\sqrt{z_{i}}\rightarrow -\sqrt{z_{i}}$ but the conifold
variables $x,$ $y,$ $z$ and $w$ remain invariant. The check of this
representation as a free solution of eq(\ref{1})\ follows directly from the
identity
\begin{equation}
\left( z_{3}^{\frac{1}{2}}+z_{3}^{-\frac{1}{2}}\right) ^{2}-\left( z_{3}^{%
\frac{1}{2}}-z_{3}^{-\frac{1}{2}}\right) ^{2}=4z_{3}^{\frac{1}{2}}z_{3}^{-%
\frac{1}{2}}=4.
\end{equation}%
By setting the three free complex holomorphic variables as $z_{i}=\left\vert
z_{i}\right\vert \exp i\theta _{i}$ , $\left\vert z_{i}\right\vert \equiv
r_{i}^{2}$, with $\theta _{i}$ angles chosen as
\begin{equation}
\theta _{1}=\left( \psi -\varphi \right) \quad ,\qquad \theta _{2}=\left(
\psi +\varphi \right) \quad ,\qquad \theta _{3}=\vartheta \quad ,
\end{equation}%
where $\psi $, $\varphi $ and $\vartheta $ are the usual three angle used in
the parameterization of the real three sphere, we get the following
realization of the conifold,%
\begin{eqnarray}
x &=&\frac{\sqrt{\mu }}{2i}r_{1}\left( r_{3}e^{\frac{i\vartheta }{2}}-\frac{1%
}{r_{3}}e^{\frac{-i\vartheta }{2}}\right) e^{\frac{i}{2}\left( \psi -\varphi
\right) }\quad ,  \notag \\
y &=&\frac{\sqrt{\mu }}{2ir_{1}}\left( r_{3}e^{\frac{i\vartheta }{2}}-\frac{1%
}{r_{3}}e^{\frac{-i\vartheta }{2}}\right) e^{\frac{-i}{2}\left( \psi
-\varphi \right) }\quad ,  \notag \\
z &=&-\frac{\sqrt{\mu }}{2}r_{2}\left( r_{3}e^{\frac{i\vartheta }{2}}+\frac{1%
}{r_{3}}e^{\frac{-i\vartheta }{2}}\right) e^{\frac{i}{2}\left( \psi +\varphi
\right) }\quad ,  \label{3} \\
w &=&\frac{\sqrt{\mu }}{2r_{2}}\left( r_{3}e^{\frac{i\vartheta }{2}}+\frac{1%
}{r_{3}}e^{\frac{-i\vartheta }{2}}\right) e^{\frac{-i}{2}\left( \psi
+\varphi \right) }\quad .  \notag
\end{eqnarray}%
For physical interpretation, it is interesting to set,%
\begin{equation}
r_{1}=\left( \varrho +\frac{1}{\varrho }\right) ,\qquad r_{2}=\left( \sigma +%
\frac{1}{\sigma }\right) ,
\end{equation}%
where together with $r_{3}$, the real variables $\varrho $ and $\sigma $
parameterize the three real non compact $T^{\ast }S^{3}$ dimensions. The
real compact variables $\vartheta ,\varphi ,\psi $, which vary in $\left[
0,4\pi \right] $, parameterize $S^{3}$; that is the real slice of the
conifold. Note that besides $SL\left( 2\right) $ manifest symmetry, the
conifold equation (\ref{1}) has moreover discrete symmetries. For instance
eq(\ref{1}) remains invariant under the following discrete change,
\begin{eqnarray}
x &\rightarrow &x^{\prime }=\left( -\right) ^{k+1}x\quad ,\qquad
y\rightarrow y^{\prime }=\left( -\right) ^{k+1}y\quad ,  \notag \\
z &\rightarrow &z^{\prime }=\left( -\right) ^{k}z\quad ,\qquad w\rightarrow
w^{\prime }=\left( -\right) ^{k}w\quad ,  \label{4}
\end{eqnarray}%
with $k$ an arbitrary integer. In the realization (\ref{3}), this
corresponds to perform the change%
\begin{eqnarray}
\kappa &\rightarrow &\kappa ^{\prime }=\frac{1}{\kappa }\quad ,\qquad
\varrho \rightarrow \varrho ^{\prime }=\frac{1}{\varrho }\quad ,  \notag \\
\sigma &\rightarrow &\sigma ^{\prime }=\frac{1}{\sigma }\quad ,\qquad
\vartheta \rightarrow \vartheta ^{\prime }=-\vartheta +2k\pi \quad ,
\label{5}
\end{eqnarray}%
where one recognizes the usual $T$ duality transformation. According to
whether $k$ is odd integer or even integer, the 1-cycles $\left( \psi
-\varphi \right) $ and $\left( \psi +\varphi \right) $ are respectively
fixed under the change (\ref{5}). Note also that $S^{3}$ is recovered from
eqs(\ref{3}) by setting $\kappa =\varrho =\sigma =1$ and $\chi =0$.%
\begin{eqnarray}
x &=&\sqrt{p}e^{\frac{i}{2}\left( \psi -\varphi \right) }\sin \left( \frac{%
\vartheta }{2}\right) \quad ,  \notag \\
y &=&\sqrt{p}e^{\frac{-i}{2}\left( \psi -\varphi \right) }\sin \left( \frac{%
\vartheta }{2}\right) \quad ,  \notag \\
z &=&-\sqrt{p}e^{\frac{i}{2}\left( \psi +\varphi \right) }\cos \left( \frac{%
\vartheta }{2}\right) \quad ,  \label{6} \\
w &=&\sqrt{p}e^{\frac{-i}{2}\left( \psi +\varphi \right) }\cos \left( \frac{%
\vartheta }{2}\right) \quad .  \notag
\end{eqnarray}%
In fact the condition $\kappa =\varrho =\sigma =1$ is required by $SU\left(
2\right) $ unimodularity which demands the identifications $y=\overline{x}$
and $w=-\overline{z}$ so that the matrix $X$ of eq(\ref{aA}) takes the form,
\begin{equation}
X=\left(
\begin{array}{cc}
x & -\overline{z} \\
z & \overline{x}%
\end{array}%
\right) \quad ,  \label{a5}
\end{equation}%
and then eq(\ref{1}) reads as $\det X=\left\vert x\right\vert
^{2}+\left\vert z\right\vert ^{2}=p$. In group theoretic language, the
automorphism symmetry leaving $\det X$ invariant corresponds to the
restriction $M^{-1}=M^{+}$ in eq(\ref{a1}); that is the group reduction%
\begin{equation}
SL\left( 2\right) \qquad \rightarrow \qquad SU\left( 2\right) \quad ,
\label{a6}
\end{equation}%
and the abelian complex subgroup $C^{\ast }$ contained in $SL\left( 2\right)
$ into the usual Cartan Weyl subgroup $U_{C}\left( 1\right) $ of $SU\left(
2\right) $. Observe also that by setting $\vartheta =\pi $ and $\psi
-\varphi =2\theta $, the 3-sphere reduces to its large large circle $%
\left\vert x\right\vert ^{2}=p$.

\subsection{Local deformations and symmetry group representations}

\qquad In the limit $\mu \rightarrow 0$, the 3-sphere $\left\vert
x\right\vert ^{2}+\left\vert z\right\vert ^{2}=\func{Re}\mu $ shrinks to a
point and the conifold $xy-zw=0$ becomes singular. Complex deformations
lifting this conic singularity are generally obtained by perturbing the
previous relation as
\begin{equation}
xy-zw=F\left( x,y,z,w\right) \quad .
\end{equation}%
According to the values of $F$, one distinguishes global deformations
captured by the zero mode $\mu =F\left( x=0,...,w=0\right) $ and local
deformations carried by%
\begin{equation}
\mathrm{T}\left( x,y,z,w\right) =F\left( x,y,z,w\right) -\mu \quad .
\end{equation}%
The function $\mathrm{T}\left( x,y,z,w\right) $ is a priori an arbitrary
holomorphic function on $T^{\ast }S^{3}$ generated by homogeneous monomials
type $x^{n_{1}}y^{n_{2}}z^{n_{3}}w^{n_{4}}$ as shown below,%
\begin{equation}
\mathrm{T}\left( x,y,z,w\right)
=\sum_{n,m}\sum_{n_{1}+n_{2}=n}%
\sum_{n_{3}+n_{4}=m}T_{n_{1},-n_{2},n_{3},-n_{4}}\text{ }%
x^{n_{1}}y^{n_{2}}z^{n_{3}}w^{n_{4}}\quad .  \label{7}
\end{equation}%
By grouping monomials with same homogeneous degrees, one can show that the
above development may be rearranged in terms of highest weight
representations of $SL\left( 2,C\right) $ conifold isometry. Observe in
passing that there are various kinds of $SU\left( 2,C\right) $ ($SL\left(
2,C\right) $) subgroups within the homogeneity group $SO\left( 8,R\right) $
of the real ambient space $R^{8}\sim C^{4}$ where lives the conifold. These
group symmetries have a nice description in the harmonic frame work eqs(\ref%
{ut},\ref{vt}). Let us discuss below some of these $SO\left( 8,R\right) $
subgroups.

\subsubsection{$SU\left( 2,C\right) \times \widetilde{SU\left( 2,C\right) }$}

First of all note that one distinguishes two special $SU\left( 2\right) $
subgroups of $SO\left( 8,R\right) $ which deal with the real slice of the
conifold. They are generated by the operator sets $\left\{ J_{0},J_{\pm
}\right\} $ and $\left\{ \widetilde{J_{0}},\widetilde{J_{\pm }}\right\} $
operating in $x-z$ and $y-w$ complex planes respectively. For the first set,
we have the geometric realization,%
\begin{eqnarray}
J_{+} &=&x\frac{\partial }{\partial \overline{z}}-z\frac{\partial }{\partial
\overline{x}}\quad ,\qquad \left( J_{+}\right) ^{\dagger }=J_{-}\quad ,
\notag \\
J_{-} &=&\overline{z}\frac{\partial }{\partial x}-\overline{x}\frac{\partial
}{\partial z}\quad ,\qquad \left( J_{0}\right) ^{\dagger }=J_{0}\quad ,
\label{m1} \\
J_{0} &=&\left( x\frac{\partial }{\partial x}+z\frac{\partial }{\partial z}%
\right) -\left( \overline{x}\frac{\partial }{\partial \overline{x}}+%
\overline{z}\frac{\partial }{\partial \overline{z}}\right) \quad ,  \notag
\end{eqnarray}%
and for the second the following one,%
\begin{eqnarray}
\widetilde{J}_{+} &=&y\frac{\partial }{\partial \overline{w}}-z\frac{%
\partial }{\partial \overline{w}}\quad ,\qquad \qquad \left( \widetilde{J}%
_{+}\right) ^{\dagger }=\widetilde{J}_{-}\quad ,  \notag \\
\widetilde{J}_{-} &=&\overline{w}\frac{\partial }{\partial y}-\overline{y}%
\frac{\partial }{\partial w}\quad ,\qquad \qquad \left( \widetilde{J}%
_{0}\right) ^{\dagger }=\widetilde{J}_{0}\quad , \\
\widetilde{J}_{0} &=&\left( y\frac{\partial }{\partial y}+w\frac{\partial }{%
\partial w}\right) -\left( \overline{y}\frac{\partial }{\partial \overline{y}%
}+\overline{w}\frac{\partial }{\partial \overline{w}}\right) \quad .  \notag
\end{eqnarray}%
These eqs are interchanged by making the substitution $\left( x,z\right)
\longleftrightarrow \left( y,w\right) $.

\subsubsection{$SL\left( 2,C\right) \times \widetilde{SL\left( 2,C\right) }$}

There are also two others $SL\left( 2,C\right) $ subgroups of $SO\left(
8,R\right) $ described by the sets $\left\{ L_{0},L_{\pm }\right\} $ and $%
\left\{ \widetilde{L_{0}},\widetilde{L_{\pm }}\right\} $; they generate the
holomorphic $SL\left( 2,C\right) $ and $\widetilde{SL\left( 2,C\right) }$
groups involved in the expansion (\ref{7}). Their geometric realizations
read as follows,%
\begin{equation}
L_{+}=x\frac{\partial }{\partial z}\quad ,\qquad L_{-}=z\frac{\partial }{%
\partial x}\quad ,\qquad L_{0}=\left( x\frac{\partial }{\partial x}-z\frac{%
\partial }{\partial z}\right) \quad ,  \label{m2}
\end{equation}%
and similar relations for $\left\{ \widetilde{L_{0}},\widetilde{L_{\pm }}%
\right\} $ by making the substitution $\left( x,z\right) \rightarrow \left(
y,w\right) $, that is
\begin{equation}
\widetilde{L_{+}}=y\frac{\partial }{\partial w}\quad ,\qquad \widetilde{L_{-}%
}=w\frac{\partial }{\partial y}\quad ,\qquad \widetilde{L_{0}}=\left( y\frac{%
\partial }{\partial y}-w\frac{\partial }{\partial w}\right) \quad ,
\end{equation}%
Under these groups, the homogeneous degree $n$ polynomials
\begin{eqnarray}
P_{n}\left( x,z\right) &=&\sum_{j=1}^{n}a_{j}x^{n-j}z^{j}\quad ,  \notag \\
\widetilde{P}_{n}\left( y,w\right) &=&\sum_{j=1}^{n}a_{j}y^{n-j}w^{j},
\end{eqnarray}%
with positive integer $n$ transform as spin $s=\frac{n}{2}$ ($\widetilde{s}=%
\frac{n}{2}$) representations of $SL\left( 2,C\right) $ ($\widetilde{%
SL\left( 2,C\right) }$). From this point of view, the development (\ref{7})
can be viewed as an expansion of local complex deformations on the complete
set of the irreducible representations $\left( \underline{\mathbf{s}}\mathbf{%
,}\widetilde{\underline{\mathbf{s}}}\right) $ of $SL\left( 2\right) \times $
$\widetilde{SL\left( 2\right) }$ as shown below%
\begin{eqnarray}
\dbigoplus\limits_{2s+1,2\widetilde{s}+1=1}^{\infty }\left( \underline{%
\mathbf{s}}\mathbf{,}\widetilde{\underline{\mathbf{s}}}\right) &=&\left(
0,0\right) \quad \oplus \quad \left( \frac{1}{2},0\right) \quad \oplus \quad
\left( 0,\frac{1}{2}\right)  \notag \\
&&\oplus \quad \left( 1,0\right) \quad \oplus \quad \left( \frac{1}{2},\frac{%
1}{2}\right) \quad \oplus \quad \left( 0,1\right) \\
&&\oplus \quad \left( \frac{3}{2},0\right) \quad \oplus \quad ...\quad .
\notag
\end{eqnarray}%
The leading singlet $\left( 0,0\right) $ corresponds just to the global
modulus $\mu $ and remaining others to local deformations giving the
parameters of $Diff\left( T^{\ast }S^{3}\right) $, the group of holomorphic
diffeomorphisms of the conifold.

\subsubsection{$SL\left( 2,C\right) _{diag}$}

There is moreover an other remarkable holomorphic $SL\left( 2,C\right) $
isometry group of the conifold. It is generated by transformations that mix
the variables $\left( x,z\right) \ $and $\left( y,w\right) $,%
\begin{eqnarray}
K_{+} &=&x\frac{\partial }{\partial w}-z\frac{\partial }{\partial y}\quad
,\qquad \qquad K_{-}=w\frac{\partial }{\partial x}-y\frac{\partial }{%
\partial z}\quad ,  \notag \\
K_{0} &=&\left( x\frac{\partial }{\partial x}+z\frac{\partial }{\partial z}%
\right) -\left( y\frac{\partial }{\partial y}+w\frac{\partial }{\partial w}%
\right) \quad .
\end{eqnarray}%
It is this symmetry group which mostly concerns us in the present study.
Upon identifying $y=\overline{x}$ and $w=\overline{z}$, the conifold reduces
to its $S^{3}$ lagrangian submanifold and $SL\left( 2,C\right) $ isometry
down to its $SU\left( 2,C\right) $ subgroup, eqs(\ref{m1}).

\subsection{Reduction to complex one dimension}

Generally speaking, the hypersurface $xy-zw=\mu $ represents a particular
non compact Calabi-Yau threefold. Typical geometries embedded in $C^{4}$
which are used in practice have in general the form%
\begin{equation}
G\left( z,w,x,y\right) =zw-H\left( y,x\right) =0\quad ,
\end{equation}%
with $H\left( y,x\right) $ some bi-holomorphic function. These geometries
have a conical Ricci flat metric and a holomorphic $\left( 3,0\right) $\
form $\Omega $ \ which can be chosen as%
\begin{equation}
\Omega \simeq \frac{dz}{z}\wedge dy\wedge dx\quad ,
\end{equation}%
where we have used $\partial G/\partial w=z,$ or equivalently as $\frac{dw}{w%
}\wedge dy\wedge dx$ by using $\left( w,y,x\right) $ as the local
coordinates. A general expression of this 3-form keeping touch with the $%
z\leftrightarrow w$ permutation symmetry is obviously given by a linear
combination type,%
\begin{equation}
\Omega \left( a,b\right) =\left( \frac{adz+bdw}{az+bw}\right) \wedge
dy\wedge dx\quad ,
\end{equation}%
where the pole singularities are located at $\left( z,w\right) =\left( \pm
b,\mp a\right) $ and where $a$ and $b$ are two constants; one of them should
be different from zero ($ab\neq 0$). Forgetting for a while about this
detail and focus on the way this holomorphic 3-form is handled in
topological string B-model on non compact Calabi-Yau threefolds.

\subsubsection{Rational curve $zw=H$}

\qquad In dealing with $\Omega $, one generally fixes the dependence in $z$
and $w$ and consider only perturbations of the function $H\left( y,x\right) $%
. In this manner, the problem reduces essentially to one complex dimension
since the Calabi-Yau threefold is viewed as a fibration over the $\left(
x,y\right) $\ plane with fiber given by the rational curve $zw=H\left(
y,x\right) $. By using Cauchy's theorem in $z$-plane, we can perform a
partial integration of the period,%
\begin{equation}
\int_{A}\Omega =\int_{\gamma _{z}\times D}\Omega \quad ,
\end{equation}%
with 3-cycles $A$ given by $\gamma _{z}\times D$ (or in general $A=\left(
\gamma _{z}\times D\right) \cup \left( \gamma _{w}\times D^{\prime }\right) $%
). We end with the complex structure\
\begin{equation}
\int_{D}dy\wedge dx\quad ,
\end{equation}%
given by the integral of the holomorphic 2-form $dy\wedge dx$ on the 2-cycle
$D$. The CY3 is then reduced to the holomorphic curve $\Sigma $ given by
\begin{equation}
H\left( y,x\right) =0\quad ,
\end{equation}%
with analytic domain $D$ and boundary $\partial D\subset \Sigma $. By
stockes theorem, the integral $\int_{D}d\left( ydx\right) $ can be reduced
further to
\begin{equation}
\int_{\partial D}ydx\quad ,
\end{equation}%
showing that the complex deformation of the holomorphic curve $H\left(
y,x\right) $ is controlled by the 1-form $ydx$. In the case of conifold we
are interested in here and where
\begin{equation}
H\left( y,x\right) =yx-\mu \quad ,
\end{equation}%
the above 1-form reads then as $\mu \frac{dx}{x}$. Note that the
perturbation of $H\left( y,x\right) $ by the local deformations $\tau \left(
x\right) =\sum_{n>0}t_{n}x^{n}$, periods of $ydx$ around the 1-cycle $\gamma
_{x}$ are given by the modes $t_{n}$,%
\begin{equation}
t_{n}\sim \int_{\gamma _{x}}\frac{dx}{x}x^{-n}\tau \left( x\right) \quad
,\qquad \gamma _{x}=\partial _{x}D\quad .  \label{tn}
\end{equation}%
These complex deformation moduli transform under the projective change $%
x\rightarrow \lambda x$ and $y\rightarrow \frac{1}{\lambda }y$ as $%
t_{n}\rightarrow \lambda ^{-n}t_{n}$. Note that what we have done for the
holomorphic variable $x$ may be equally done for the dual variable $y$.
Instead of eq(\ref{tn}), one has rather
\begin{equation}
\widetilde{t}_{-n}\sim \int_{\gamma _{y}}\frac{dy}{y}y^{-n}\widetilde{\tau }%
\left( y\right) \quad .
\end{equation}%
To have both of modes $t_{n}$ and $\widetilde{t}_{-n}$, one should have a
local complex deformation type $H\left( y,x\right) =\mu +\tau \left(
x\right) +\widetilde{\tau }\left( y\right) $ letting understand that $\tau
\left( x\right) $ and $\widetilde{\tau }\left( y\right) $ are just the two
leading perturbation terms of a two variable holomorphic function
\begin{equation}
t\left( x,y\right) =\tau \left( x\right) +\widetilde{\tau }\left( y\right) +O%
\left[ \tau \widetilde{\tau },\tau ^{2},\widetilde{\tau }^{2}\right] \quad .
\label{txy}
\end{equation}%
where $O\left[ \tau \widetilde{\tau },\tau ^{2},\widetilde{\tau }^{2}\right]
$ stands for higher perturbation orders.

\subsubsection{Beyond eq(\protect\ref{txy})}

One may also consider perturbations involving, in addition to $x$ and $y$,
the $z$ and $w$ variables as well. This is particularly interesting for the
conifold case where the two holomorphic variable isodoublets $\left(
x,z\right) \equiv u^{\alpha }$ and $\left( y,w\right) \equiv v_{\alpha }$
play a symmetric role. In this case the conifold equation reads as
\begin{equation}
u^{\alpha }v_{\alpha }=\mu \quad ,
\end{equation}%
and the holomorphic 3-form invariant under $SL\left( 2\right) $ isometry
group may be written as,%
\begin{eqnarray}
\Omega &=&\frac{1}{2}\frac{\left( \varepsilon _{\alpha \beta }\eta ^{\alpha
}du^{\beta }\right) }{\eta .u}\wedge \left( dv_{\gamma }\wedge dv_{\delta
}\varepsilon ^{\gamma \delta }\right)  \notag \\
&=&\frac{\eta _{1}dx+\eta _{2}dz}{\eta _{1}x+\eta _{2}z}\wedge dy\wedge
dw\quad ,
\end{eqnarray}%
where $\eta _{\alpha }=\left( \eta _{1},\eta _{2}\right) $ is a constant
isodoublet and where the scalar $\eta .u$ stands for $SL\left( 2,C\right) $
invariant product
\begin{equation}
\eta _{\delta }u^{\delta }=\varepsilon _{\delta \gamma }u^{\delta }\eta
^{\gamma }\quad .
\end{equation}%
The point $u_{\delta }=\eta _{\delta }$, corresponding to $x=\eta ^{1}=\eta
_{2}$ and $z=\eta ^{2}=-\eta _{1}$, is a pole singularity of $\Omega $.
Performing a partial integration of the period $\int_{A}\Omega $ by using
Cauchy's theorem in the plane
\begin{equation}
z^{\prime }=\eta _{1}x+\eta _{2}z\quad ,
\end{equation}%
one gets the restriction $u_{\delta }=\eta _{\delta }$, $\eta ^{\alpha
}v_{\alpha }=\mu $ and ends with
\begin{equation}
\int_{D}\left( dv_{\gamma }\wedge dv_{\delta }\right) \varepsilon ^{\gamma
\delta }\quad ,
\end{equation}%
describing the complex structure of the holomorphic 2-form $\left(
dv_{\gamma }\wedge dv_{\delta }\right) \varepsilon ^{\gamma \delta }$ on the
2-cycle $D$. By stockes theorem, this integral can be also brought to
\begin{equation}
\int_{\partial D}\left( v_{\gamma }dv_{\delta }\right) \varepsilon ^{\gamma
\delta }\quad ,
\end{equation}%
showing that part of complex deformations of the conifold is controlled by
the 1-form
\begin{equation}
\left( v_{\gamma }dv_{\delta }\right) \varepsilon ^{\gamma \delta }\quad ,
\end{equation}%
with $\eta ^{\alpha }v_{\alpha }=\mu $. Solving the constraint eq $\eta
^{\alpha }v_{\alpha }=\mu $ as
\begin{equation}
\eta ^{\alpha }=\frac{\mu }{\theta .v}\theta ^{\gamma }\quad ,
\end{equation}%
one may rewrite the complex structure as follows,
\begin{equation}
\frac{\left( \theta _{\gamma }dv_{\delta }\right) \varepsilon ^{\gamma
\delta }}{\theta .v}=\frac{d\left( \theta .v\right) }{\theta .v}=\frac{%
\left( \theta _{1}dv_{2}-\theta _{2}dv_{1}\right) }{\theta _{1}v_{2}-\theta
_{2}v_{1}}\quad ,
\end{equation}%
where $\theta ^{\alpha }=\left( \theta ^{1},\theta ^{2}\right) $ is a
constant isospinor. In the $\left( x,y,z,w\right) $ complex holomorphic
variable language, the local complex deformations read infinitesimally as
\begin{equation}
T\left( x,y,z,w\right) \simeq \tau \left( x\right) +\widetilde{\tau }\left(
y\right) +\upsilon \left( z\right) +\widetilde{\upsilon }\left( w\right)
\quad ,
\end{equation}%
with the mode expansions,%
\begin{eqnarray}
\tau \left( x\right) &=&\sum_{n>0}t_{n}x^{n},\qquad \widetilde{\tau }\left(
y\right) =\sum_{n>0}t_{-n}y^{n}\quad ,  \notag \\
\upsilon \left( x\right) &=&\sum_{n>0}s_{n}z^{n},\qquad \widetilde{\upsilon }%
\left( y\right) =\sum_{n>0}s_{-n}w^{n}\quad .  \label{ab2}
\end{eqnarray}%
Up on thinking about $y$ and $w$ as $y\sim \frac{1}{x}$ and $z\sim \frac{1}{w%
}$, eqs(\ref{ab2}) can be put into the following simplest form,%
\begin{eqnarray}
\tau \left( x\right) +\widetilde{\tau }\left( \frac{1}{x}\right) &\sim
&t\left( x\right) =\sum_{n\neq 0}t_{n}x^{n}\quad ,  \notag \\
\upsilon \left( z\right) +\widetilde{\upsilon }\left( \frac{1}{z}\right)
&\sim &s\left( x\right) =\sum_{n\neq 0}s_{n}z^{n}\quad .  \label{ab3}
\end{eqnarray}%
The holomorphic functions $\tau ,$ $\widetilde{\tau }$, $\upsilon $ and $%
\widetilde{\upsilon }$ generate the leading terms of the conifold local
complex deformations. The expansion (\ref{7}) describing the full set of
local complex deformations of $T^{\ast }S^{3}$ can be thought of as,%
\begin{equation}
T\left( x,y,z,w\right) =\tau \left( x\right) +\widetilde{\tau }\left(
y\right) +\upsilon \left( z\right) +\widetilde{\upsilon }\left( w\right) +%
\mathcal{O}\left[ higher\right] \quad ,  \label{ab4}
\end{equation}%
where the higher terms are given by couplings generated by monomials type
\begin{equation}
\left[ \tau \left( x\right) \right] ^{m_{1}}\left[ \widetilde{\tau }\left(
y\right) \right] ^{m_{2}}\left[ \upsilon \left( z\right) \right] ^{m_{3}}%
\left[ \widetilde{\upsilon }\left( w\right) \right] ^{m_{4}}\quad ,
\label{ab5}
\end{equation}%
with $m_{i}$ positive integers. The simplest non linear terms are given by
the bilinears
\begin{equation}
\tau \left( x\right) \widetilde{\tau }\left( y\right)
=\sum_{n,m>0}t_{n}t_{-m}x^{n}y^{n}\quad ,
\end{equation}%
and
\begin{equation}
\upsilon \left( z\right) \widetilde{\upsilon }\left( w\right)
=\sum_{n,m>0}s_{n}s_{-m}z^{n}w^{n}\quad .
\end{equation}%
More details on conifold local complex deformations and applications for 2D $%
c=1$ non critical string and topological string B model will given in the
harmonic formulation we want to study now.

\section{Conifold\textbf{\ in harmonic framework}}

\qquad In this section, we develop the harmonic set up of conifold complex
deformation analysis as it is the appropriate formalism in which $SL\left(
2,C\right) $ isometry of the conifold is manifest. This formalism applies as
well to the study of local deformations of $S^{3}$ preserving manifestly its
$SU\left( 2,C\right) $ symmetry of $S^{3}$ and to local complex deformations
of cotangeant bundle $T^{\ast }P^{1}$ of complex one dimensional projective
space $P^{1}\sim S^{2}$.

To that purpose, we first describe harmonic variables as generally used in $%
4D $ $\mathcal{N}=2$ supersymmetric theories. Then, we use the covariant
harmonic analysis to study local complex deformations of conifold and some
specific submanifolds. After that, we give the full classification of these
deformations in terms of $SU\left( 2,C\right) $ and $SL\left( 2,C\right) $
representations and establish a dictionary giving a\textit{\ 1 to 1
correspondence} between Fourrier expansion on circle and harmonic analysis
on 3-sphere.

\subsection{Harmonic variables}

\qquad Roughly speaking, harmonic variables $u_{\alpha }^{\pm }$, with $%
u_{\alpha }^{-}$ being the complex conjugate of $u^{+\alpha }$ ($u_{\alpha
}^{-}=\overline{u^{+\alpha }}$) are commuting $SU\left( 2,C\right) $
isospinors ($s=\frac{1}{2}$) carrying two pairs of indices namely $\alpha
=1,2$ and the charges $\pm $. The first index ($\alpha =1,2$) is related to
the usual basic $SU\left( 2,C\right) $\ spin projection $s_{z}=\pm \frac{1}{2%
}$ by the relation,
\begin{equation}
s_{z}=\alpha -\frac{3}{2}\quad ,
\end{equation}%
while the second index ($\pm $) refers to the fundamental charges of the
Cartan Weyl subgroup $U_{C}\left( 1\right) $ of the $SU\left( 2,C\right) $\
group. Recall that as a Lie group, $SU\left( 2\right) $ can be split as%
\begin{equation}
SU\left( 2\right) =U_{C}\left( 1\right) \times \left[ SU\left( 2\right)
/U_{C}\left( 1\right) \right] )\quad .
\end{equation}%
The $u_{\alpha }^{\pm }$ variables give a simple way to parameterize
functions living on the \textit{unit} real 3-sphere including its defining
equations which read as,%
\begin{eqnarray}
u^{+\gamma }u_{\gamma }^{-} &=&1\quad ,  \notag \\
u^{+\gamma }u_{\gamma }^{+} &=&0\quad ,  \label{s} \\
u_{\gamma }^{-}u^{-\gamma } &=&0\quad ,  \notag
\end{eqnarray}%
where $f^{\gamma }g_{\gamma }=-f_{\gamma }g^{\gamma }$ stands for $%
\varepsilon _{\gamma \delta }f^{\gamma }g^{\delta }$ and will be often set
as $f.g$. This definition is useful for studies involving harmonic expansion
of function living on the sphere. \newline
Local development may be also considered in harmonic frame work; all one has
to do is to single out a given point on the sphere, say $u_{\alpha }^{\pm
}=a_{\alpha }^{\pm }$ satisfying eqs(\ref{s}), and use the identity
\begin{equation}
\varepsilon _{\alpha \beta }=\left( u_{\alpha }^{+}u_{\beta }^{-}-u_{\alpha
}^{-}u_{\beta }^{+}\right) \quad ,
\end{equation}%
to rewrite previous equations as follows,%
\begin{equation}
\left( a^{+}.u^{+}\right) \left( a^{-}.u^{-}\right) -\left(
a^{+}.u^{-}\right) \left( a^{-}.u^{+}\right) =1\quad .
\end{equation}%
For $u_{\alpha }^{\pm }=a_{\alpha }^{\pm }$, we recover $%
a^{+}.a^{+}=a^{-}.a^{-}=0$ and $a^{+}.a^{-}=-a^{-}.a^{+}=1$. \newline
The use of $u_{\alpha }^{\pm }$ variables allows to avoid the complexity of $%
SU\left( 2\right) $\ tensor calculus and keeps $SU\left( 2\right) $\
symmetry manifest. In harmonic setting, $SU\left( 2\right) $ irreducible
tensors
\begin{equation}
T^{\left( \alpha _{1}...\alpha _{n}\right) }\quad ,
\end{equation}%
are described by the highest weight function,
\begin{equation}
T^{+n}=u_{(\alpha _{1}}^{+}...u_{\alpha _{n})}^{+}T^{\left( \alpha
_{1}...\alpha _{n}\right) }\quad ,
\end{equation}%
carrying $n$ Cartan charges and satisfying the $SU\left( 2\right) $ highest
weight state condition%
\begin{eqnarray}
D^{0}T^{+n} &=&nT^{+n}\quad ,  \notag \\
D^{++}T^{+n} &=&0\quad .
\end{eqnarray}%
Local properties of harmonic functions are captured by the harmonic
distributions
\begin{equation}
\frac{1}{\left( a^{+}.u^{+}\right) ^{n}}\quad ,\qquad n>0\quad ,
\end{equation}%
and
\begin{equation}
\frac{1}{\left( a^{-}.u^{-}\right) ^{n}}\quad ,\qquad n>0\quad ,
\end{equation}%
with pole singularities at $a^{+}=u^{+}$ and $a^{-}=u^{-}$ respectively.
Cauchy's theorem of complex analysis reads in terms of harmonic variables as,%
\begin{equation}
\int_{\gamma _{a}}\frac{a^{+}.du^{+}}{a^{+}.u^{+}}\sim 1\quad ,
\end{equation}%
where $\gamma _{a}$ is a real contour surrounding the pole $a^{+}=u^{+}$;%
\begin{equation}
a^{+}.u^{+}|_{a^{+}=u^{+}}=0\quad .
\end{equation}%
These features allow remarkable applications of harmonic variables in
mathematical physics. Recall in passing that harmonic variables have been
used to solve several problems in classical and quantum field theory
requiring covariant $SU\left( 2\right) $ tensor calculus.

Notice also that, in addition to the usual complex conjugation denoted here
as $\left( \overline{X}\right) $, harmonic variables involve an extra
conjugation $\left( \ast \right) $ reversing the sign of the charges of the $%
U\left( 1\right) $ Cartan subgroup of $SU\left( 2\right) $ symmetry. In the
following table, we collect the general features of the conjugations $\left(
-\right) $ and $\left( \ast \right) $\ as well as their combination $\left(
\overline{\ast }\right) $ which for convenience we denote it as $\left( \sim
\right) $.%
\begin{equation}
\begin{tabular}{|l|l|l|l|}
\hline
field variables \ \ $X$ & $\ \ \ \ \ \left( \overline{X}\right) $ & $\ \ \ \
\left( X^{\ast }\right) $ & $\overline{X}^{\ast }\equiv \widetilde{X}$ \\
\hline
$\ \ \ \ \ \ \ \ w^{+\alpha }$ & $\ \ \ \ \ w_{\alpha }^{-}$ & $\ \ \ \
w^{-\alpha }$ & $\ \ \ \ w_{\alpha }^{+}$ \\ \hline
$\ \ \ \ \ \ \ \ w_{\alpha }^{-}$ & $\ \ \ \ \ w^{+\alpha }$ & $\ \ \ \
-w^{-\alpha }$ & $\ \ \ \ -w_{\alpha }^{+}$ \\ \hline
$\ \ \ \ F^{q}\left( u^{+},v^{-}\right) $ & $\overline{F}^{-q}\left(
u^{-},v^{+}\right) $ & $F^{-q}\left( u^{-},v^{+}\right) $ & $\ \widetilde{%
F^{q}}\left( u^{+},v^{-}\right) $ \\ \hline
$\ \ \ \ \widetilde{F^{q}}\left( u^{+},v^{-}\right) $ & $F^{-q}\left(
u^{-},v^{+}\right) $ & $\overline{F}^{-q}\left( u^{-},v^{+}\right) $ & $%
\left( -\right) ^{q}F^{q}\left( u^{+},v^{-}\right) $ \\ \hline
\end{tabular}%
\end{equation}%
As one sees, $\left( \sim \right) $ is not a standard conjugation since $%
\left( \overline{\ast }\right) ^{2}\neq \left( \overline{\ast }\right) $.
Like in 4D $\mathcal{N}=2$ SYM, reality condition will be taken here also in
the sense of $\left( \sim \right) $. Note finally that only harmonic
function $F^{q}=F^{q}\left( u\right) $ that carry an even number of Cartan
charges ($q=2n$) that can be subject to reality condition. Harmonic
functions $F^{2n+1}$ with an odd integer number of charge are necessary
complex.

\subsubsection{Harmonic variables and $4D$ $\mathcal{N}=2$ supersymmetric
theories}

\qquad To our knowledge, harmonic variables have been first introduced for
solving the problem of a manifestly superspace formulation of $4D$ $\mathcal{%
N}=2$ extended Super Yang Mills $\cite{33}$ and $4D$ $\mathcal{N}=2$
supergravity theories $\cite{37}$-$\cite{39}$. This method has been extended
as well to their reductions down to $2D$ $\mathcal{N}=4$ supersymmetric
theories $\cite{53}$-$\cite{57}$. Then, they have been used for approaching
different matters; in particular for studying $4D$ euclidean Yang Mills and
gravitational instantons $\cite{41}$-$\cite{46}$ and recently in dealing
with the analysis of singularities of so called "hyperKahler" Calabi-Yau
manifolds $\cite{42}$.\newline
In the harmonic superspace formulation of $4D$ $\mathcal{N}=2$ extended
Super Yang Mills, the ordinary superspace with $SU\left( 2\right) $
R-symmetry,
\begin{equation}
z^{M}=\left( x^{\mu },\theta _{a}^{\alpha },\overline{\theta }_{\overline{a}%
}^{\alpha }\right) \quad ,\qquad \alpha =1,2\quad ,
\end{equation}%
gets mapped into the harmonic superspace
\begin{equation}
z^{M}=\left( Y^{m},\theta _{a}^{-},\overline{\theta }_{\overline{a}%
}^{-},u_{\alpha }^{\pm }\right) \quad ,
\end{equation}%
with%
\begin{equation}
Y^{m}=\left( y^{\mu },\theta _{a}^{+},\overline{\theta }_{\overline{a}%
}^{+}\right) \quad ,
\end{equation}%
and
\begin{eqnarray}
y^{\mu } &=&x^{\mu }+i\left( \theta ^{+}\sigma ^{\mu }\overline{\theta }%
^{-}+\theta ^{-}\sigma ^{\mu }\overline{\theta }^{+}\right) \quad , \\
\theta _{a}^{+} &=&u_{\alpha }^{+}\theta _{a}^{\alpha },\qquad \overline{%
\theta }_{\overline{a}}^{+}=u_{\alpha }^{+}\overline{\theta }_{\overline{a}%
}^{\alpha }\quad .  \notag
\end{eqnarray}%
$4D$ $\mathcal{N}=2$ matter superfields described by hypermultiplets are
nicely represented in harmonic superspace. The typical Fayet-Iliopoulos
hypermultiplet is described by a harmonic superfunction $Q^{+}\left(
Y^{m}\right) $ carrying one positive Cartan-Weyl charge and has the
following $\theta $-expansion,%
\begin{equation}
Q^{+}\left( y\mathbf{,}\theta ^{+},\overline{\theta }^{+},u\right)
=u_{\alpha }^{+}q^{\alpha }\left( y,u\right) +\theta ^{+a}\psi _{a}\left(
y,u\right) +\overline{\theta }_{\overline{a}}^{+}\chi ^{\overline{a}}\left(
y,u\right) +...\quad .
\end{equation}%
The gauge multiplet is described by the prepotential $V^{++}\left( y\mathbf{,%
}\theta ^{+},\overline{\theta }^{+}\right) $ carrying two positive
Cartan-Weyl charge; it reads in the Wess-Zumino gauge as
\begin{equation}
V^{++}\left( y\mathbf{,}\theta ^{+},\overline{\theta }^{+}\right) =\theta
^{+}\sigma ^{\mu }\overline{\theta }^{+}\mathcal{A}_{\mu }\left( x\right) +%
\overline{\theta }^{+2}\theta ^{+a}\mathcal{\lambda }_{a}\left( x\right)
+\theta ^{+2}\overline{\theta }_{\overline{a}}^{+}\mathcal{\lambda }^{%
\overline{a}}\left( x\right) +\theta ^{+2}\overline{\theta }^{+2}\mathcal{D}%
\left( x\right) \quad ,
\end{equation}%
and it it used to covariantize the harmonic derivatives%
\begin{equation}
D^{++}=u^{+}\frac{\partial }{\partial u^{-}}\quad ,\qquad \widetilde{D^{++}}%
=D^{++}\quad ,
\end{equation}%
which becomes then
\begin{equation}
\mathcal{D}^{++}=D^{++}+V^{++}\quad ,\qquad \widetilde{V^{++}}=V^{++}\quad .
\end{equation}%
With these tools, one can go ahead and write down the superfield action $%
\mathcal{S}=\mathcal{S}\left[ Q^{+},V^{++}\right] $ for $\mathcal{N}=2$ SYM$%
_{4}$%
\begin{equation}
\mathcal{S}=\int d^{4}xd^{2}\theta ^{+}d^{2}\overline{\theta }^{+}\mathcal{L}%
^{4+}\left[ Q^{+},\widetilde{Q}^{+},V^{++}\right] \quad .  \label{sl}
\end{equation}%
The harmonic superspace lagrangian density $\mathcal{L}^{4+}\left[ Q^{+},%
\widetilde{Q}^{+},V^{++}\right] $ carries four positive Cartan charges. The
structure of its matter superfield dependence is,%
\begin{equation}
\mathcal{L}^{4+}\sim \widetilde{Q}^{+}\mathcal{D}^{++}Q^{+}+\mathcal{L}%
_{int}^{4+}\quad ,
\end{equation}%
with $\mathcal{L}_{int}^{4+}$ giving the matter self couplings; i.e $%
\mathcal{L}_{int}^{4+}\left( \widetilde{Q}^{+},Q^{+}\right) $. For more
details on eq(\ref{sl}) and its quantization, see $\cite{33,39}$.

\subsubsection{Harmonic analysis for conifold}

The harmonic formalism for conifold is based on spinor representations of $%
SL\left( 2,C\right) $ group; it goes beyond the $SU\left( 2,C\right) $
harmonic analysis developed for $4D$, $\mathcal{N}=2$ supersymmetric gauge
theories. The latter is recovered from conifold harmonic formalism as a
special case by imposing reality condition and appropriate constraint eqs to
be specified later. \newline
Conifold harmonic analysis is used here to solve technical difficulties due
to $SL\left( 2,C\right) $ tensor calculus. Explicit examples will be given
in forthcoming sections. It has moreover the advantage to shed more light on
local complex deformations for conifold $T^{\ast }S^{3}$ and its subspace $%
T^{\ast }P^{1}$ by still keeping their isometries manifest. \newline
To parameterize the conifold, we use the following set harmonic variables:
\newline
(\textbf{i}) A complex pair of variables
\begin{equation}
U^{+\gamma }=\left( U^{+1},U^{+2}\right) \quad ,
\end{equation}%
transforming under $SU_{u}\left( 2,C\right) $ as isospinor doublet $|s=\frac{%
1}{2},\quad s_{z}=\pm \frac{1}{2}>$; but as a degenerate one component state
\begin{equation}
|j=\frac{1}{2},\quad j_{z}=+\frac{1}{2}>
\end{equation}%
under the $SL\left( 2,C\right) $ symmetry group. They are related to the
previous small harmonic variables as
\begin{equation}
U^{+\gamma }=l\text{ }u^{+\gamma }\quad .
\end{equation}%
(\textbf{ii}) A second complex pair of variables with opposite charge,%
\begin{equation}
V_{\gamma }^{-}=\left( V_{1}^{-},V_{2}^{-}\right) \quad ,
\end{equation}%
transforming under $SU_{v}\left( 2,C\right) $ as isospinor doublet; but
under $SL\left( 2,C\right) $ as a one component state as shown below,
\begin{equation}
|j=\frac{1}{2},\quad j_{z}=-\frac{1}{2}>\quad .
\end{equation}%
The vector $\left( U^{+},V^{-}\right) $ constitutes then an $SL\left(
2,C\right) $ doublet of complex holomorphic variables. The two other partner
pairs namely,%
\begin{equation}
U_{\gamma }^{-}=\left( U_{1}^{-},U_{2}^{-}\right) \quad ,\qquad V^{+\gamma
}=\left( V^{+1},V^{+2}\right) \quad ,
\end{equation}%
should be thought of as the anti-holomorphic variables. They are obtained
from the previous ones by complex conjugation,%
\begin{equation}
U_{\gamma }^{-}=\overline{U^{+\gamma }}\quad ,\qquad V^{+\gamma }=\overline{%
V_{\gamma }^{-}}\quad .
\end{equation}%
To fix the ideas, we recall that one should think of $U^{+\gamma }$ and $%
V_{\gamma }^{-}$ as respectively associated with the complex holomorphic
pairs $\left( x,z\right) $ and $\left( y,w\right) $,%
\begin{equation}
U^{+\gamma }=\left( x,z\right) \quad ,\qquad V_{\gamma }^{-}=\left(
y,w\right) \quad .
\end{equation}%
The variables $U_{\gamma }^{-}$ and $V^{+\gamma }$ correspond then to the
anti-holomorphic partners $\left( \overline{x},-\overline{z}\right) $ and $%
\left( \overline{y},-\overline{w}\right) $,%
\begin{equation}
U_{\gamma }^{-}=\left( \overline{x},-\overline{z}\right) \quad ,\qquad
V^{+\gamma }=\left( \overline{y},-\overline{w}\right) \quad .
\end{equation}%
Using this correspondence, and by help of the $SL\left( 2,C\right) $
invariant metric $\varepsilon _{\alpha \beta }$ ($\varepsilon
_{21}=-\varepsilon _{12}=1$) for spinors, the conifold eq(\ref{1}) can be
easily shown to read as
\begin{eqnarray}
U^{+\gamma }V_{\gamma }^{-} &=&\mu \quad ,  \label{8} \\
U^{+\gamma }U_{\gamma }^{+} &=&0\quad ,  \label{9a} \\
V^{-\gamma }V_{\gamma }^{-} &=&0\quad .  \label{9b}
\end{eqnarray}%
As one sees from these relations, the harmonic variables $U^{+\gamma }$ and $%
V_{\gamma }^{-}$ scales as $\sqrt{\mu }$ that is like a lenght $L$. For
convenience, it is interesting to use the rescaling
\begin{equation}
U^{+\gamma }=l_{1}u^{+\gamma }\quad ,\qquad V_{\gamma }^{-}=l_{2}v_{\gamma
}^{-}\quad ,\qquad l_{1}l_{2}=\mu \quad ,  \label{sc}
\end{equation}%
where the small harmonic variables $u^{+\gamma }$ and $v_{\gamma }^{-}$ are
dimensionless and where $l_{1}$ and $l_{2}$ are two complex numbers whose
module scale as length. This scaling property has the remarkable
consequences. \newline
(\textbf{a}) As far as singularities are concerned, the exact way to define
the conifold is as
\begin{equation}
U^{+\gamma }V_{\gamma }^{-}=l_{1}l_{2}\quad ,
\end{equation}%
where $\mu =0$ is reached either by taking $l_{1}=0$, $l_{2}\neq 0$ or $%
l_{2}=0$, $l_{1}\neq 0$ or $l_{1}=l_{2}=0$. \newline
(\textbf{b}) Thinking about the ambient complex space $C^{4}$ as the product
$C_{u}^{2}\times C_{v}^{2}$ with
\begin{equation*}
U^{\pm }\in C_{u}^{2}\sim R_{u}^{4}\text{\qquad\ and \qquad }V^{\mp }\in
C_{v}^{2}\sim R_{v}^{4}\quad ,
\end{equation*}%
one sees that the real 3-sphere is obtained by identifying $V^{-}$ and $%
U^{-} $ as shown below,%
\begin{equation}
V^{-}=\left( \frac{l_{1}l_{2}}{r^{2}}\right) U^{-}\quad ,  \label{cor}
\end{equation}%
where $r$ is some real fixed number. In this case, the real slice of the
conifold reads as follows,
\begin{equation}
U^{+\gamma }U_{\gamma }^{-}=r^{2}\quad .
\end{equation}%
(\textbf{c}) The scaling (\ref{sc}) has also the effect of mapping the
conifold real slice $S^{3}$,%
\begin{equation}
U^{+\gamma }U_{\gamma }^{-}=\func{Re}\left( \mu \right) =p\quad ,
\end{equation}%
with radius $\sqrt{\func{Re}\left( \mu \right) }$\ to the unit sphere $%
u^{+\gamma }u_{\gamma }^{-}=1$. In general under the change (\ref{sc}), the
generic conifold eqs(\ref{8}-\ref{9b}) maps to,
\begin{eqnarray}
u^{+\gamma }v_{\gamma }^{-} &=&1\quad ,  \notag \\
u^{+\gamma }u_{\gamma }^{+} &=&0\quad ,  \label{10} \\
v_{\alpha }^{-}v^{-\alpha } &=&0\quad ,  \notag
\end{eqnarray}%
where now the real slice is given bu the unit sphere.

\subsection{Conifold isometries}

As far as isometries of $T^{\ast }S^{3}$ are concerned, we have the
following:\newline
(\textbf{1}) The $SL_{u}\left( 2,C\right) $ isometry subgroup factor%
\footnote{%
Here hermiticity is taken in the sense of the combined conjugation $\sim
\equiv \left( \overline{\ast }\right) $.} generated by the general
coordinate change,%
\begin{equation}
u^{+\gamma }\rightarrow \Lambda ^{++}v^{-\gamma }\quad ,\qquad v_{\gamma
}^{-}\rightarrow v_{\gamma }^{-}\quad ,  \label{ha1}
\end{equation}%
where $\Lambda ^{++}$ is priori an arbitrary complex function of $u^{+}$ and
$v^{-}$; i.e
\begin{equation}
\Lambda ^{++}=\Lambda ^{++}\left( u^{+},v^{-}\right) \quad .
\end{equation}%
The leading terms of the harmonic expansion of this complex holomorphic
function reads, for the case of the charge is conserved, as%
\begin{equation}
\Lambda ^{++}=\Lambda _{\left( \alpha \beta \right) }u^{+\alpha }u^{+\beta
}+\Lambda _{\left( \alpha _{1}\alpha _{2}\beta _{1}\beta _{2}\right)
}u^{+\alpha _{1}}u^{+\alpha _{2}}v^{-\beta _{1}}v^{-\beta _{2}}+...\quad .
\end{equation}%
The general expression will be discussed later. Note that global $%
SL_{u}\left( 2,C\right) $ invariance is described by the group parameter $%
\Lambda _{0}^{++}$ constrained as
\begin{equation}
\frac{\partial \Lambda _{0}^{++}}{\partial v^{-}}=0\quad ,\qquad u^{+\alpha }%
\frac{\partial \Lambda _{0}^{++}}{\partial u^{+\alpha }}=2\Lambda
_{0}^{++}\quad ,
\end{equation}%
with harmonic expansion given,%
\begin{equation}
\Lambda _{0}^{++}\left( u^{+},v^{-}\right) =\Lambda _{\left( \alpha \beta
\right) }u^{+\alpha }u^{+\beta }\quad .  \label{ha2}
\end{equation}%
In this relation there is no $v^{-}$ dependence and where one recognizes the
$SL_{u}\left( 2,C\right) $ triplet of complex parameters
\begin{equation}
\Lambda _{\left( \alpha \beta \right) }\equiv \left( \Lambda _{11},\Lambda
_{\left( 12\right) },\Lambda _{22}\right) \quad .
\end{equation}%
To make contact with the complex analysis of previous section, the harmonic
transformations (\ref{ha1}-\ref{ha2}) should be associated with eqs(\ref{m2}%
). \newline
(\textbf{2}) The $SL_{v}\left( 2,C\right) $ isometry subgroup factor of the
conifold reads as,
\begin{equation}
u^{+\gamma }\rightarrow u^{+\gamma }\quad ,\qquad v_{\gamma }^{-}\rightarrow
v_{\gamma }^{-}=\Gamma ^{--}u_{\gamma }^{+}\quad ,  \label{ha3}
\end{equation}%
where, like before, the parameter $\Gamma ^{--}$ is an arbitrary holomorphic
function in the harmonic variables $u^{+}$ and $v^{-}$; i.e
\begin{equation}
\Gamma ^{--}=\Gamma ^{--}\left( u^{+},v^{-}\right) \quad .
\end{equation}%
Its leading terms preserving manifestly global $SL_{v}\left( 2,C\right) $
symmetry read as,
\begin{equation}
\Gamma _{0}^{--}=v_{\alpha }^{-}v_{\beta }^{-}\Gamma ^{\left( \alpha \beta
\right) }\quad ,\qquad \frac{\partial \Gamma _{0}^{--}}{\partial u^{+\gamma }%
}=0\quad ,\qquad -v^{-\alpha }\frac{\partial \Gamma _{0}^{--}}{\partial
v^{-\alpha }}=-2\Gamma _{0}^{--}\quad ,  \label{ha4}
\end{equation}%
In the language of section 2, this symmetry should be associated with the
symmetry group generated by $\left\{ \widetilde{L}_{0},\widetilde{L}_{\pm
}\right\} $, see also (\ref{m2}).

\subsubsection{Generators}

In eq(\ref{10}) defining $T^{\ast }S^{3}$, there are two kinds of harmonic
variables namely $u^{+}$ and $v^{-}$; but no complex conjugate $u^{-}$ nor $%
v^{+}$. Each one of these holomorphic harmonic variables belongs to one of
the two $C^{2}$ factors of the complex space $C^{4}\sim C_{u}^{2}\times
C_{v}^{2}$. For later use, let us make a comment regarding these two sectors.%
\newline
(\textbf{i}) In the u-sector, the $u^{+\gamma }$ holomorphic harmonic
variable (belonging to $C_{u}^{2}$) together with its complex conjugate $%
u_{\gamma }^{-}=\overline{u^{+\gamma }}$ parameterize the real $S^{3}$ slice
of $T^{\ast }S^{3}$,%
\begin{eqnarray}
U^{+\gamma }U_{\gamma }^{-} &=&\mathrm{p}u^{+\gamma }u_{\gamma }^{-}=\mathrm{%
p}\quad ,  \label{11} \\
U^{+\gamma }U_{\gamma }^{+} &=&\mathrm{p}u^{+\gamma }u_{\gamma }^{+}=0\quad ,
\label{12a} \\
U^{-\gamma }U_{\gamma }^{-} &=&\mathrm{p}u^{-\gamma }u_{\gamma }^{-}=0\quad .
\label{12b}
\end{eqnarray}%
with $\mathrm{p}=\left( \func{Re}\mu \right) $. Both capital harmonic
variables $U_{\alpha }^{\pm }$; and the corresponding small $u_{\alpha
}^{\pm }$ ones are rotated by the same $SU_{u}\left( 2,C\right) $ algebra
whose generators read as,%
\begin{eqnarray}
D_{u}^{++} &=&u^{+\alpha }\frac{\partial }{\partial u^{-\alpha }}\quad ,
\notag \\
D_{u}^{--} &=&u^{-\alpha }\frac{\partial }{\partial u^{+\alpha }}\quad ,
\label{13} \\
D_{u}^{0} &=&u^{+\alpha }\frac{\partial }{\partial u^{+\alpha }}-u^{-\alpha }%
\frac{\partial }{\partial u^{-\alpha }}\quad ,  \notag
\end{eqnarray}%
or equivalently, by using capital harmonic variables, as,%
\begin{eqnarray}
D_{U}^{++} &=&U^{+\alpha }\frac{\partial }{\partial U^{-\alpha }}\equiv
D_{u}^{++}\quad ,  \notag \\
D_{U}^{--} &=&U^{-\alpha }\frac{\partial }{\partial U^{+\alpha }}\equiv
D_{u}^{--}\quad ,  \label{130} \\
D_{U}^{0} &=&U^{+\alpha }\frac{\partial }{\partial U^{+\alpha }}-U^{-\alpha }%
\frac{\partial }{\partial U^{-\alpha }}\equiv D_{u}^{0}\quad .  \notag
\end{eqnarray}%
The identification between the harmonic differential operators $D_{U}^{++}$,
$D_{U}^{--},$\ $D_{U}^{0}$\ and $D_{u}^{++}$, $D_{u}^{--},$\ $D_{u}^{0}$
respectively follows the fact that with the coordinate change $U^{\pm \alpha
}=lu^{\pm \alpha }$, the scaling parameter $l$ is an $SU_{u}\left(
2,C\right) $ singlet which does not depend on harmonic variables. The
identification is directly seen on the following typical tranformation,%
\begin{equation}
D_{u}^{++}=\left( D_{u}^{++}U^{-\beta }\right) \frac{\partial }{\partial
U^{-\beta }}+\left( D_{u}^{++}l\right) \frac{\partial }{\partial l}%
=D_{U}^{++}\quad .
\end{equation}%
The differential harmonic operators $D_{u}^{q}$, with $q=0,$ $\pm 2$, carry
only Cartan-Weyl charges and no free $SU_{u}\left( 2,C\right) $ spinor
indices. Direct computations shows that they satisfy the usual $SU_{u}\left(
2,C\right) $ commutation relations,%
\begin{eqnarray}
\left[ D_{u}^{++},D_{u}^{--}\right] &=&D_{u}^{0}\quad ,  \notag \\
\left[ D_{u}^{0},D_{u}^{++}\right] &=&2D_{u}^{++}\quad ,  \label{14} \\
\left[ D_{u}^{0},D_{u}^{--}\right] &=&-2D_{u}^{--}\quad .  \notag
\end{eqnarray}%
Similar expressions are also valid for $D_{U}^{++},$ $D_{U}^{--}$\ and $%
D_{U}^{0}$; they show that $SU_{u}\left( 2,C\right) $ symmetry remains
invariant under global scaling of the harmonic variables. Acting by $%
D_{u}^{++},$ $D_{u}^{--}$ and $D_{u}^{0}$ on eqs(\ref{11}),
\begin{equation}
D_{u}^{++}\left( U^{+\gamma }U_{\gamma }^{-}\right) \quad ,
\end{equation}%
and so on, one gets, on one hand, eq(\ref{12a}),
\begin{equation}
D_{u}^{++}\left( U^{+\gamma }U_{\gamma }^{-}\right) =\mathrm{p}u^{+\gamma
}u_{\gamma }^{+}=0\quad ,
\end{equation}%
and on the other hand $D_{u}^{++}\mathrm{p}$. Consistency requires then
\begin{equation}
D_{u}^{++}\mathrm{p}=0\quad .
\end{equation}%
So the complex deformation parameter $\mathrm{p}$ should be an $SU_{u}\left(
2,C\right) $ invariant. In general the relations defining this invariance
read as,%
\begin{equation}
D_{u}^{++}\mathrm{p}=D_{u}^{--}\mathrm{p}=D_{u}^{0}\mathrm{p}=0\quad .
\label{15}
\end{equation}%
This means also that $SU_{u}\left( 2,C\right) $ rotations commute with the
global volume fluctuations of the 3-sphere $S^{3}$. Since these deformations
are generated by $\partial _{p}\equiv \frac{\partial }{\partial p};$ we then
have,
\begin{equation}
\left[ D_{u}^{++},\frac{\partial }{\partial p}\right] =\left[ D_{u}^{--},%
\frac{\partial }{\partial p}\right] =\left[ D_{u}^{0},\frac{\partial }{%
\partial p}\right] =0\quad .
\end{equation}%
Details on variations generated by the local complex deformations beyond eq(%
\ref{15}) will be given after making the following comment.\newline
(\textbf{ii}) In the v-sector, the holomorphic harmonic variables $v_{\gamma
}^{-}$ belong to the second $C^{2}$ ($C_{v}^{2}$) copy of $C^{4}\sim
C_{u}^{2}\times C_{v}^{2}$, where lives the conifold $T^{\ast }S^{3}$.
Together with their complex conjugates $v^{+\gamma }=\overline{v_{\gamma
}^{-}}$; these $v^{\pm }$ harmonic variables parameterize an other $S^{3}$
sphere which is actually embedded in $C_{v}^{2}$,
\begin{eqnarray}
V^{+\gamma }V_{\gamma }^{-} &=&\mathrm{q}v^{+\gamma }v_{\gamma }^{-}=\mathrm{%
q}\quad ,  \notag \\
\varepsilon _{\gamma \delta }v^{+\gamma }v^{+\delta } &=&0,\qquad
\varepsilon ^{\gamma \delta }v_{\alpha }^{-}v_{\beta }^{-}=0\quad .
\label{16}
\end{eqnarray}%
Similarly as in u-sector, the capital harmonic variables $V_{\alpha }^{\pm }$
and the small $v_{\alpha }^{\pm }$ ones are rotated by the same $%
SU_{v}\left( 2,C\right) $ algebra,%
\begin{eqnarray}
D_{v}^{++} &=&v^{+\alpha }\frac{\partial }{\partial v^{-\alpha }}\quad ,
\notag \\
D_{v}^{--} &=&v^{-\alpha }\frac{\partial }{\partial v^{+\alpha }}\quad ,
\label{17} \\
D_{v}^{0} &=&v^{+\alpha }\frac{\partial }{\partial v^{+\alpha }}-v^{-\alpha }%
\frac{\partial }{\partial v^{-\alpha }}\quad ,  \notag
\end{eqnarray}%
satisfying the commutation relations,%
\begin{eqnarray}
\left[ D_{v}^{++},D_{v}^{--}\right] &=&D_{v}^{0}\quad ,  \notag \\
\left[ D_{v}^{0},D_{v}^{++}\right] &=&2D_{v}^{++}\quad ,  \label{18} \\
\left[ D_{v}^{0},D_{v}^{--}\right] &=&-2D_{v}^{--}\quad .  \notag
\end{eqnarray}%
Like before, it is not difficult to check form eqs(\ref{16}), that the
complex deformation parameter $\mathrm{q}$ is an $SU_{v}\left( 2,C\right) $
invariant,
\begin{equation}
D_{v}^{++}\mathrm{q}=D_{v}^{--}\mathrm{q}=D_{v}^{0}\mathrm{q}=0\quad .
\end{equation}%
Concerning the $SL\left( 2,C\right) $ symmetry of eq(\ref{10}) rotating the
holomorphic $u^{+\gamma }$ and $v_{\gamma }^{-}$ variables, we have the
generators,
\begin{eqnarray}
\nabla ^{++} &=&u^{+\alpha }\frac{\partial }{\partial v^{-\alpha }}\quad ,
\notag \\
\nabla ^{--} &=&v^{-\alpha }\frac{\partial }{\partial u^{+\alpha }}\quad ,
\label{19} \\
\nabla ^{0} &=&u^{+\alpha }\frac{\partial }{\partial u^{+\alpha }}%
-v^{-\alpha }\frac{\partial }{\partial v^{-\alpha }}\quad ,  \notag
\end{eqnarray}%
satisfying,%
\begin{eqnarray}
\left[ \nabla ^{++},\nabla ^{--}\right] &=&\nabla ^{0}\quad ,  \notag \\
\left[ \nabla ^{0},\nabla ^{++}\right] &=&2\nabla ^{++}\quad ,  \label{20} \\
\left[ \nabla ^{0},\nabla ^{--}\right] &=&-2\nabla ^{--}\quad .  \notag
\end{eqnarray}%
Similarly as before, we have here also the constraint eqs
\begin{equation}
\nabla ^{++}\mu =\nabla ^{--}\mu =\nabla ^{0}\mu =0\quad ,
\end{equation}%
showing that $\mu $ is invariant under the holomorphic $SL\left( 2,C\right) $
symmetry. Moreover, global fluctuations of the conifold with respect to the
variation of $\mu $\ commute with $SL\left( 2,C\right) $ rotations. So we
have,%
\begin{equation}
\left[ \nabla ^{++},\frac{\partial }{\partial \mu }\right] =\left[ \nabla
^{--},\frac{\partial }{\partial \mu }\right] =\left[ \nabla ^{0},\frac{%
\partial }{\partial \mu }\right] =0\quad .  \label{21}
\end{equation}

\subsubsection{Antiholomorphic sector}

Note in passing that along with $\nabla ^{++},$ $\nabla ^{--}$ and $\nabla
^{0}$, we have also the anti-holomorphic differential operator given by,
\begin{eqnarray}
\overline{\nabla }^{++} &=&v^{+\alpha }\frac{\partial }{\partial u^{-\alpha }%
}\quad ,\qquad  \notag \\
\overline{\nabla }^{--} &=&u^{-\alpha }\frac{\partial }{\partial v^{+\alpha }%
}\quad ,\qquad  \label{22} \\
\overline{\nabla }^{0} &=&\left[ \overline{\nabla }^{++},\overline{\nabla }%
^{--}\right] \quad .  \notag
\end{eqnarray}%
These operators which obey an $\overline{SL\left( 2,C\right) }$ algebra;
deals with the complex conjugate sector $SL\left( 2,C\right) $; they don't
concern us in present study since all the harmonic functions
\begin{equation}
F=F\left( u^{+},v^{-}\right) \quad ,
\end{equation}%
we will encounter in what follows have no $u^{-}$ and $v^{+}$ dependence.
They are then are annihilated by these operators,
\begin{equation}
\overline{\nabla }^{++}F=0\quad ,\qquad \overline{\nabla }^{--}F=0\quad .
\end{equation}%
To summarize, the description of the conifold harmonic frame work involves\
the harmonic variables $u_{\alpha }^{+}$ and $v_{\alpha }^{-}$ rotated under
$SL\left( 2,C\right) $ symmetry; but no $u_{\alpha }^{-}$ and $v_{\alpha
}^{+}$. There are two sector for $SL\left( 2,C\right) $ isometry in harmonic
space; the first denoted $SL_{u}\left( 2,C\right) $ with group parameter
function $\Lambda ^{++}$ and the second denoted $SL_{v}\left( 2,C\right) $
with group parameter function $\Gamma ^{--}$. Under reality condition
\begin{equation}
\Lambda ^{++}=\widetilde{\Lambda ^{++}}\quad ,\qquad \Gamma ^{--}=\widetilde{%
\Gamma ^{--}}\quad ,
\end{equation}%
these groups reduce respectively to $SU_{u^{+}}\left( 2,C\right) $ and $%
SU_{u^{-}}\left( 2,C\right) $.

\subsection{$T^{\ast }P^{1}$ as a submanifold of $T^{\ast }S^{3}$}

Here we show that, like for the relation between the three sphere $S^{3}$
and complex one dimension projective space $P^{1}\sim S^{2}$, we have quite
similar correspondence between $T^{\ast }P^{1}$ and $T^{\ast }S^{3}$. Recall
that the 3-sphere $S^{3}$ may be thought of as a non trivial fibration of a
circle $S^{1}$ over $S^{2}$,%
\begin{equation}
S^{3}\sim S^{1}\propto S^{2}\quad .  \label{s12}
\end{equation}%
In the same way, there is an analogous link between the cotangeant bundle of
the two sphere $T^{\ast }P^{1}$ and conifold $T^{\ast }S^{3}$. More
precisely, one may think about $T^{\ast }S^{3}$ as given by the fibration,%
\begin{equation}
T^{\ast }S^{3}\sim C^{\ast }\propto T^{\ast }P^{1}\quad .  \label{S23}
\end{equation}%
This correspondence can be made more precise in group theoretic language
where the above fibrations take well known expressions. The fibration (\ref%
{s12}) corresponds to the factorisation of $SU\left( 2,C\right) $ group in
terms of its $U\left( 1,C\right) $ abelian subgroup and the $\frac{SU\left(
2,C\right) }{U\left( 1,C\right) }$ coset subgroup as shown below,%
\begin{equation}
SU\left( 2,C\right) \sim U\left( 1,C\right) \times \frac{SU\left( 2,C\right)
}{U\left( 1,C\right) }\quad .
\end{equation}%
Similarly the fibration (\ref{S23}) corresponds to the following
factorisation of the group $SL\left( 2,C\right) $,%
\begin{equation}
SL\left( 2,C\right) \sim GL\left( 1,C\right) \times \frac{SL\left(
2,C\right) }{GL\left( 1,C\right) }\quad .
\end{equation}%
With this correspondence in mind the derivation of cotangeant bundle $%
T^{\ast }P^{1}$ from the previous harmonic analysis of $T^{\ast }S^{3}$
becomes a simple matter. It is obtained by fixing $GL\left( 1,C\right)
\simeq C^{\ast }$ symmetry subgroup of conifold harmonic equations,%
\begin{equation}
U^{+\alpha }V_{\alpha }^{-}=\mu \quad ,\qquad U^{+\alpha }U_{\alpha
}^{+}=0\quad ,\qquad V^{-\alpha }V_{\alpha }^{-}=0\quad .
\end{equation}%
Recall that the $C^{\ast }$ action generating the projective transformations
is given by,
\begin{equation}
U^{+\gamma }\rightarrow \lambda U^{+\gamma }\quad ,\qquad V_{\gamma
}^{-}\rightarrow \frac{1}{\lambda }V_{\gamma }^{-}\quad ,\qquad \lambda \in
C^{\ast }\quad .  \label{24}
\end{equation}%
In the harmonic analysis, the fixing of $GL\left( 1,C\right) $ symmetry is
achieved by making the identification,%
\begin{equation}
U^{+\gamma }\equiv \lambda U^{+\gamma }\quad ,\qquad V_{\gamma }^{-}\equiv
\frac{1}{\lambda }V_{\gamma }^{-}\quad ,
\end{equation}%
which corresponds to taking the harmonic variables $\left(
U_{1}^{+},U_{2}^{+},V_{1}^{-},V_{2}^{-}\right) $ in the weighted projective
space $WP_{\left( +1,+1,-1,-1\right) }^{3}$. The complex two dimension
holomorphic variety $T^{\ast }P^{1}$ embedded in $WP_{\left(
+1,+1,-1,-1\right) }^{3}$ can be then defined as
\begin{equation}
T^{\ast }P^{1}=T^{\ast }S^{3}/C^{\ast }\quad .  \label{23}
\end{equation}%
The complex holomorphic $T^{\ast }P^{1}$ geometry is a complex two dimension
Calabi-Yau manifold embedded in $WP_{\left( +1,+1,-1,-1\right) }^{3}$.
\newline
In the harmonic frame work, the complex two dimension Calabi-Yau manifold $%
T^{\ast }P^{1}$ is defined, like for the conifold, by
\begin{equation}
U^{+\gamma }V_{\gamma }^{-}=\mu \quad ,\qquad \varepsilon _{\gamma \delta
}U^{+\gamma }U^{+\delta }=0\quad ,\qquad \varepsilon ^{\gamma \delta
}V_{\alpha }^{-}V_{\beta }^{-}=0\quad ,  \label{25}
\end{equation}%
but now with $\left( U_{1}^{+},U_{2}^{+},V_{1}^{-},V_{2}^{-}\right) $
belonging to $WP_{\left( +1,+1,-1,-1\right) }^{3}$. \newline
More generally, functions $F^{q}\left( U^{+},V^{-}\right) $ living on $%
T^{\ast }P^{1}$ with $q$ an integer should be covariant objects; they are
harmonic functions carrying a well defined $C^{\ast }$\ charge $q$. This
means that contrary to functions on conifold, functions $F^{q}\left(
U^{+},V^{-}\right) $ on $T^{\ast }P^{1}$\ are homogeneous harmonic functions
constrained as,%
\begin{equation}
F^{q}\left( \lambda U^{+},\frac{1}{\lambda }V^{-}\right) =\lambda ^{q}\unit{F%
}^{q}\left( U^{+},V^{-}\right) \quad .
\end{equation}%
This constraint equation can be also put in the following equivalent form,
\begin{equation}
\left[ \nabla ^{0},F^{q}\left( u^{+},v^{-}\right) \right] =qF^{q}\left(
u^{+},v^{-}\right) \quad ,
\end{equation}%
where
\begin{equation}
\nabla ^{0}=\left( u^{+\alpha }\frac{\partial }{\partial u^{+\alpha }}%
-v^{-\alpha }\frac{\partial }{\partial v^{-\alpha }}\right) \quad ,
\end{equation}%
is the charge operator of the $SL\left( 2,C\right) $ isometry. Note that
similar conclusions are also valid for the real slices $S^{2}$ of $T^{\ast
}P^{1}$. In particular, the defining equation of the real two sphere $S^{2}$
which is just $S^{3}/U\left( 1\right) $ is obtained from the harmonic
equations of the three sphere,%
\begin{equation}
u^{+\gamma }u_{\gamma }^{-}=1\quad ,\qquad u^{+\gamma }u_{\gamma
}^{+}=0\quad ,\qquad u_{\gamma }^{-}u^{-\gamma }=0\quad ,  \label{26}
\end{equation}%
by requiring moreover the identification%
\begin{equation}
u^{+\gamma }\equiv e^{i\varphi }u^{+\gamma }\quad ,\qquad u_{\gamma
}^{-}=e^{-i\varphi }u_{\gamma }^{-}\quad ,
\end{equation}%
with $\varphi \in \left[ 0,2\pi \right] $. Harmonic functions on $S^{2}$\
are then homogeneous functions obeying
\begin{equation}
F^{q}\left( e^{i\varphi }u^{+},e^{-i\varphi }u^{-}\right) =e^{iq\varphi
}F^{q}\left( u^{+},u^{-}\right) \quad .
\end{equation}%
This property may be also stated as
\begin{equation}
\left[ D_{u}^{0},F^{q}\left( u^{+},u^{-}\right) \right] =qF^{q}\left(
u^{+},u^{-}\right) \quad .
\end{equation}

\section{Harmonic expansion of complex deformations}

\qquad So far we have considered global deformations of the conifold
singularity $U^{+\gamma }V_{\gamma }^{-}=0$ in harmonic space which becomes
then $U^{+\gamma }V_{\gamma }^{-}=\mu $. In this section we consider the
case of local complex deformations where $\mu $ get replaced by an arbitrary
function
\begin{equation}
\xi =\xi \left( U^{+\gamma },V_{\gamma }^{-}\right) .
\end{equation}%
Then we study the classification of these deformations by using harmonic
space. This manifestly covariant harmonic analysis will be used later for
completing partial results on $S^{3}$ quantum cosmology model of
Gukov-Sarakin and Vafa $\cite{30}$; in particular in the derivation of the
three following: \newline
(\textbf{i}) the manifestly $SL\left( 2,C\right) $ invariant conifold
partition function
\begin{equation}
\mathcal{Z}_{top}=\mathcal{Z}_{top}\left( T^{\ast }S^{3}\right) ,
\end{equation}%
(\textbf{ii}) the manifestly $SU\left( 2,C\right) $ invariant of the
Hartle-Hawking probability density
\begin{equation}
\varrho =\left\vert \Psi \left( S^{3}\right) \right\vert ^{2},
\end{equation}%
and (\textbf{iii}) quantum cosmology correlation functions of the
fluctuation fields.\newline
Recall that in the standard formulation, the classification of local complex
deformations
\begin{equation}
\left\{ t_{n},\widetilde{t}_{n};\text{ }n>0\right\}
\end{equation}%
of the conifold has been considered from different point of views; in
particular in connection with the study of the ground ring of $c=1$ non
critical string $\cite{16}$ and in relation with the computation of the
partition function $\mathcal{Z}_{top}\left( t,\widetilde{t}\right) $ of
topological string B model on conifold $\cite{15}$.

To begin note that along with the global deformations generated by the
variation of the\ global modulus $\mu $, there are infinitely many local
deformations of conifold. These transformations, which are no longer
isometries of the conifold, are complex deformations captured by harmonic
functions depending on the local coordinates $U^{+\gamma }$ and $V_{\gamma
}^{-}$ of the conifold. In harmonic framework, the set
\begin{equation}
\mathcal{J}=\left\{ \mathrm{\xi }:T^{\ast }S^{3}\longrightarrow C\right\}
\label{j}
\end{equation}%
of conifold local complex deformations is an infinite set generated by
arbitrary harmonic functions
\begin{equation}
\xi =\xi \left( U^{+\gamma },V_{\gamma }^{-}\right) ,\qquad U^{+\gamma
}V_{\gamma }^{-}=\mu .
\end{equation}%
Under these local deformations, eqs(\ref{8}-\ref{10}) get mapped to,%
\begin{equation}
U^{+\gamma \prime }V_{\gamma }^{-\prime }=\mu +\xi \left( U^{+\gamma
},V_{\gamma }^{-}\right) ,  \label{27}
\end{equation}%
where the new harmonic variables,%
\begin{equation}
U^{+\prime }=U^{+\prime }\left( U^{+},V^{-}\right) ,\qquad V^{-\prime
}=V^{-\prime }\left( U^{+},V^{-}\right) ,
\end{equation}%
parameterize the locally deformed conifold. As noted earlier, the parameter $%
\mu $\ may be usually absorbed in $\xi $ as a zero mode as shown on $%
U^{+\gamma \prime }V_{\gamma }^{-\prime }=f$ \ with $f=\mu +\xi $. For later
use, we shall keep however this splitting. By comparing the manifolds $%
U^{+\gamma }V_{\gamma }^{-}=\mu $ and $U^{+\gamma \prime }V_{\gamma
}^{-\prime }=f$, one sees that the deformation corresponds to varying the
global complex parameter $\mu $ by local moduli and are generated by the
scaling,%
\begin{equation}
U^{+\prime }=\Lambda U^{+},\qquad =V^{-\prime }=\Gamma V^{-},
\end{equation}%
where $\Lambda =\Lambda \left( U^{+},V^{-}\right) $ and $\Gamma =\Gamma
\left( U^{+},V^{-}\right) $ are two harmonic functions living on conifold;
they are related to the function $\xi $\ as
\begin{equation}
\Lambda \Gamma =\mu +\xi .
\end{equation}%
On the conifold real slice, we have $\Gamma =\widetilde{\Lambda }$ and so $%
\Lambda \widetilde{\Lambda }=p+\xi $ with the restriction $\widetilde{\xi }%
=\xi $.

\subsection{Local deformations}

First observe that the harmonic functions $\xi \left( U^{+\gamma },V_{\gamma
}^{-}\right) $ scales like $\mu $ but are no longer invariant under the $%
SL\left( 2,C\right) $ conifold isometry group. In harmonic differential
operator language, this means that we have,%
\begin{equation}
\left[ \nabla ^{++},\xi \left( U^{+},V^{-}\right) \right] \neq 0,\qquad %
\left[ \nabla ^{--},\xi \left( U^{+},V^{-}\right) \right] \neq 0  \label{28}
\end{equation}%
and so
\begin{equation}
\left[ \nabla ^{0},\xi \left( U^{+},V^{-}\right) \right] \neq 0,
\end{equation}%
where the differential operators $\nabla ^{++}$, $\nabla ^{0}$ and $\nabla
^{--}$ are as in eqs(\ref{19}). To classify these local complex
deformations,\ it is interesting to consider first local deformations of $%
T^{\ast }P^{1}$; then come back to those of $T^{\ast }S^{3}$.

\subsubsection{Local deformations of $T^{\ast }P^{1}$}

Local complex deformation of $T^{\ast }P^{1}$ are special deformations of
the conifold; they are given by subset $\mathcal{J}_{0}^{\left( 0\right) }$
of harmonic function $\zeta ^{0}=\zeta ^{0}\left( U^{+},V^{-}\right) $
invariant under projective transformations
\begin{equation}
\left( U^{+},V^{-}\right) \longrightarrow \left( \lambda U^{+},\frac{1}{%
\lambda }V^{-}\right) .
\end{equation}%
Note that under above scaling, we have in general the following sections,%
\begin{equation}
\mathcal{J}_{0}^{\left( q\right) }=\left\{ \quad \xi |_{T^{\ast
}P^{1}}\equiv \zeta ^{q}\right\} ,\qquad \text{ }q\in Z.
\end{equation}%
As we will see later, the conifold deformations $\xi \left(
U^{+},V^{-}\right) $ have in general the following decomposition
\begin{equation}
\xi \left( U^{+},V^{-}\right) =\sum_{n=-\infty }^{\infty }z^{q}\zeta
^{q}\left( U^{+},V^{-}\right) ,\qquad U^{+\gamma }V_{\gamma }^{-}=\mu ,
\end{equation}%
where $\zeta ^{q}\left( U^{+},V^{-}\right) $ are homogeneous harmonic
functions constrained as follows:%
\begin{equation}
\zeta ^{q}\left( \lambda U^{+\gamma },\frac{1}{\lambda }V_{\gamma
}^{-}\right) =\lambda ^{q}\zeta ^{q}\left( U^{+\gamma },V_{\gamma
}^{-}\right) ,\qquad \text{ }q\in Z,  \label{29}
\end{equation}%
for any non zero complex number $\lambda $. Local complex deformations of $%
T^{\ast }P^{1}$ are then associated with the restriction,%
\begin{equation}
\xi |_{T^{\ast }P^{1}}=\zeta ^{0}\left( U^{+\gamma },V_{\gamma }^{-}\right) .
\end{equation}%
In the $SL\left( 2,C\right) $ harmonic differential operator language, the
symmetry property
\begin{equation}
\zeta ^{0}\left( \lambda U^{+\gamma },\frac{1}{\lambda }V_{\gamma
}^{-}\right) =\zeta ^{0}\left( U^{+\gamma },V_{\gamma }^{-}\right)
\end{equation}%
means that local complex deformations of $T^{\ast }P^{1}$ are constrained as
follows,%
\begin{equation}
\left[ \nabla ^{0},\zeta ^{0}\left( U^{+\gamma },V_{\gamma }^{-}\right) %
\right] =0,  \label{29c}
\end{equation}%
but we still have,%
\begin{equation}
\nabla ^{++}\zeta ^{0}\left( U^{+\gamma },V_{\gamma }^{-}\right) \neq
0,\qquad \nabla ^{--}\zeta ^{0}\left( U^{+\gamma },V_{\gamma }^{-}\right)
\neq 0.  \label{29d}
\end{equation}%
By comparing with the global parameter $\mu $, one sees that the special
local deformations $\zeta ^{0}\left( U^{+\gamma },V_{\gamma }^{-}\right) $
carries the same $C^{\ast }$ charge as $\mu $ eq(\ref{29c}); but have a non
trivial dependence on the harmonic variables eqs(\ref{29d}). The general
solution of the constraint eq(\ref{29c}),%
\begin{equation}
\mu +\zeta ^{0}\left( U^{+},V^{-}\right) =\zeta _{0}^{0}+\sum_{n=1}^{\infty
}\zeta ^{\left( \alpha _{1}...\alpha _{n},\beta _{1}...\beta _{n}\right)
}U_{(\alpha _{1}}^{+}...U_{\alpha _{n}}^{+}V_{\beta _{1}}^{-}...V_{\beta
_{n})},  \label{ts}
\end{equation}%
where the zero mode $\zeta _{0}^{0}=\mu $ and where the non zero harmonic
modes $\zeta ^{\left( \alpha _{1}...\alpha _{n},\beta _{1}...\beta
_{n}\right) }$ are $SL\left( 2,C\right) $ tensors given by,%
\begin{equation}
\zeta ^{\left( \alpha _{1}...\alpha _{n},\beta _{1}...\beta _{n}\right)
}=\int_{T^{\ast }P^{1}}U^{(+\alpha _{1}}...U^{+\alpha _{n}}V^{-\beta
_{1}}...V^{-\beta _{n})}\zeta ^{0}\left( U^{+},V^{-}\right) .
\end{equation}%
In getting this relation, we have used the following harmonic space identity
for $T^{\ast }P^{1}$,%
\begin{equation}
\int_{T^{\ast }P^{1}}(U^{+})^{(m}(V^{-})^{n)}(U^{+})_{(k}(V^{-})_{l)}=\frac{%
(-1)^{n}m!n!}{(m+n+1)!}\delta _{(j_{1}}^{(i_{1}}\cdots \delta
_{j_{k+l)}}^{i_{m+n)}}.
\end{equation}%
For more details on the properties of harmonic variables, harmonic
distributions and harmonic integration on $T^{\ast }P^{1}$; see appendix
section. \newline
Note that in the harmonic expansion (\ref{ts}) and because of the
conservation of $C^{\ast }$ charge, harmonic variables $U_{\alpha _{i}}^{+}$
and $V_{\beta _{j}}^{-}$ are usually coupled and appear everywhere in pairs.
Now, using the above features of harmonic variables, one may classify local
complex deformation parameters
\begin{equation}
\zeta ^{\left( \alpha _{1}...\alpha _{n},\beta _{1}...\beta _{n}\right)
},\qquad n\geq 1,
\end{equation}%
in terms of the parameters of $SDiff\left( T^{\ast }P^{1}\right) $, the
group of area preserving diffeomorphisms of $T^{\ast }P^{1}$. Indeed
restricting, in a first step, the set $\mathcal{J}_{0}$ of local complex
deformations of $T^{\ast }P^{1}$ to the subset $\mathcal{J}_{0}^{\left(
n\right) }$ generated by the following constraint eqs,%
\begin{eqnarray}
\left( \nabla ^{++}\right) ^{n+1}\zeta ^{0} &=&0,  \label{cst} \\
\nabla ^{0}\zeta ^{0} &=&0.  \label{csa}
\end{eqnarray}%
where $n$ is a positive integer, we get the two following: \newline
(\textbf{1}) The cardinal of $\mathcal{J}_{0}^{\left( n\right) }$ is exactly
equal to the dimension of the $SL\left( n+1,C\right) $ group.
\begin{equation}
order\left( \mathcal{J}_{0}^{\left( N\right) }\right) =\dim \left[ SL\left(
N+1,C\right) \right] =N^{2}-1.
\end{equation}%
To see this identity, it is enough to check it on the $n=1$ and $n=2$
leading terms. For $n=1$, the solution $\zeta ^{0}{}_{\left( 1\right) }$ of
the constraint eqs (\ref{cst}-\ref{csa}) reads as%
\begin{equation}
\zeta ^{0}{}_{\left( 1\right) }\left( U^{+\gamma },V_{\gamma }^{-}\right)
=\zeta ^{\left( \alpha ,\beta \right) }U_{(\alpha }^{+}V_{\beta )}^{-},
\end{equation}%
and so we have,%
\begin{eqnarray}
order\left( \mathcal{J}_{0}^{\left( 2\right) }\right) &=&order\left\{ \zeta
^{\left( 1,1\right) },\zeta ^{\left( 1,2\right) },\zeta ^{\left( 2,2\right)
}\right\}  \notag \\
&=&3=\left( 1+1\right) ^{2}-1.
\end{eqnarray}%
For the case $n=2$, the solution $\zeta ^{0}{}_{\left( 2\right) }$ of the
previous constraint eqs is given by%
\begin{equation}
\zeta _{\left( 2\right) }^{0}\left( U^{+\gamma },V_{\gamma }^{-}\right)
=\zeta ^{\left( \alpha ,\beta \right) }U_{(\alpha }^{+}V_{\beta )}^{-}+\zeta
^{\left( \alpha \gamma ,\beta \delta \right) }U_{(\alpha }^{+}U_{\gamma
}^{+}V_{\beta }^{-}V_{\delta )}^{-},
\end{equation}%
together with
\begin{equation}
order\left( \mathcal{J}_{0}^{\left( 3\right) }\right) =3+5=8=\left(
2+1\right) ^{2}-1.
\end{equation}%
More generally, we have for generic case $n=N$,\ the solution $\zeta
^{0}{}_{\left( N\right) }$,%
\begin{equation}
\zeta ^{0}\left( U^{+},V^{-}\right) =\sum_{j=1}^{N}\zeta ^{\left( \alpha
_{1}...\alpha _{j}\beta _{1}...\beta _{j}\right) }U_{(\alpha
_{1}}^{+}...U_{\alpha _{j}}^{+}V_{\beta _{1}}^{-}...V_{\beta _{j})},
\label{30}
\end{equation}%
and the order of the set $\mathcal{J}_{0}^{\left( N\right) }$ is given by
\begin{equation}
order\left( \mathcal{J}_{0}^{\left( N\right) }\right) =\sum_{j=1}^{N}\left(
2j+1\right)
\end{equation}%
which is also equal to $\left( N+1\right) ^{2}-1$. \newline
(\textbf{2}) Since the set $\mathcal{J}_{0}$ corresponds just to the limit $%
\mathcal{J}_{0}^{\left( +\infty \right) }$, we have then the correspondence $%
\mathcal{J}_{0}\sim SL\left( \infty ,C\right) $ which is nothing but the
volume preserving diffeomorphism group $SDiff\left( T^{\ast }P^{1}\right) $.
\newline
In the end of this discussion note that the constraint eqs (\ref{cst}-\ref%
{csa}) which may be rewritten as%
\begin{eqnarray}
\left[ \nabla ^{++},\left( \nabla ^{++}\right) ^{n}\xi ^{0}\right] &=&0, \\
\left[ \nabla ^{0},\left( \nabla ^{++}\right) ^{n}\xi ^{0}\right]
&=&2n\left( \nabla ^{++}\right) ^{n}\xi ^{0},
\end{eqnarray}%
transform in spin $s$\ representation of $SL\left( 2,C\right) $ and the
solutions $\zeta _{\left( n-j\right) }^{0}$ with $1\leq j\leq n-1$ are just
kernels of of the harmonic operator $\left( \nabla ^{++}\right) ^{n}$. In
the following table, we give the representation group structure of the
harmonic modes $\zeta ^{\left( \alpha _{1}...\alpha _{n},\beta _{1}...\beta
_{n}\right) }$,%
\begin{equation}
\begin{tabular}{|l|l|l|l|l|l|}
\hline
Parameters & spin $SU_{u}\left( 2\right) $ & spin $SU_{v}\left( 2\right) $ &
spin $SL\left( 2\right) $ & $C^{\ast }$ charge & scale dim \\ \hline
$U^{+\alpha }$ & $\ \ \ \ \ \ \ \ \frac{1}{2}$ & $\ \ \ \ \ \ \ \ 0$ & $\ \
\ \ \ \ \ \frac{1}{2}$ & $\ \ \ \ \ \ \ 1$ & $\ \ \ \ \ \ \ 1$ \\ \hline
$V^{-\alpha }$ & $\ \ \ \ \ \ \ \ 0$ & $\ \ \ \ \ \ \ \ \frac{1}{2}$ & $\ \
\ \ \ \ \ \frac{1}{2}$ & $\ \ \ \ -1$ & $\ \ \ \ \ \ \ 1$ \\ \hline
$\mu \equiv \zeta _{\left( 0,0\right) }$ & $\ \ \ \ \ \ \ \ 0$ & $\ \ \ \ \
\ \ \ 0$ & $\ \ \ \ \ \ \ 0$ & $\ \ \ \ \ \ \ 0$ & $\ \ \ \ \ \ \ 2$ \\
\hline
$\zeta ^{\left( \alpha \beta \right) }\equiv \zeta _{\left( 1,1\right) }$ & $%
\ \ \ \ \ \ \ \ \frac{1}{2}$ & $\ \ \ \ \ \ \ \ \frac{1}{2}$ & $\ \ \ \ \ \
\ 1$ & $\ \ \ \ \ \ \ 0$ & $\ \ \ -0$ \\ \hline
$\zeta _{\left( n,n\right) }$ & $\ \ \ \ \ \ \ \ \frac{n}{2}$ & $\ \ \ \ \ \
\ \ \frac{n}{2}$ & $\ \ \ \ \ \ \ n$ & $\ \ \ \ \ \ \ 0$ & $\ \ \ 2-2n$ \\
\hline
\end{tabular}%
\text{,}  \label{31}
\end{equation}%
with $n$ an arbitrary positive integer and where we have set $\zeta ^{\left(
\alpha _{1}...\alpha _{n},\beta _{1}...\beta _{n}\right) }\equiv \zeta
_{\left( n,n\right) }$.

\subsubsection{Local deformations fo $T^{\ast }S^{3}$}

In the conifold case there is no constraint eqs type (\ref{29},\ref{29c})
and so the set eq(\ref{j}) of local complex deformations of $T^{\ast }S^{3}$
is richer than in $T^{\ast }P^{1}$. For $T^{\ast }S^{3}$, there no
projective symmetry restricting the harmonic function $\xi \left(
U^{+},V^{-}\right) $. However, because of the fibration $T^{\ast }S^{3}\sim
C^{\ast }\times T^{\ast }P^{1}$, we can think about
\begin{equation}
\xi \left( U^{+},V^{-}\right)
\end{equation}%
with $U^{+}$ and $V^{-}$ on conifold as
\begin{equation}
\zeta \left( z,U^{+\prime },V^{-\prime }\right)
\end{equation}%
where now $z$ and $\left( U^{+\prime },V^{-\prime }\right) $ are the local
coordinates $C^{\ast }$ and $T^{\ast }P^{1}$ respectively. In this
parameterization, conifold is embedded in the non compact projective space $%
WP_{\left( -1,1,1,-1,-1\right) }^{4}$ and so we have the correspondence,
\begin{equation}
\xi \left( T^{\ast }S^{3}\right) \qquad \Longleftrightarrow \qquad \zeta
\left( \frac{1}{\lambda }z,\lambda U^{+\prime },\frac{1}{\lambda }V^{-\prime
}\right) =\zeta \left( z,U^{+\prime },V^{-\prime }\right) .
\end{equation}%
Dropping the primes on $U^{+\prime }$ and $V^{-\prime }$; then expanding $%
\zeta \left( z,U^{+},V^{-}\right) $ in a Laurent series in power of $z$, we
get the following,%
\begin{equation}
\zeta \left( z,U^{+},V^{-}\right) =\sum_{q=-\infty }^{\infty }z^{q}\zeta
^{q}\left( U^{+},V^{-}\right) ,  \label{31a}
\end{equation}%
where $\zeta ^{q}\left( U^{+},V^{-}\right) $ are harmonic functions
(sections) on $T^{\ast }P^{1}$ given by,%
\begin{equation}
\zeta ^{q}\left( U^{+},V^{-}\right) =\int_{\gamma _{0}}\frac{dz}{2i\pi }%
z^{-q-1}\zeta \left( z,U^{+},V^{-}\right) ,
\end{equation}%
with $\gamma _{0}$ a contour path surrounding the origin in the $C^{\ast }$
plane. As one sees, the Laurent modes $\zeta ^{q}\left( U^{+},V^{-}\right) $
transform under the projective transformation $z\rightarrow \frac{1}{\lambda
}z,$ $U^{+}\rightarrow \lambda U^{+}$ and $V^{-}\rightarrow \frac{1}{\lambda
}V^{-}$ as,%
\begin{equation}
\zeta ^{q}\left( \lambda U^{+},\frac{1}{\lambda }V^{-}\right) =\lambda
^{q}\zeta ^{q}\left( U^{+},V^{-}\right) ,\qquad q\in Z,  \label{32}
\end{equation}%
in agreement with results on homogeneous functions $\zeta ^{q}\left(
U^{+},V^{-}\right) $ living on $T^{\ast }P^{1}$ and
\begin{equation}
\nabla ^{0}\zeta ^{q}\left( U^{+},V^{-}\right) =q\zeta ^{q}\left(
U^{+},V^{-}\right) .  \label{34}
\end{equation}%
Besides the case $q=0$ which is associated with local complex deformations
of $T^{\ast }P^{1}$, the solution of eq(\ref{34}) depends on the sign of $q$%
. For $q$ positive, say $q=+m$, the general solution is given by the
following harmonic expansion,
\begin{eqnarray}
\zeta ^{m}\left( U^{+},V^{-}\right) &=&\mu ^{\frac{m}{2}}u_{(\alpha
_{1}}^{+}...u_{\alpha _{m})}^{+}\zeta ^{\left( \alpha _{1}...\alpha
_{m}\right) }  \notag \\
&&\mu ^{\frac{m+2}{2}}u_{(\alpha _{1}}^{+}...u_{\alpha _{m+1}}^{+}v_{\beta
_{1})}^{-}\zeta ^{\left( \alpha _{1}...\alpha _{m}\alpha _{m+1},\beta
_{1}\right) }+...,  \label{35}
\end{eqnarray}%
where we have used the scaling (\ref{sc}) and where the $SL\left( 2,C\right)
$ representation structure of $\zeta ^{\left( \alpha _{1}...\alpha
_{m}\alpha _{m+j},\beta _{1}...\beta _{j}\right) }$ is as in previous table;
see also table below. \newline
Moreover setting the convention notation,
\begin{equation}
\zeta _{\left( m+j,j\right) }\equiv \zeta ^{\left( \alpha _{1}...\alpha
_{m+j},\beta _{1}...\beta _{j}\right) },
\end{equation}%
eq(\ref{35}) reads as
\begin{equation}
\zeta ^{m}\left( U^{+},V^{-}\right) =\sum_{j=0}^{\infty }\zeta _{\left(
m+j,j\right) }U^{(+m+j}V^{-j)}.
\end{equation}%
For $q$ a negative integer, say $q=-m$, we have the harmonic expansion,%
\begin{eqnarray}
\zeta ^{-m}\left( U^{+},V^{-}\right) &=&\mu ^{\frac{m}{2}}v_{\beta
_{1}}^{-}...v_{\beta _{m}}^{-}\zeta ^{\left( \beta _{1}...\beta _{m}\right) }
\notag \\
&&+\sum_{j\geq 1}\mu ^{\frac{m+2j}{2}}u_{\alpha _{1}}^{+}...u_{\alpha
_{j}}^{+}v_{\beta _{1}}^{-}...v_{\beta _{m+j}}^{-}\zeta ^{\left( \alpha
_{1}...\alpha _{j},\beta _{1}...\beta _{m+j}\right) }.  \label{36}
\end{eqnarray}%
Similarly, setting $\zeta _{\left( j,m+j\right) }\equiv \zeta ^{\left(
\alpha _{1}...\alpha _{j},\beta _{1}...\beta _{m+j}\right) }$, the previous
expansion of $\zeta ^{-m}\left( U^{+},V^{-}\right) $ reads also as%
\begin{equation}
\zeta ^{-m}\left( U^{+},V^{-}\right) =\sum_{j=0}^{\infty }\zeta _{\left(
j,m+j\right) }U^{(+j}V^{-m-j)}.
\end{equation}%
Putting these expansions altogether, we have the general result on the
harmonic expansion of local complex deformations of conifold,%
\begin{eqnarray}
\zeta \left( z,U^{+},V^{-}\right) &=&\sum_{j=0}^{\infty }\zeta _{\left(
j,j\right) }U^{(+j}V^{-j)}  \notag \\
&&+\sum_{q=1}^{\infty }z^{q}\left( \sum_{j=0}^{\infty }\zeta _{\left(
q+j,j\right) }U^{(+q+j}V^{-j)}\right) \\
&&+\sum_{q=1}^{\infty }z^{-q}\left( \sum_{j=0}^{\infty }\zeta _{\left(
j,q+j\right) }U^{(+j}V^{-q-j)}\right) ,  \notag
\end{eqnarray}%
which may be put into the following formal condensed form,%
\begin{equation}
\zeta \left( z,U^{+},V^{-}\right) =\sum_{q=-\infty }^{\infty }z^{q}\left(
\sum_{j=0}^{\infty }\left[ \zeta _{\left( q+j,j\right)
}U^{(+q+j}V^{-j)}+\zeta _{\left( j,q+j\right) }U^{(+j}V^{-q-j)}\right]
\right) .
\end{equation}%
Therefore the local complex deformations of conifold are classified as
follows%
\begin{equation}
\begin{tabular}{|l|l|l|l|l|l|}
\hline
& spin s$_{u}$ & spin s$_{v}$ & spin s & $C^{\ast }$ charge & Scale dim \\
\hline
$\zeta _{\left( n,m\right) }$ & $\ \ \frac{n}{2}$ & $\ \ \frac{m}{2}$ & $\ \
\frac{n+m}{2}$ & $\ \ \ n-m$ & $2-\left( \left\vert n\right\vert +\left\vert
m\right\vert \right) $ \\ \hline
\end{tabular}%
\text{,}
\end{equation}%
where $n$ and $m$ are arbitrary positive integers and where we have set
\begin{equation}
\zeta ^{\left( \alpha _{1}...\alpha _{j},\beta _{1}...\beta _{m+j}\right)
}\equiv \zeta _{\left( j,m+j\right) }
\end{equation}%
and
\begin{equation}
\zeta ^{\left( \alpha _{1}...\alpha _{m}\alpha _{m+j},\beta _{1}...\beta
_{j}\right) }\equiv \zeta _{\left( m+j,,j\right) }.
\end{equation}%
A noted in section 2, a particularly interesting subclass of conifold
complex deformations are those associated with eqs(\ref{ab1}-\ref{ab5}). In
the harmonic language, these deformations correspond to%
\begin{equation}
\zeta \left( z,U^{+},V^{-}\right) \sim \zeta \left( z,U^{+}\right) +%
\widetilde{\zeta }\left( z,V^{-}\right) ,  \label{39}
\end{equation}%
with
\begin{equation}
\zeta \left( z,U^{+}\right) =\sum_{q\geq 1}z^{q}\zeta ^{q}\left( U^{+}\right)
\end{equation}%
and
\begin{equation}
\widetilde{\zeta }\left( z,V^{-}\right) =\sum_{q\geq 1}z^{-q}\zeta
^{-q}\left( V^{-}\right)
\end{equation}%
where $\zeta ^{\pm q}$ obey similar relations as before. These expansions
recall the usual Fourrier expansion on the circle namely%
\begin{equation}
t\left( \theta \right) =\sum_{n>0}t_{n}e^{in\theta
}+\sum_{n>0}t_{-n}e^{-in\theta }.
\end{equation}%
In fact there a \textit{1 to 1} correspondence between Fourrier analysis on
the unit circle $S^{1}$ and the harmonic expansion of the unit $S^{3}$
sphere. In what follows, we give a dictionary allowing the passage between
the two formalisms.

\subsection{$S^{1}\longleftrightarrow S^{3}$ dictionary}

One of the lessons one learns from above harmonic analysis on the 3-sphere
and the one developed in the appendix, is that there is a one to one
correspondence between Fourrier expansion on $S^{1}$ and harmonic analysis $%
S^{3}$. This analysis extends also $T^{\ast }S^{1}$ and $T^{\ast }S^{3},$%
\begin{equation}
T^{\ast }S^{1}\qquad \longleftrightarrow \qquad T^{\ast }S^{3}\text{.}
\end{equation}%
In the following table we collect some relevant correspondences which will
be used in forthcoming sections. Other useful relations will be given at
proper times.

\begin{equation}
\begin{tabular}{|l|l|l|}
\hline
& $S^{1}\sim U\left( 1\right) $ & $S^{3}\sim S^{1}\times S^{2}$ \\ \hline
{\small basis}$\quad n\in N$ & ${\small b}_{n}{\small =e}^{in\theta }{\small %
,}$ & ${\small e}^{in\theta }{\small b}_{n}{\small =e}^{in\theta }{\small u}%
^{(+n)}{\small ,\quad {}}$ \\ \hline
& ${\small b}_{-n}{\small =e}^{-in\theta }$ & ${\small e}^{-in\theta }%
{\small b}_{-n}{\small =e}^{-in\theta }{\small u}^{(-n)}{\small ,\quad }$ \\
\hline
{\small charge operator} & ${\small Q=}\frac{\partial }{i\partial \theta }$
& ${\small D}_{u}^{0}{\small =}\left( u^{+\alpha }\frac{\partial }{\partial
u^{+\alpha }}-u^{-\beta }\frac{\partial }{\partial u^{-\beta }}\right) $ \\
\hline
{\small eigenvalue eqs} & ${\small Qb}_{\pm n}{\small =\pm nb}_{\pm n}$ & $%
{\small D}_{u}^{0}{\small b}_{\pm n}{\small =\pm nb}_{\pm n}$ \\ \hline
{\small Expansion} & ${\small F}\left( \theta \right) {\small =}\sum_{n\neq
0}{\small F}_{n}{\small e}^{in\theta }$ & ${\small \zeta }\left(
z,u^{+}\right) {\small +}\widetilde{\zeta }\left( z,v^{-}\right) {\small =}$
\\ \hline
&  & $\sum_{n>0}\left( {\small e}^{in\theta }{\small b}_{n}\zeta _{\left(
n,0\right) }+{\small e}^{-in\theta }{\small b}_{-n}\zeta _{\left( 0,n\right)
}\right) $ \\ \hline
&  & ${\small \zeta }_{\left( n,0\right) }{\small =\zeta }_{\left( \alpha
_{1}...\alpha _{n}\right) }{\small ,\quad \zeta }_{\left( 0,n\right) }%
{\small =\zeta }^{\left( \beta _{1}...\beta _{n}\right) }$ \\ \hline
{\small integral measure} & $\int_{S^{1}}\frac{d\theta }{2\pi }{\small =1}$
& $\int_{S^{3}}{\small d}^{3}{\small u=}\int_{S^{1}}\frac{d\theta }{2\pi }%
\int_{S^{2}}{\small d}^{2}{\small u=1}$ \\ \hline
{\small Distributions} & ${\small \delta }_{n,m}{\small =}\int_{S^{1}}\frac{%
d\theta }{2\pi }{\small b}_{n}{\small b}_{-m}$ & ${\small \delta }^{\left(
n,m\right) }{\small =}\int_{S^{2}}{\small d}^{2}{\small ub}_{n}{\small b}%
_{-m}$ \\ \hline
{\small Modes, }${\small n>0}$ & ${\small F}_{n}{\small =}\int_{S^{1}}\frac{%
d\theta }{2\pi }{\small e}^{-in\theta }{\small F}\left( \theta \right) $ & $%
{\small \zeta }_{\left( n,0\right) }{\small =}\int_{S^{2}}{\small d}^{2}%
{\small u\zeta }^{+n}\left( u\right) {\small b}_{-n}{\small ,}$ \\ \hline
& ${\small F}_{-n}{\small =}\int_{S^{1}}\frac{d\theta }{2\pi }{\small e}%
^{in\theta }{\small F}\left( \theta \right) $ & ${\small \zeta }_{\left(
0,n\right) }{\small =}\int_{S^{2}}{\small d}^{2}{\small u\zeta }^{-n}\left(
u\right) {\small b}_{n}$ \\ \hline
{\small products} & $\int_{S^{1}}\frac{d\theta }{2\pi }{\small F}^{2}\left(
\theta \right) $ & $\int_{S^{3}}{\small d}^{3}{\small u}\left[ \zeta \left(
z,u^{+},v^{-}\right) \right] ^{2}{\small =}$ \\ \hline
& ${\small =}\sum_{n}{\small F}_{n}{\small F}_{-n}$ & $\sum_{n}\int_{S^{2}}%
{\small d}^{2}{\small u\zeta }^{+n}{\small \zeta }^{-n}{\small =}\sum_{n}%
{\small \zeta }_{\left( n,0\right) }{\small \zeta }_{\left( 0,n\right) }$ \\
\hline
\end{tabular}
\label{di}
\end{equation}%
where we have set,%
\begin{eqnarray}
{\small u}^{(+n)} &{\small =}&{\small u}^{(+\alpha _{1}}{\small ...u}%
^{+\alpha _{n})},  \notag \\
{\small u}^{(-n)} &{\small =}&{\small u}_{(\beta _{1}}^{-}{\small ...u}%
_{\beta _{n})}^{-}.
\end{eqnarray}%
By scaling the variables, one can immediately write down the corresponding
relations for real spheres with generic radii. \newline
More details on the properties of harmonic analysis, differential and
integral calculus as well as harmonic distributions may be found in $\cite%
{36,58}$ and appendix section. In what follows, we shall use the results on
the table and send to appendix for technical details.

\section{Topological string on conifold}

We start by reviewing the structure of the partition function $\mathcal{Z}%
_{top}$ of the B model topological string on conifold using complex
holomorphic analysis. Then we reconsider the building of $\mathcal{Z}_{top}$%
; but this time using harmonic frame work. By comparing results for the
restriction of $T^{\ast }S^{3}$ down to $T^{\ast }S^{1}$ and using
dictionary eqs(\ref{di}), we conjecture the general structure of $\mathcal{Z}%
_{top}=\exp \left( \mathcal{F}\left( \zeta ,\widetilde{\zeta }\right)
\right) $ preserving manifestly $SL\left( 2,C\right) $ symmetry. For genus
zero for instance, the leading terms of free energy reads as,%
\begin{eqnarray}
\mathcal{F}_{0}\left( \zeta ,\widetilde{\zeta }\right) &=&-\frac{1}{g_{s}^{2}%
}\sum_{n>0}\frac{\mu ^{n-2}}{n}\int_{T^{\ast }S^{2}}\left( \zeta ^{n}%
\widetilde{\zeta }^{-n}\right)  \notag \\
&&+\frac{1}{g_{s}^{2}}\sum_{n_{1}+n_{2}+n_{3}=0}p^{\frac{\left\vert
n_{1}\right\vert +\left\vert n_{2}\right\vert +\left\vert n_{3}\right\vert -2%
}{2}-2}\left( \int_{T^{\ast }S^{2}}\zeta ^{n_{1}}\zeta ^{n_{2}}\zeta
^{n_{3}}\right) \\
&&+...  \notag
\end{eqnarray}%
where for positive integers $\zeta ^{+n}=\zeta ^{+n}\left( u^{+}\right) $
and $\zeta ^{-n}=\zeta ^{-n}\left( v^{-}\right) $.

\subsection{$\mathcal{Z}_{top}$ Partition function}

There is a nice correspondence between $SU\left( 2\right) \times SU\left(
2\right) $ symmetry classifying the deformation parameters of the conifold,
\begin{equation}
xy-zw=\mu +T\left( x,y,z,w\right) \text{\quad },
\end{equation}%
and the $SU\left( 2\right) \times SU\left( 2\right) $ symmetry of the
conformal theory $c=1$ at the self dual radius. Following the study of $\cite%
{15,30}$, we distinguish two basic sectors. \newline
The first one concern restricting $T\left( x,y,z,w\right) $ to local complex
deformations to,
\begin{equation}
T_{momentum}\left( x,y\right) =t\left( x,y\right) \text{\quad },
\end{equation}%
with $t\left( x,y\right) $ taken as,%
\begin{equation}
t\left( x,y\right) =\tau \left( x\right) +\widetilde{\tau }\left( y\right) +%
\text{ ...\quad }.
\end{equation}%
Local deformations involve the $x$ and $y$\ variables only; i.e setting
\begin{equation}
z=w=0\text{\quad }.
\end{equation}%
This restriction corresponds to turning on momentum modes in the $c=1$ non
critical string theory. \newline
The second sector concern the local complex deformations $T\left(
x,y,z,w\right) $ taken as
\begin{equation}
T_{winding}\left( z,w\right) =s\left( z,w\right) \text{\quad },
\end{equation}%
where now,%
\begin{equation}
s\left( z,w\right) =\sigma \left( z\right) +\widetilde{\sigma }\left(
w\right) +\text{ ...\quad },
\end{equation}%
involving $z$ and $w$\ variables only; i.e
\begin{equation}
x=y=0\text{\quad }.
\end{equation}%
This restriction corresponds to turning on winding modes.\newline
Arbitrary complex deformations of the conifold generated by arbitrary $%
T=T\left( x,y,z,w\right) $ correspond then to turning all modes; that is
momentum modes
\begin{equation}
\left\{ \tau _{n},\widetilde{\tau }_{n}\right\}
\end{equation}%
and winding ones
\begin{equation}
\left\{ \sigma _{n},\widetilde{\sigma }_{n}\right\} .
\end{equation}%
In this general case however, there is only partial results and so needs
more investigation. The harmonic set up we will develop later give precious
informations for this issue; but to make comparisons let us develop a little
bit the usual Laurrent complex holomorphic analysis by focusing in a first
step on momentum and winding modes sectors separately. Then explore the way
to couple them. Once we do this, we turn back to harmonic analysis formalism.

In the case where complex deformations $T\left( x,y,z,w\right) $ are taken
as $T_{momentum}\left( x,y\right) $, one is mainly restricting local complex
deformations of $T^{\ast }S^{3}$ to the complex one dimension $T^{\ast
}S^{1} $ region of the conifold,
\begin{equation}
xy=\mu ,\qquad z=w=0,
\end{equation}%
which is then deformed as,
\begin{equation}
xy=\mu +\tau \left( x\right) +\widetilde{\tau }\left( y\right) \text{ +
corrections\quad }.  \label{ms1}
\end{equation}%
On the circle $S^{1}$ describing the real slice of the cotangeant bundle $%
T^{\ast }S^{1}$ and obtained by setting
\begin{equation}
y=\overline{x},
\end{equation}%
the previous deformed equation reduces to,
\begin{equation}
\left\vert x\right\vert ^{2}=p+\tau \left( x\right) +\overline{\tau \left(
x\right) }\text{ + corrections\quad },  \label{ms2}
\end{equation}%
where $p=\func{Re}\left( \mu \right) $ is the squared radius of the circle $%
S^{1}$.

A similar analysis may be done for the winding sector. Local complex
deformations of $T^{\ast }S^{3}$ is restricted to a second complex one
dimension $T^{\ast }S^{1}$ region of the conifold namely%
\begin{equation}
zw=\mu ,\qquad x=y=0\text{\quad }.
\end{equation}
Local complex deformations of the corresponding real slice geometry reads as
\begin{equation}
\left\vert z\right\vert ^{2}=p+\sigma \left( z\right) +\overline{\sigma
\left( z\right) }\text{ + corrections\quad }.
\end{equation}%
where $w=\overline{z}$ and $x=y=0$.

\subsubsection{Geometric interpretation}

To make contact with the parametrisation of $S^{3}$ analysis eqs(\ref{3}-\ref%
{6}), which we recall below for convenience,%
\begin{eqnarray}
x\left( \vartheta ,\psi ,\varphi \right) &=&\sqrt{p}e^{\frac{i}{2}\left(
\psi -\varphi \right) }\sin \frac{\vartheta }{2},\qquad y\left( \vartheta
,\psi ,\varphi \right) =\sqrt{p}e^{\frac{-i}{2}\left( \psi -\varphi \right)
}\sin \frac{\vartheta }{2},  \notag \\
z\left( \vartheta ,\psi ,\varphi \right) &=&-\sqrt{p}e^{\frac{i}{2}\left(
\psi +\varphi \right) }\cos \frac{\vartheta }{2},\qquad w\left( \vartheta
,\psi ,\varphi \right) =\sqrt{p}e^{\frac{-i}{2}\left( \psi +\varphi \right)
}\cos \frac{\vartheta }{2},  \label{s3}
\end{eqnarray}%
the previous two restrictions of $S^{3}$ down to $S^{1}$ correspond to the
following: \newline
Either fixing the degrees of freedom $\vartheta ,$ $\varphi $ and $\psi $\
angles as,
\begin{equation}
\vartheta =\pi ,\qquad z=0,\quad w=0\text{\quad },
\end{equation}%
for the first circle%
\begin{equation}
x=\sqrt{p}e^{\frac{i}{2}\left( \psi -\varphi \right) },\quad y=\sqrt{p}e^{%
\frac{-i}{2}\left( \psi -\varphi \right) }\text{\quad }.
\end{equation}%
Or by fixing $\vartheta $ like%
\begin{equation}
\vartheta =0,\qquad x=y=0\text{\quad },
\end{equation}%
leading to the second circle%
\begin{equation}
z=\sqrt{p}e^{\frac{i}{2}\left( \psi +\varphi \right) },\quad w=\sqrt{p}e^{%
\frac{-i}{2}\left( \psi +\varphi \right) }\text{\quad }.
\end{equation}%
To get an explicit expression of $\mathcal{Z}_{top}\left( \tau ,\widetilde{%
\tau }\right) $, one uses a set of approximations. First thinks about $%
T_{momentum}\left( x,y\right) $ as given by the following perturbative
development,%
\begin{equation}
t\left( x,y\right) =\tau \left( x\right) +\widetilde{\tau }\left( y\right) +%
\mathcal{O}\left( \frac{\tau \widetilde{\tau }}{\mu },\frac{\tau ^{2}}{\mu },%
\frac{\widetilde{\tau }^{2}}{\mu }\right) ,
\end{equation}%
where, at leading order, left and right moving modes are decoupled.
Implementation of couplings requires going beyond the leading order of the
perturbation; they correspond to non linear corrections in the field
parameters $\tau \left( x\right) $ and $\widetilde{\tau }\left( y\right) $.
The one holomorphic variable functions $\tau \left( x\right) $ and $%
\widetilde{\tau }\left( y\right) $, supposed smaller with respect to $\mu $,
\begin{equation}
\tau \left( x\right) <\mu ,\qquad \widetilde{\tau }\left( y\right) <\mu
,\qquad \tau \left( x\right) \widetilde{\tau }\left( y\right) <\mu ^{2}\quad
,
\end{equation}%
for any $x,y\in T^{\ast }S^{1}\subset T^{\ast }S^{3}$, generate special
infinitesimal local complex deformations of the conifold. They have the
following Laurent expansions%
\begin{equation}
\tau \left( x\right) =\sum_{n=1}^{\infty }t_{n}x^{n},\qquad \widetilde{\tau }%
\left( y\right) =\sum_{n=1}^{\infty }t_{-n}y^{n}\quad .
\end{equation}%
On the real slice of $T^{\ast }S^{1}$, the complex variable $y$ get
identified with $\overline{x}$. Solving the reduced one dimensional geometry
$\left\vert x\right\vert ^{2}=p$ as
\begin{equation}
x=\sqrt{p}e^{i\theta },\qquad y=\sqrt{p}e^{-i\theta }\quad ,
\end{equation}%
then substituting in
\begin{equation}
t\left( x,\overline{x}\right) =\tau \left( x\right) +\overline{\tau }\left(
\overline{x}\right) +O\left( 2\right) \quad ,
\end{equation}%
the leading fluctuations that are linear in $t_{n}$ and $t_{-n}$ take the
form,
\begin{equation}
t\left( p,\theta \right) =\sum_{n=1}^{\infty }p^{\frac{\left\vert
n\right\vert }{2}}\left( t_{n}+t_{-n}\right) e^{in\theta }\quad ,  \label{fl}
\end{equation}%
where now $t_{-n}=\overline{t_{n}}$.

\subsubsection{Partition function}

The B model partition function $\mathcal{Z}_{top}\left( \tau ,\widetilde{%
\tau }\right) $ on the conifold is a holomorphic a functional depending on
the complex deformation moduli $\mu $ and $t_{+n}$, $\widetilde{t}_{+n}$;
and reads as usual as
\begin{equation}
\mathcal{Z}_{top}\left( \tau ,\widetilde{\tau }\right) =\exp \mathcal{F}%
\left( \tau ,\widetilde{\tau }\right)
\end{equation}%
where $\mathcal{F}\left( \tau ,\widetilde{\tau }\right) $ is the free
energy. Because of the correspondence between B model topological string on
conifold and $c=1$ non critical string, $\mathcal{F}\left( \tau ,\widetilde{%
\tau }\right) \equiv \mathcal{F}\left( t\right) $ is given by the free
energy $\mathcal{F}_{c=1}\left( t\right) $ and has the genus $g$ expansion,%
\begin{equation}
\mathcal{F}\left( t\right) =\left( \frac{\mu }{g_{s}}\right) ^{2}\mathcal{F}%
_{0}\left( t\right) +\sum_{g=1}^{\infty }\left( \frac{g_{s}}{\mu }\right)
^{2g-2}\mathcal{F}_{g}\left( t\right) \quad ,  \label{eq1}
\end{equation}%
where $g_{s}$ stands for the string coupling constant and where the genus $g$
free energy component $\mathcal{F}_{g}\left( t\right) $\ has the following
structure $\cite{30}$,%
\begin{equation}
\mathcal{F}_{g}\left( t\right) =\sum_{m\geq 2}P_{g}^{m}\left( n_{i}\right)
\dprod\limits_{i=1}^{n}\left( \mu ^{\frac{\left\vert n_{i}\right\vert }{2}%
-1}t_{n_{i}}\right) \quad .
\end{equation}%
In this relation, $P_{g}^{m}\left( n_{i}\right) $ is a polynomial in momenta
$n_{i}$ and has as degree $d=d\left( m,g\right) $ depending on $m$ and $g$.
Following $\cite{48,30}$, we have, amongst others, the two useful
informations on $P_{g}^{m}\left( n_{i}\right) $. First $d\left( 2,g\right)
=4g-1$ and second for $n$ a positive definite integer,%
\begin{equation}
P_{g=1}^{m=2}\left( n\right) =\frac{1}{24}\left( n-1\right) \left(
n^{2}-n-1\right) ,\qquad g=0,1,...\quad .
\end{equation}%
Restricting the analysis to the leading genus term, the free energy $%
\mathcal{F}\left( t\right) $ reduces mainly to the genus zero factor
\begin{equation}
g_{s}^{-2}\mu ^{2}\mathcal{F}_{0}\left( t\right)
\end{equation}%
which, in QFT$_{4}$ language, corresponds to particular F-terms in the
effective action of 4D supersymmetric Yang Mills theory. Genus $g\geq 1$
components given by
\begin{equation}
\mathcal{F}_{1}\left( t\right) +O\left[ \left( \frac{g_{s}}{\mu }\right) ^{2}%
\right]
\end{equation}%
are understood as gravitational corrections. \newline
In the super yang Mills approximation, the contribution of the genus zero
term in the free energy series read as,%
\begin{equation}
\mu ^{2}\mathcal{F}_{0}\left( t\right) =-\sum_{n>0}\frac{\mu ^{n}}{n}%
t_{n}t_{-n}+\frac{L_{3}\left( t\right) }{3!}+\frac{L_{4}\left( t\right) }{4!}%
+O\left( t^{5}\right) ,  \label{eq3}
\end{equation}%
where, for commodity, we have set%
\begin{eqnarray}
L_{3}\left( t\right) &=&\sum_{n_{1}+n_{2}+n_{3}=0}\mu ^{\frac{\left\vert
n_{1}\right\vert +\left\vert n_{2}\right\vert +\left\vert n_{3}\right\vert -2%
}{2}}t_{n_{1}}t_{n_{2}}t_{n_{3}},  \notag \\
L_{4}\left( t\right) &=&\sum_{n_{1}+n_{2}+n_{3}+n_{4}=0}\left( 1-\max
\left\{ \left\vert n_{i}\right\vert \right\} \right) \mu ^{\frac{\left\vert
n_{1}\right\vert +\left\vert n_{2}\right\vert +\left\vert n_{3}\right\vert
+\left\vert n_{4}\right\vert -4}{2}}t_{n_{1}}t_{n_{2}}t_{n_{3}}t_{n_{4}}.
\label{eq4}
\end{eqnarray}%
Before going ahead let us make three remarks which turn out to be helpful
when we consider the introduction of harmonic analysis to approach partition
function $\mathcal{Z}_{top}$. \newline
The first remark deals with rewriting free energy $\mathcal{F}\left(
t\right) $ as an integral over the circle $S^{1}$.%
\begin{equation}
\mathcal{F}\left[ p,t\right] =\int_{S^{1}}\frac{d\theta }{2\pi }\mathcal{H}%
\left[ t\left( p,\theta \right) \right] ,
\end{equation}%
where $\mathcal{H}\left[ t\left( p,\theta \right) \right] $ is a free energy
density on $S^{1}$. \newline
The second remark concerns implementation of winding modes and the third one
deals with how these deformations can be made $SU\left( 2\right) $ covariant.

\paragraph{(\textbf{1}) \textbf{Free energy density }$\mathcal{H}$\textbf{%
\qquad\ \ }\newline
}

Expansion $t\left( p,\theta \right) $ as
\begin{equation}
t\left( p,\theta \right) =\sum_{n\neq 0}^{\infty }f_{n}\left( p,\theta
\right)
\end{equation}%
with $f_{n}\left( p,\theta \right) =p^{\frac{\left\vert n\right\vert }{2}%
}t_{n}e^{in\theta }$. Then using the scaling operator $p\frac{\partial }{%
\partial p}$, which acts on $f_{n}\left( p,\theta \right) $ as
\begin{equation}
p\frac{\partial }{\partial p}f_{n}=\frac{\left\vert n\right\vert }{2}f_{n},
\end{equation}%
and the specific property of integrals on circle; in particular
\begin{equation}
\delta _{n,m}=\int_{S^{1}}\frac{d\theta }{2\pi }e^{i\left( n-m\right) \theta
},
\end{equation}%
one can re-express the energy libre $\mathcal{F}_{0}\left[ p,t\right] $ as a
functional integral over $S^{1}$. Indeed for genus zero terms, one can show
that free energy $\mu ^{2}\mathcal{F}_{0}\left[ p,t\right] $ can be
rewritten as
\begin{equation}
\int_{S^{1}}\mathcal{H}_{0}\left[ T\left( p,\theta \right) \right]
\end{equation}%
with
\begin{equation}
\mathcal{H}_{0}\left[ T\left( p,\theta \right) \right] =\int_{S^{1}}T\left(
p,\theta \right) \frac{2}{D}T\left( p,\theta \right) +\frac{1}{p}%
\int_{S^{1}}T^{3}\left( p,\theta \right) +O\left( \frac{1}{p^{2}}%
T^{4}\right) ,  \label{fr}
\end{equation}%
where $D=p\frac{\partial }{\partial p}$. The higher terms
\begin{equation}
O\left( \frac{1}{p^{2}}T^{4}\right)
\end{equation}%
are corrections suppressed by the perturbation condition $T<p$. This
condition is naturally ensured for the case $p\rightarrow \infty $.

\paragraph{(\textbf{2}) \textbf{Modes couplings}\qquad\ \newline
}

Local complex deformations of the conifold may be classified into specific
subsets. Generally speaking, conifold local deformations using perturbation
theory ideas should read as
\begin{equation}
xy-zw=\mu +LCD
\end{equation}%
where the $LCD$ stands for \ "the full set of local complex deformations".
In perturbation theory, $LCD$ has the following structure,%
\begin{eqnarray}
LCD &=&+\sum_{n>0}\left( t_{n}x^{n}+t_{-n}y^{n}+s_{n}z^{n}+s_{-n}w^{n}\right)
\notag \\
&&+\text{ }MMC\text{ }+\text{ }WMC\text{ }+\text{ }MWC.  \label{627}
\end{eqnarray}%
In this relation, the term $MMC$ refers to "momentum modes couplings" and $%
WMC$ to "winding modes couplings" and finally $MWC$ refers to
"momentum-winding modes interaction". \newline
Observe that in the pure momentum sector ($z=w=0$),
\begin{equation}
xy-zw=\mu +\sum_{n>0}\left( t_{n}x^{n}+t_{-n}y^{n}\right) +MMC,  \label{e1}
\end{equation}%
local complex deformations
\begin{equation}
t_{n}x^{n}+t_{-n}y^{n}
\end{equation}%
as well as $MMC$ couplings type
\begin{equation}
t_{n}t_{-m}x^{n}y^{m}  \label{e3}
\end{equation}%
are all of them invariant under
\begin{equation}
x\rightarrow xe^{2i\pi },\qquad y\rightarrow ye^{-2i\pi }\quad .
\end{equation}%
This is just the symmetry of the real slice ($y=\overline{x}$, $z=w=0$) of $%
T^{\ast }S^{1}$ we have considered earlier.\newline
For winding modes sector, we have quite similar relations; in particular%
\begin{equation}
xy-zw=\mu +\sum_{n>0}\left( s_{n}z^{n}+s_{-n}w^{n}\right) +WMC\quad ,
\end{equation}%
with leading winding mode couplings given by,%
\begin{equation}
s_{n}s_{-m}z^{n}w^{m}\quad .
\end{equation}%
Along with these two sectors; that is momentum sector (M-sector) and winding
sector (W-sector) there are also couplings involving both sectors.

\paragraph{(\textbf{3}) $SL\left( 2\right) $ \textbf{covariance\qquad\ }%
\newline
}

In a manifestly $SL\left( 2,C\right) $ covariant formalism, the M-sector and
W-sector should a priori be related under $SL\left( 2,C\right) $
transformations. This means thal along with $MMC$ and $WMC$, we should have
moreover couplings involving both momentum and winding modes. To get all of
these couplings, we proceed as follows:\newline
First think about the zero order perturbation of the conifold as
corresponding just to the global deformation generated by $\mu $, that is
global deformation of the conic singularity $xy-zw=0$ leading then to $%
xy-zw=\mu $. \newline
In the leading perturbation order in $\tau \left( x\right) $, $\widetilde{%
\tau }\left( y\right) $, $\sigma \left( x\right) $ and $\widetilde{\sigma }%
\left( y\right) $, conifold local deformations read as,%
\begin{equation}
xy-zw=\mu +\sum_{n>0}\left( \upsilon _{\left( n,0\right) }+\upsilon _{\left(
0,n\right) }\right) ,  \label{pe}
\end{equation}%
where, for later use, we have set
\begin{eqnarray}
\upsilon _{\left( n,0\right) } &=&t_{n}x^{n}+s_{n}z^{n},  \notag \\
\upsilon _{\left( 0,n\right) } &=&t_{-n}y^{n}+s_{-n}w^{n}.
\end{eqnarray}%
Then think about the two components $\upsilon _{\left( n,0\right) }$ and $%
\upsilon _{\left( 0,n\right) }$\ respectively as just the upper and down
components of a $\left( n+1\right) $ components vector multiplet $\mathbf{%
\upsilon }_{\frac{n}{2}}$; that is,%
\begin{equation}
\mathbf{\upsilon }_{\frac{n}{2}}=\left(
\begin{array}{c}
\upsilon _{\left( n,0\right) } \\
\upsilon _{\left( n-1,1\right) } \\
... \\
\upsilon _{\left( 1,n-1\right) } \\
\upsilon _{\left( 0,n\right) }%
\end{array}%
\right) .
\end{equation}%
Clearly $\upsilon _{\left( n,0\right) }$ and $\upsilon _{\left( 0,n\right) }$
are just the highest and lowest components of a spin $s=\frac{n}{2}$
multiplet of the group $SL\left( 2,C\right) $ generated by the set $\left\{
\mathcal{K}_{0},\mathcal{K}_{\pm }\right\} $ given by the step operators $%
\mathcal{K}_{\pm }$,
\begin{eqnarray}
\mathcal{K}_{+} &=&x\frac{\partial }{\partial w}-z\frac{\partial }{\partial y%
}\quad ,  \notag \\
\mathcal{K}_{-} &=&w\frac{\partial }{\partial x}-y\frac{\partial }{\partial z%
}\quad ,
\end{eqnarray}%
together with the Cartan-Weyl charge operator $\mathcal{K}_{0}$,%
\begin{equation}
\mathcal{K}_{0}=\left( x\frac{\partial }{\partial x}+z\frac{\partial }{%
\partial z}\right) -\left( y\frac{\partial }{\partial y}+w\frac{\partial }{%
\partial w}\right) \quad .
\end{equation}%
By direct computations, we have the following remarkable relations defining
respectively highest and lowest weight representations of $SL\left(
2,C\right) $,%
\begin{eqnarray}
\mathcal{K}_{+}\upsilon _{\left( n,0\right) } &=&0\quad ,  \notag \\
\mathcal{K}_{0}\upsilon _{\left( n,0\right) } &=&n\upsilon _{\left(
n,0\right) }\quad ,
\end{eqnarray}%
and%
\begin{eqnarray}
\mathcal{K}_{-}\upsilon _{\left( 0,n\right) } &=&0\quad ,  \notag \\
\mathcal{K}_{0}\upsilon _{\left( 0,n\right) } &=&-n\upsilon _{\left(
0,n\right) }\quad .
\end{eqnarray}%
To get the remaining $\upsilon _{\left( n-k,k\right) }$ components of the
spin $s=\frac{n}{2}$ multiplet, one applies successively $\mathcal{K}_{-}$
on $\upsilon _{\left( n,0\right) }$ or $\mathcal{K}_{+}$ on $\upsilon
_{\left( 0,n\right) }$. For generic states
\begin{equation}
\upsilon _{\left( n-k,k\right) }\sim \mathcal{K}_{-}^{k}\upsilon _{\left(
n,0\right) }\quad ,
\end{equation}%
we have the following solution,%
\begin{equation}
\upsilon _{\left( n-k,k\right) }\sim \left( t_{n,k}x^{n-k}w^{k}+\left(
-\right) ^{k}s_{n,k}z^{n-k}y^{k}\right) ,\qquad 0\leq k\leq n\quad ,
\label{36p}
\end{equation}%
where now the complex numbers $t_{n,k}$ and $s_{n,k}$\ carry both momentum
and winding modes. A manifestly $SL\left( 2,C\right) $ covariant formulation
of the infinitesimal local complex deformations of the conifold should be
then as in eq(\ref{36p}). The leading perturbation order should involve a
development in terms of $SL\left( 2,C\right) $ representations in agreement
with known results on $SDiff(S^{3})$. Highest and lowest weight components $%
\upsilon _{\left( n,0\right) }$ of $SL\left( 2,C\right) $ multiplets $%
\mathbf{\upsilon }_{\frac{n}{2}}$ are associated with linear infinitesimal
deformations in $t_{n}$ and $s_{n}$. Couplings involving both momentum and
winding ones are described by the states $\upsilon _{\left( n-k,k\right) }$
with $1\leq k\leq n-1$. For the special case $n=1$, something special
happens; there is no coupling between momentum and winding modes. This is a
strong point in favor of harmonic analysis to be consider below. Higher
perturbation orders describe non linear momentum modes couplings, non linear
winding mode couplings and non linear momentum-winding modes interactions.
Non zero connected amplitudes $<\mathcal{T}_{r_{1}}...\mathcal{T}_{r_{k}}>$
are defined as usual as,%
\begin{equation}
<\mathcal{T}_{r_{1}}...\mathcal{T}_{r_{k}}>=\frac{\delta ^{k}\mathcal{F}%
\left( t\right) }{\delta t_{r_{1}}...\delta t_{r_{k}}}|_{t=0},\qquad \text{ }%
\sum_{i=1}^{k}r_{i}=0,
\end{equation}%
where for the case of momentum modes the energy libre is as in eqs(\ref{eq1}-%
\ref{eq4}).

\subsection{$\mathcal{Z}_{top}$ in harmonic frame work}

We first show how harmonic variables $U^{+\gamma }$ and $V_{\gamma }^{-}$
describe momentum and winding mode units. Then we give the manifestly
invariant expression of the free energy
\begin{equation}
\mathcal{F}_{top}=\mathcal{F}_{top}\left( U^{+},V^{-}\right)
\end{equation}%
of B model topological string on conifold.

\subsubsection{Harmonic variables as unit momentum and winding modes}

In the harmonic framework where the harmonic variables $U^{+\gamma }$ and $%
V_{\gamma }^{-}$ form two $SL\left( 2,C\right) $ isodoublets, momentum and
winding modes get an interesting representation in terms of conifold
isometry. Unit momentum modes $m=\pm 1$ and units of winding $\omega =\pm 1$
combine as%
\begin{equation}
\left(
\begin{array}{c}
m=1 \\
\omega =1%
\end{array}%
\right) ,\qquad \qquad \left(
\begin{array}{c}
m=-1 \\
\omega =-1%
\end{array}%
\right) ,
\end{equation}%
and are respectively associated with the components of two fundamental
isodoublets
\begin{equation}
U^{+\gamma },\qquad \qquad V_{\gamma }^{-}
\end{equation}%
This feature is directly seen on these harmonic variables by using the
complex coordinates $\left( x,y,z,w\right) $,
\begin{equation}
\left(
\begin{array}{c}
U^{+1} \\
U^{+2}%
\end{array}%
\right) =\left(
\begin{array}{c}
x \\
z%
\end{array}%
\right) ,\qquad \qquad \left(
\begin{array}{c}
V_{1}^{-} \\
V_{2}^{-}%
\end{array}%
\right) =\left(
\begin{array}{c}
y \\
w%
\end{array}%
\right) .
\end{equation}%
The trivial (linear) monomials $x$ and $y$ are associated with momentum mode
$\left\vert m\right\vert =1$ and the monomials $z$ and $w$ with winding
modes $\left\vert \omega \right\vert =1$. As pointed out previously, this is
a special property of a general feature of $SL\left( 2,C\right) $
multiplets. \newline
For the $SL\left( 2,C\right) $ isotriplets, we have%
\begin{equation}
U^{+(\alpha _{1}}U^{+\alpha _{2})}=\left(
\begin{array}{c}
x^{2} \\
xz \\
z^{2}%
\end{array}%
\right) ,\qquad V_{(\beta _{1}}^{-}V_{\beta _{2})}^{-}=\left(
\begin{array}{c}
y^{2} \\
yw \\
w^{2}%
\end{array}%
\right) ,
\end{equation}%
where the monomials $x^{2}$ and $y^{2}$ are associated with momentum modes $%
\left\vert m\right\vert =2$ and $z^{2}$ and $w^{2}$ with the windings $%
\left\vert \omega \right\vert =2$. The crossed terms
\begin{equation}
xz,\qquad \text{and }\qquad yw
\end{equation}%
have mode charges as,
\begin{equation}
\left( \left\vert m\right\vert ,\left\vert \omega \right\vert \right)
=\left( 1,1\right) \quad ;
\end{equation}%
they describe couplings between units of momenta and windings.

\subsubsection{Expression of $\mathcal{Z}_{top}$ in harmonic analysis}

Besides the physical property given above, the power of harmonic frame work
in dealing with topological string partition function comes moreover from an
other remarkable feature of conifold. With harmonic variable coordinates,
all goes as if we are dealing with linear perturbation in $T^{\ast }S^{1}$.
\newline
Under local complex deformations of eq $U^{+\gamma }V_{\gamma }^{-}=\mu $,
which gets then mapped to the three complex dimension manifold,%
\begin{equation}
U^{+\gamma }V_{\gamma }^{-}=\mu +\xi \left( U^{+},V^{-}\right) \quad ,
\label{def}
\end{equation}%
the function $\xi \left( U^{+},V^{-}\right) $ has a remarkable harmonic
expansion recalling the one used in dealing with the subspace $T^{\ast
}S^{1} $. Indeed using the fibration property $T^{\ast }S^{3}$ as $C^{\ast }$
fibered over the base $T^{\ast }P^{1}$ and considering local coordinates
\begin{equation}
\left( \sigma ,U^{+1},U^{+2},V_{2}^{-},V_{2}^{-}\right)
\end{equation}%
of $WP_{\left( -1,1,1,-1,-1\right) }^{4}$, local complex deformations%
\begin{equation}
\xi \left( U^{+},V^{-}\right) \equiv \varphi \left( \sigma
,U^{+},V^{-}\right)
\end{equation}%
may first be expanded in Laurent series as,
\begin{equation}
\varphi \left( \sigma ,U^{+},V^{-}\right) =\zeta ^{0}+\sum_{n=1}^{\infty
}\sigma ^{n}\zeta ^{+n}+\sum_{n=1}^{\infty }\sigma ^{-n}\zeta ^{-n}\quad ,
\label{man}
\end{equation}%
where the Laurent modes
\begin{equation}
\zeta ^{0,\pm n}=\zeta ^{0,\pm n}\left( U^{+},V^{-}\right)
\end{equation}%
are harmonic functions on $T^{\ast }P^{1}$ constrained as follows%
\begin{equation}
\zeta ^{q}\left( \lambda U^{+},\frac{1}{\lambda }V^{-}\right) =\lambda
^{q}\zeta ^{q}\left( U^{+},V^{-}\right) ,\qquad \lambda \in C^{\ast }\quad .
\end{equation}%
The expansion (\ref{man}) should be compared with the right hand of eq(\ref%
{627}),%
\begin{eqnarray}
\mu +LCD &=&\mu +\sum_{n>0}\left(
t_{n}x^{n}+t_{-n}y^{n}+s_{n}z^{n}+s_{-n}w^{n}\right)  \notag \\
&&+\text{ }MMC\text{ }+\text{ }WMC\text{ }+\text{ }MWC\quad .  \label{ord}
\end{eqnarray}%
The two relations describe exactly the same thing; but expressed in two
different coordinate systems. The difference is that in eq(\ref{man}), $%
SL\left( 2,C\right) $ symmetry is manifest while it is not in eq(\ref{627}).
By analyzing eq(\ref{man}) and eq(\ref{627}), we can conjecture the the
explicit expression of $\mathcal{Z}_{top}\left( \mu ,\xi \right) $
preserving manifestly SL$\left( 2,C\right) $ conifold isometry.

\begin{theorem}
:\quad\ \textbf{Expression of }$\mathcal{Z}_{top}\left( \mu ,\xi \right) $
\textbf{preserving manifestly} SL$\left( 2\right) $ \textbf{isometry}\newline
Denoting by $\mathcal{Z}_{T^{\ast }S^{1}}$ and $\mathcal{Z}_{T^{\ast }S^{3}}$
the two following:\newline
(\textbf{i}) $\mathcal{Z}_{T^{\ast }S^{1}}=\mathcal{Z}_{top}\left( \mu
,t\right) $, the partition function of B model topological string on the
locally deformed conifold
\begin{equation}
xy-zw=\mu +t\left( x,y\right) \quad ,\qquad t\left( x,y\right) =T\left(
x,y,z,w\right) |_{z=w=0}\quad ,
\end{equation}%
where the $t\left( x,y\right) $ deformations are restricted to $T^{\ast
}S^{1}$ with $S^{1}$\ being the large circle of $S^{3}$.\newline
(\textbf{ii}) $\mathcal{Z}_{T^{\ast }S^{3}}=\mathcal{Z}_{top}\left( \mu ,\xi
\right) $ the harmonic space partition function of B model topological
string of the manifestly SL$\left( 2,C\right) $ covariant locally deformed
conifold,
\begin{equation}
U^{+\gamma }V_{\gamma }^{-}=\mu +\xi \left( U^{+},V^{-}\right) \quad ,
\end{equation}%
with local deformations generated by $\xi \left( U^{+},V^{-}\right) $,%
\newline
Then we have the following 1-1 correspondence,%
\begin{equation}
\mathcal{Z}_{T^{\ast }S^{3}}\left( \mu ,\xi \right) \qquad
\longleftrightarrow \qquad \mathcal{Z}_{T^{\ast }S^{1}}\left( \mu ,t\right)
\quad .  \label{cj}
\end{equation}%
In above eqs, $t\left( x,y\right) $ lives on $T^{\ast }S^{1}$ and expands in
general as%
\begin{eqnarray}
t\left( x,y\right) &=&\sum_{n,m=0}^{\infty }t_{n,m}x^{n}y^{m}  \notag \\
&=&t_{0,0}+\sum_{n=1}^{\infty }t_{n,0}x^{n}+\sum_{m=1}^{\infty
}t_{0,m}y^{m}+\sum_{n,m\geq 1}^{\infty }t_{n,m}x^{n}y^{m}\quad ,
\end{eqnarray}%
while $\xi $ lives on $T^{\ast }S^{3}$ and is given by the harmonic
expansion
\begin{equation}
\xi =\zeta ^{0}+\sum_{n=1}^{\infty }\left( \sigma ^{n}\zeta ^{+n}+\sigma
^{-n}\zeta ^{-n}\right) \quad ,
\end{equation}%
with $\zeta ^{\pm n}=\zeta ^{\pm n}\left( U^{+},V^{-}\right) $ are
homogeneous functions of degree $\pm n$ living on $T^{\ast }S^{2}$ and whose
harmonic developments read as
\begin{eqnarray}
\zeta ^{+n}\left( U^{+},V^{-}\right) &=&\sum_{j=0}^{\infty }U_{(\alpha
_{1}}^{+}..U_{\alpha _{j+n}}^{+}V_{\beta _{1}}^{-}..V_{\beta _{j})}^{-}\zeta
^{\left( \alpha _{1}...\alpha _{j+n}\beta _{1}...\beta _{j}\right) }\quad ,
\notag \\
\zeta ^{-n}\left( U^{+},V^{-}\right) &=&\sum_{j=0}^{\infty }U_{(\alpha
_{1}}^{+}..U_{\alpha _{j}}^{+}V_{\beta _{1}}^{-}..V_{\beta _{j+n})}^{-}\zeta
^{\left( \alpha _{1}...\alpha _{j}\beta _{1}...\beta _{j+n}\right) }\quad .
\end{eqnarray}%
$\zeta ^{0}=\zeta ^{0}\left( U^{+},V^{-}\right) $ is obtained from above
relations by setting $n=0$.
\end{theorem}

The conjectured relation (\ref{cj}) allows to derive the manifestly $%
SL\left( 2,C\right) $ expression of the partition function of topological
string B model on conifold. Higher order perturbation theory described by
the terms $MMC,$ $WMC$ and $MWC$ of eq(\ref{ord}) are trivially captured by
the harmonic space function $\xi $. But what have become the usual
difficulties of standard formalism? The answer is that they are still
present; but they have taken an other form. A way to see it is to split $\xi
=\varphi \left( \sigma ,U^{+},V^{-}\right) $ in basic blocks as follows,
\begin{eqnarray}
\varphi \left( \sigma ,U^{+},V^{-}\right) &=&\left[ \zeta \left( \sigma
,U^{+}\right) +\delta \eta \left( \sigma ,U^{+},V^{-}\right) \right]  \notag
\\
&&+\left[ \widetilde{\zeta }\left( \sigma ,V^{-}\right) +\delta \widetilde{%
\eta }\left( \sigma ,U^{+},V^{-}\right) \right] \quad .  \label{lea}
\end{eqnarray}%
Then use the fibration $T^{\ast }S^{3}\simeq T^{\ast }S^{1}\times T^{\ast
}S^{2}$ to expand these functions; first in Laurent series in $\sigma $ as,
\begin{eqnarray}
\zeta \left( \sigma ,U^{+}\right) &=&\sum_{n>0}\sigma ^{n}\zeta ^{+n}\left(
U^{+}\right) \quad ,  \notag \\
\widetilde{\zeta }\left( \sigma ,V^{-}\right) &=&\sum_{n>0}\sigma ^{-n}\zeta
^{-n}\left( V^{-}\right) \quad ,  \notag \\
\delta \eta \left( \sigma ,U^{+},V^{-}\right) &=&\sum_{n>0}\sigma ^{n}\delta
\eta ^{+n}\left( U^{+},V^{-}\right) \quad , \\
\delta \widetilde{\eta }\left( \sigma ,U^{+},V^{-}\right)
&=&\sum_{n>0}\sigma ^{-n}\delta \eta ^{-n}\left( U^{+},V^{-}\right) .  \notag
\end{eqnarray}%
Then develop the Laurent modes in harmonic series on $T^{\ast }S^{2}$. We
have for $\zeta ^{+n}\left( U^{+}\right) $%
\begin{equation}
\zeta ^{+n}\left( U^{+}\right) =U^{+(\alpha _{1}}...U^{+\alpha _{n})}\zeta
_{\left( \alpha _{1}...\alpha _{n}\right) }
\end{equation}%
which up on using the convention notations
\begin{eqnarray}
U_{\alpha }^{+} &=&\sqrt{p}u_{\alpha }^{+},  \notag \\
U^{\left( +n\right) } &=&U^{+(\alpha _{1}}...U^{+\alpha _{n})} \\
u^{\left( +n\right) } &=&u^{+(\alpha _{1}}...u^{+\alpha _{n})}  \notag
\end{eqnarray}%
and
\begin{eqnarray}
U^{\left( +n\right) } &=&p^{\frac{n}{2}}u^{\left( +n\right) } \\
\zeta _{\left( \alpha _{1}...\alpha _{n}\right) } &\equiv &\zeta _{\left(
n,0\right) },  \notag
\end{eqnarray}%
it can be rewritten as%
\begin{equation}
\zeta ^{+n}\left( U^{+}\right) =p^{\frac{n}{2}}u^{\left( +n\right) }\zeta
_{\left( n,0\right) }.
\end{equation}%
We also have,
\begin{eqnarray}
\delta \eta ^{+n}\left( U^{+},V^{-}\right) &=&\sum_{j=1}^{\infty }U_{(\alpha
_{1}}^{+}..U_{\alpha _{j+n}}^{+}V_{\beta _{1}}^{-}..V_{\beta _{j})}^{-}\text{
}\delta \eta ^{\left( \alpha _{1}...\alpha _{j+n}\beta _{1}...\beta
_{j}\right) }  \notag \\
&=&\sum_{j=1}^{\infty }U^{(+n+j}V^{-j)}\text{ }\delta \eta _{\left(
n+j,j\right) },
\end{eqnarray}%
Similar relations may be written down for both for $\widetilde{\zeta }\left(
\sigma ,V^{-}\right) $ and $\delta \eta \left( \sigma ,U^{+},V^{-}\right) $
using first Laurent expansion on $T^{\ast }S^{1}$ and second harmonic
development on $T^{\ast }S^{2}$. In particular we have
\begin{eqnarray}
\zeta ^{-n}\left( V^{-}\right) &=&V_{(\beta _{1}}^{-}..V_{\beta _{n})}^{-}%
\widetilde{\zeta }^{\left( \beta _{1}...\beta _{n}\right) }  \notag \\
&=&p^{\frac{n}{2}}v^{\left( -n\right) }\widetilde{\zeta }_{\left( 0,n\right)
}.
\end{eqnarray}%
Note that $\zeta ^{+n}\left( U^{+}\right) $ and $\zeta ^{-n}\left(
V^{-}\right) $\ are the leading terms in eq(\ref{lea}); they involve only $%
SL\left( 2,C\right) $ irreducible representations with spins $\left(
s,0\right) =\left( n+1,0\right) $ and $\left( 0,s\right) =\left(
0,n+1\right) $\ captured by the completely symmetric tensors $\zeta _{\left(
\alpha _{1}...\alpha _{n}\right) }$ and $\widetilde{\zeta }^{\left( \beta
_{1}...\beta _{n}\right) }$. The extra terms $\delta \eta ^{+n}\left(
U^{+},V^{-}\right) $ and $\delta \widetilde{\eta }^{-n}\left(
U^{+},V^{-}\right) $ involve infinite towers of $SL\left( 2,C\right) $
representations and are the sources of difficulties raised above. \newline
Integration on $T^{\ast }S^{2}$ of monomials,
\begin{equation}
M^{\left( \mathrm{n,m}\right) }=\dprod\limits_{n_{i}}\delta \eta
^{+n_{i}}\left( U^{+},V^{-}\right) \dprod\limits_{m_{j}}\delta \widetilde{%
\eta }^{-m_{j}}\left( U^{+},V^{-}\right) ,
\end{equation}%
involves computing infinite traces of product of $SL\left( 2,C\right) $
representations. This is a technical difficulty which may be overcome by
using approximation methods.

\subsubsection{Small local complex deformations}

A way to deal with the local complex deformations of conifold,%
\begin{equation}
U^{+\gamma }V_{\gamma }^{-}=\mu +\varphi \left( \sigma ,U^{+},V^{-}\right) ,
\label{deff}
\end{equation}%
with $\varphi \left( \sigma ,U^{+},V^{-}\right) =\varphi $\ as in eq(\ref%
{lea}), is to think about $\varphi $ as a small perturbation around the
global deformation parameter $\mu $. In this view the splitting
\begin{equation}
\varphi =\zeta +\widetilde{\zeta }+\delta \eta +\delta \widetilde{\eta }
\end{equation}%
may be interpreted as follows. The term
\begin{equation}
\zeta \left( \sigma ,U^{+}\right) +\widetilde{\zeta }\left( \sigma
,V^{-}\right)
\end{equation}%
is thought of as generating the leading perturbation order and
\begin{equation}
\delta \eta \left( \sigma ,U^{+},V^{-}\right) +\delta \widetilde{\eta }%
\left( \sigma ,U^{+},V^{-}\right) =\mathcal{O}\left[ \zeta \widetilde{\zeta }%
,\left( \zeta \right) ^{2},\left( \widetilde{\zeta }\right) ^{2}\right]
\end{equation}%
as corresponding to higher perturbation orders in $\zeta $ and $\widetilde{%
\zeta }$. As such eq(\ref{deff}) can be put into the following form,%
\begin{eqnarray}
U^{+\gamma }V_{\gamma }^{-} &=&\mu +\sum_{n>0}p^{\frac{n}{2}}\left( \sigma
^{n}u^{\left( +n\right) }\zeta _{\left( n,0\right) }+\sigma ^{-n}v^{\left(
-n\right) }\widetilde{\zeta }_{\left( 0,n\right) }\right)  \notag \\
&&+\sum_{n,m>0}p^{\frac{n+m}{2}}\left( \sigma ^{n-m}u^{\left( +n\right)
}v^{\left( -m\right) }\zeta _{\left( n,0\right) }\widetilde{\zeta }_{\left(
0,m\right) }\right) \\
&&+\text{ higher corrections.}  \notag
\end{eqnarray}%
On the real slice of the conifold obtained by setting $\left\vert \sigma
\right\vert =1$ ($\sigma =e^{i\theta }$) and $V^{-}=U^{-}=\overline{U^{+}}$,
the above eq takes the following reduced form,
\begin{eqnarray}
U^{+\gamma }U_{\gamma }^{-} &=&\mu +\sum_{n>0}p^{\frac{n}{2}}\left[
e^{in\theta }u^{\left( +n\right) }\zeta _{n}+e^{-in\theta }u^{\left(
-n\right) }\overline{\zeta }_{-n}\right]  \notag \\
&&+\sum_{n,m>0}p^{\frac{n+m}{2}}\left( e^{i\left( n-m\right) \theta
}u^{\left( +n\right) }v^{\left( -m\right) }\zeta _{n}\overline{\zeta }%
_{-m}\right) \\
&&+\text{ higher corrections,}  \notag
\end{eqnarray}%
where we have set%
\begin{equation*}
\zeta _{n}=\zeta _{\left( n,0\right) },\qquad \qquad \overline{\zeta }_{-n}=%
\widetilde{\zeta }_{\left( 0,n\right) }.
\end{equation*}%
The first term may be also put in the form
\begin{equation}
\sum_{n\neq 0}^{\infty }p^{\frac{\left\vert n\right\vert }{2}}e^{in\theta
}u^{\left( n\right) }\zeta _{n}
\end{equation}%
and should be compared to
\begin{equation}
t\left( p,\theta \right) =\sum_{n\neq 0}^{\infty }p^{\frac{\left\vert
n\right\vert }{2}}t_{n}e^{in\theta }
\end{equation}%
describing local complex deformations of the large circle $S^{1}$ of the
conifold. Therefore we have the $1\rightarrow \left( n+1\right) $
correspondence between the momentum mode $t_{\pm n}$ on the large circle $%
S^{1}$ of the 3-sphere and the modes $u^{\left( \pm n\right) }\zeta _{\pm n}$
on the full 3-sphere,%
\begin{equation}
t_{\pm n}\qquad \longleftrightarrow \qquad u^{\left( \pm n\right) }\zeta
_{\pm n}.  \label{sub}
\end{equation}%
To each section $t_{\pm n}$ on the large circle $S^{1}$ of the 3-sphere with
$U\left( 1\right) $ charge $\pm n$, corresponds a homogeneous harmonic
function $\zeta ^{\pm n}$ on the two sphere $S^{2}$ with Cartan Weyl charges
$\pm n$.\

\subsubsection{Free energy in harmonic set up}

Using the correspondence eqs(\ref{cj},\ref{sub}) as well as the dictionary
we have given in section 4, one can write down the explicit expression of
the manifestly $SU\left( 2,C\right) $ conifold free energy $\mathcal{F}%
\left( \zeta ,\overline{\zeta }\right) $. By help of table (\ref{di}), the $%
SU\left( 2,C\right) $ covariantization of the identities
\begin{equation}
\sum_{n>0}^{\infty }\frac{2p^{n}}{n}t_{-n}t_{n}
\end{equation}%
and
\begin{equation}
\sum_{n_{1}+n_{2}+n_{3}=0}p^{\frac{\left\vert n_{1}\right\vert +\left\vert
n_{2}\right\vert +\left\vert n_{3}\right\vert }{2}%
-1}t_{n_{1}}t_{n_{2}}t_{n_{3}}
\end{equation}%
read respectively as%
\begin{equation}
\sum_{n>0}^{\infty }\frac{2p^{n}}{^{n}}\left( \int_{S^{2}}\zeta ^{-n}\left(
u^{-}\right) \zeta ^{n}\left( u^{+}\right) \right) ,
\end{equation}%
and
\begin{equation}
\sum_{n_{1}+n_{2}+n_{3}=0}p^{\frac{\left\vert n_{1}\right\vert +\left\vert
n_{2}\right\vert +\left\vert n_{3}\right\vert }{2}-1}\left(
\int_{S^{2}}\zeta ^{n_{1}}\zeta ^{n_{2}}\zeta ^{n_{3}}\right) .
\end{equation}%
Using harmonic integration rule on the 3-sphere described in appendix, these
quantities can be also rewritten as
\begin{equation}
\int_{S^{3}}\overline{\zeta }\left( \sigma ,U^{-}\right) \frac{2}{\nabla ^{0}%
}\zeta \left( \sigma ,U^{+}\right) ,
\end{equation}%
and
\begin{equation}
\frac{1}{p}\int_{S^{3}}\left( \Upsilon \left( p,u^{\pm }\right) \right) ^{3}.
\end{equation}%
The $SL\left( 2,C\right) $ manifestly partition function for B model
topological string on conifold restricted to real slice reads then as
\begin{equation*}
\mathcal{Z}_{top}\left( \zeta ,\overline{\zeta }\right) =\exp \mathcal{F}%
\left( \zeta ,\overline{\zeta }\right) .
\end{equation*}%
The free energy $\mathcal{F}\left( \zeta ,\overline{\zeta }\right) $ is
obtained from $\mathcal{F}\left( t,\overline{t}\right) $ eq(\ref{eq1}-\ref%
{eq4}) by substituting everywhere
\begin{equation*}
t_{\pm n}
\end{equation*}%
moduli by the homogeneous harmonic function
\begin{equation*}
u^{\left( \pm n\right) }\zeta _{\pm n}=\zeta ^{\pm n}\left( u^{\pm }\right)
\end{equation*}%
on the 2-sphere, eq(\ref{sub}). For instance, the genus zero contribution to
the total free energy $\mathcal{F}\left( \zeta ,\overline{\zeta }\right) $
namely $g_{s}^{-2}\mu ^{2}\mathcal{F}_{0}\left( \zeta ,\overline{\zeta }%
\right) $ reads as follows,
\begin{equation}
\mu ^{2}\mathcal{F}_{0}\left( \zeta ,\overline{\zeta }\right) =-\sum_{n>0}%
\frac{\mu ^{n}}{n}\int_{S^{2}}\left( \zeta ^{n}\overline{\zeta }^{-n}\right)
+\int_{S^{2}}L\left( \zeta ,\overline{\zeta }\right) ,
\end{equation}%
where the function\ $L\left( \zeta ,\overline{\zeta }\right) $ is an $%
SU\left( 2,C\right) $ invariant function involving specific monomials in $%
\zeta $ and $\overline{\zeta }$ and whose two leading terms may be read as
\begin{equation}
L\left( \zeta ,\overline{\zeta }\right) \simeq \frac{L_{3}\left( \zeta ,%
\overline{\zeta }\right) }{3!}+\frac{L_{4}\left( \zeta ,\overline{\zeta }%
\right) }{4!}+O\left[ \left( \zeta ,\overline{\zeta }\right) ^{5}\right]
\end{equation}%
with,%
\begin{equation}
L_{3}\left( \zeta ,\overline{\zeta }\right) =\sum_{n_{1}+n_{2}+n_{3}=0}p^{%
\frac{\left\vert n_{1}\right\vert +\left\vert n_{2}\right\vert +\left\vert
n_{3}\right\vert -2}{2}}\left( \int_{S^{2}}\zeta ^{n_{1}}\left( u\right)
\zeta ^{n_{2}}\left( u\right) \zeta ^{n_{3}}\left( u\right) \right) ,
\end{equation}%
and, by setting setting $C_{n_{1},n_{2},n_{3},n_{4}}=\left( 1-\max \left\{
\left( \left\vert n_{1}\right\vert ,\left\vert n_{2}\right\vert ,\left\vert
n_{3}\right\vert ,\left\vert n_{4}\right\vert \right) \right\} \right) $, we
have as well,
\begin{equation}
L_{4}\left( \zeta ,\overline{\zeta }\right)
=\sum_{n_{1}+n_{2}+n_{3}+n_{4}=0}C_{n_{1},n_{2},n_{3},n_{4}}p^{\frac{%
\left\vert n_{1}\right\vert +\left\vert n_{2}\right\vert +\left\vert
n_{3}\right\vert +\left\vert n_{4}\right\vert -4}{2}}\left(
\int_{S^{2}}\zeta ^{n_{1}}\left( u\right) \zeta ^{n_{2}}\left( u\right)
\zeta ^{n_{3}}\left( u\right) \zeta ^{n_{4}}\left( u\right) \right) .
\end{equation}%
With the explicit expression of the free energy $\mathcal{F}\left[ \zeta ,%
\overline{\zeta }\right] $, one can compute the manifestly $SU\left(
2,C\right) $ covariant correlation function on arbitrary points of $S^{3}$.
Non zero connected amplitudes $<\mathcal{\digamma }^{n_{1}}...\mathcal{%
\digamma }^{n_{r}}>$ in harmonic framework are defined as usual as,%
\begin{equation}
<\mathcal{\digamma }^{n_{1}}\left( u_{1}^{\pm }\right) ...\mathcal{\digamma }%
^{n_{r}}\left( u_{k}^{\pm }\right) >=\frac{\delta ^{k}\mathcal{F}\left[
\zeta ,\overline{\zeta }\right] }{\delta \zeta ^{-n_{1}}\left( u_{1}\right)
...\delta \zeta ^{-n_{k}}\left( u_{k}\right) }|_{\zeta ,\overline{\zeta }=0}.
\end{equation}%
This relation is non trivial except for the case where the Cartan-Weyl
charges $n_{i}$ add to zero; i.e%
\begin{equation}
\sum_{i=1}^{k}n_{i}=0.
\end{equation}%
This feature is require by $SU\left( 2,C\right) $ invariance. Having these
results at hand, we turn now to study correlations functions in the GSV
quantum cosmology on $S^{3}$ \cite{30}.

\section{Correlation functions in GSV quantum cosmology}

After a brief review on the GSV $S^{3}$ quantum cosmological toy model, we
compute the correlation functions of conformal fields
\begin{equation}
\Phi _{i}=\Phi \left( U_{i}^{\pm }\right)
\end{equation}
at generic points $U_{i}^{\pm }$ on the 3-sphere
\begin{equation}
U_{i}^{+\alpha }U_{\alpha i}^{-}=p.
\end{equation}
This computation, which involves both momenta and winding corrections, is
done by using harmonic framework and convariantized Hartle-Hawking
probability density $\varrho =\left\vert \Psi \right\vert ^{2}$ preserving
manifestly $SU\left( 2,C\right) $ isometry of $S^{3}$.

\subsection{The GSV model}

In the GKV quantum cosmology toy model, the four dimensional space time
representing the cosmological real world is identified with a number $N$ of $%
D3$ branes wrapping the three sphere $S^{3}$; the real slice of the
conifold. To motivate this construction, it is interesting to recall the
following steps.\newline
Start with a flux compactification of $10D$ type IIB superstring on $T^{\ast
}S^{3}\times S^{2}\times S^{1}$ with the RR 5-form field strength flux,
\begin{equation}
F_{5}=F_{3}\wedge \omega _{2}\quad ,
\end{equation}%
threading through $T^{\ast }S^{3}\times S^{2}$ with $\omega _{2}$ being the
unit volume 2-form on $S^{2}$ and $F_{3}$ a real 3-form on $T^{\ast }S^{3}$.
\newline
Then consider a symplectic basis $\left\{ A^{i},B_{j}\right\} $ of $%
H_{3}\left( T^{\ast }S^{3}\right) $ homology with intersection numbers%
\begin{eqnarray}
A^{i}\cap A^{j} &=&0,\qquad B_{i}\cap B_{j}=0\quad ,  \notag \\
A^{i}\cap B_{j} &=&\delta _{j}^{i},\qquad i,j=0,1,...,h^{2,1}\quad ,
\end{eqnarray}%
with $h^{2,1}=h^{2,1}\left( T^{\ast }S^{3}\right) $ being the complex
dimension of the moduli space of the complex deformations of conifold.
\newline
After; choose an integral basis $\left\{ \mathbf{a}_{i},\mathbf{b}%
^{j}\right\} $ of 3-cocycles in $H^{3}\left( T^{\ast }S^{3},Z\right) $
cohomology to decompose the $F_{3}$ field strength as
\begin{equation}
F_{3}=\sum_{i=0}^{h^{2,1}}\left( P^{i}\mathbf{a}_{i}+Q_{i}\mathbf{b}%
^{i}\right) \quad ,
\end{equation}%
with $P^{i}$ and $Q_{i}$ are respectively the magnetic and electric fluxes.
The moduli\ space of the complex deformations of conifold is parameterized
by the periods of the holomorphic 3-form $\Omega $ on the 3-cycles $A^{i}$
and $B_{j}$ as shown below%
\begin{equation}
X^{i}=\int_{A_{i}}\Omega \quad ,\qquad \frac{\partial \mathcal{F}_{0}\left(
X\right) }{\partial X^{i}}=\int_{B^{i}}\Omega \quad ,
\end{equation}%
with $X^{i}=\left( \func{Re}X^{i}\right) +i\left( \func{Im}X^{i}\right) $
projective coordinates.\newline
Then write down the Hartlee-Hawking wave function $\Psi \left( X,\overline{X}%
\right) $ associated with the conifold fluctuations. To do so one has to
overcome the difficulty due to the fact that the projective $X^{i}$ and $%
\overline{X^{i}}$ variables do not commute in the BPS minisuperspace $\cite%
{19,21}$. This non commutative behaviour is solved by restricting the moduli
space to the subspace parameterized by%
\begin{equation}
X^{i}=P^{i}+\frac{i}{\pi }\Phi ^{i},\qquad P^{i}=\func{Re}\left(
X^{i}\right) ,
\end{equation}%
where the attractor mechanism has been used to fix the conifold complex
structure as $\func{Re}\left( X^{i}\right) =P^{i}$ and the potential $%
\mathcal{F}_{0}$ in terms of electric charge as
\begin{equation}
\func{Re}\left( C\frac{\partial \mathcal{F}_{0}\left( X\right) }{\partial
X^{i}}\right) =Q_{i}.
\end{equation}%
For the complex structure $\mu $ for example, the attractor mechanism eqs $%
P^{i}=\func{Re}\left( CX^{i}\right) $ and $Q_{i}=\func{Re}\left( C\frac{%
\partial \mathcal{F}_{0}\left( X\right) }{\partial X^{i}}\right) $ lead to
\begin{equation}
\frac{2}{g_{s}}\left( \func{Re}\mu \right) =N=P^{0},
\end{equation}%
with $N>>1$ so that $\func{Re}\mu /g_{s}>>1$. An overall rescaling of the
charge $P^{0}$ by the inverse of the topological string coupling constant $%
g_{s}$ ($P^{0}\rightarrow \frac{1}{g_{s}}P^{0}$), we bring it to the form $%
\left( \func{Re}\mu \right) =N$ and so,
\begin{equation}
\mu =N+\frac{i}{\pi }\Phi ^{0},\qquad \mathcal{F}_{0}\left( \mu \right) =%
\frac{i\mu }{\pi g_{s}}\ln \left( \frac{2\mu }{\Lambda g_{s}}\right) ,
\end{equation}%
In terms of the commutative coordinate $\Phi ^{i}$, the Hartlee-Hawking wave
function reads therefore as follows,%
\begin{equation}
\Psi _{\left( P^{i},Q_{j}\right) }\left( \Phi ^{i}\right) =\mathcal{Z}%
_{top}\left( X^{0},...,X^{h^{2,1}}\right) e^{\sum_{j=0}^{h^{2,1}}\frac{1}{2}%
Q_{j}\Phi ^{j}},
\end{equation}%
where $\mathcal{Z}_{top}\left( X\right) $ is the partition function of the
B-model topological string on conifold.\newline
Next consider a conifold geometry in the limit of a large $S^{3}$ with a
radius$\sqrt{p}$ greater than some given scale $l_{0}$ beyond which
perturbation theory for higher genus corrections holds,%
\begin{equation}
\sqrt{p}>l_{0}=\Lambda \sqrt{g_{s}}\quad .
\end{equation}%
This condition follows from implementation of momentum modes contribution
and the perturbative study of the higher genus corrections to the propagator
\begin{equation}
<t_{n}t_{-n}>
\end{equation}%
where it has been observed that the right perturbation parameter is
\begin{equation}
\frac{n^{2}g_{s}}{\mu }
\end{equation}%
rather than $\frac{g_{s}}{\mu }$. The parameter $\Lambda $\ is a cut off
parameter restricting local deformations $\sum_{n=-\infty }^{\infty }p^{%
\frac{\left\vert n\right\vert }{2}}t_{n}e^{in\theta }$ of the round 3-sphere
into,%
\begin{equation}
t\left( x,\overline{x}\right) =\sum_{n=-\Lambda }^{\Lambda }p^{\frac{%
\left\vert n\right\vert }{2}}t_{n}e^{in\theta }\quad .  \label{v1}
\end{equation}%
The four dimensional cosmological real world is identified with a number $N$
of $D3$ branes wrapping the three sphere $S^{3}$. Local fluctuations deform
the shape of the three sphere and induce perturbations $\delta g_{\mu \nu
}^{\left( 0\right) }$ of round metric $g_{\mu \nu }^{\left( 0\right) }$
which becomes then,
\begin{equation}
g_{\mu \nu }\left( t\right) =g_{\mu \nu }^{\left( 0\right) }+\frac{\delta
g_{\mu \nu }^{\left( 0\right) }}{\delta t}\delta t+O\left( t^{2}\right) .
\end{equation}%
Local complex deformations (\ref{v1}) deform the holomorphic form $\Omega
_{0}$\ to $\Omega \left( t\right) ,$
\begin{equation}
\Omega =f\left( t\right) \Omega _{0}
\end{equation}%
where $f\left( t\right) $ is a scale factor. On the real slice of the
conifold, these infinitesimal deformations induce fluctuations of the real
part $\func{Re}\left( \Omega _{0}\right) $. The new real part $\func{Re}%
\left( \Omega \right) $ is related to $\mathbf{\func{Re}}\left( \Omega
_{0}\right) $ as,
\begin{equation}
\frac{\func{Re}\Omega }{\mathbf{\func{Re}}\Omega _{0}}\mathbf{=}\exp \left[
\Phi \left( t,\overline{t}\right) \right] .
\end{equation}%
where
\begin{equation}
\Phi =\Phi \left( t,\overline{t}\right)
\end{equation}%
is the conformal factor. It is this field $\Phi \left( t,\overline{t}\right)
$ which is at central focus in GSV cosmology; it is assumed that its
fluctuations would be observable. Local perturbations of the conifold, and
so of the 3-sphere, should be felt on the D3 brane world as metric
inhomogeneities and matter fluctuations. Note also that in this picture
supersymmetry is supposed to be weakly broken so that mini-superspace
approximation for stringy Hartle-Hawking wave function is still valid.

In what follows, we use harmonic set up considered in previous sections to
compute the correlation functions of conformal factor in case where both
momentum and winding corrections are implemented.

\subsection{Harmonic differential geometry on $T^{\ast }S^{3}$}

We begin by giving the explicit expression of the\ holomorphic volume of
conifold in harmonic set up. Then we consider its real reduction on the
three sphere and study field fluctuations.

\subsubsection{Holomorphic 3-form in harmonic space}

In the local coordinate system $\left( \sigma
,U^{+1},U^{+2},V_{1}^{-},V_{2}^{-}\right) $ of $WP_{\left(
-1,+1,+1,-1,-1\right) }^{4}$ where conifold
\begin{eqnarray}
U^{+\alpha }V_{\alpha }^{-} &=&\mu ,\qquad \sigma ,  \notag \\
U^{+\alpha } &\equiv &\lambda U^{+\alpha },\qquad V_{\alpha }^{-}\equiv
\frac{1}{\lambda }V_{\alpha }^{-},\qquad \sigma \equiv \frac{1}{\lambda }%
\sigma ,
\end{eqnarray}%
is embedded\footnote{%
Recall that in projective capital letters, conifold eq is given by $%
U^{+\alpha }U_{\alpha }^{-}=\mu $ and $\sigma $ free. For simplicity we will
also use projective small letters $u^{+\alpha }v_{\alpha }^{-}=1$. The
passage between the two coordinates is $U^{+\alpha }=\sqrt{\mu }u^{+\alpha }$%
, $V_{\alpha }^{-}=\sqrt{\mu }v_{\alpha }^{-}$.}, the globally defined
holomorphic 3-form $\Omega _{3}$ of conifold is given by,
\begin{equation}
\Omega _{3}=\frac{d\sigma }{\sigma }dU^{+\alpha }dV^{-\beta }\varepsilon
_{\alpha \beta }.
\end{equation}%
Clearly this holomorphic 3-form is $SL\left( 2,C\right) $ invariant and its
integration over a 3-cycle $A\sim \mathcal{C}_{\sigma =0}\times D$ of $%
H_{3}\left( T^{\ast }S^{3},Z\right) ,$ with $\partial D\neq 0,$ gives the
complex structure $\mu $; i.e
\begin{equation}
\mu \sim \int_{A}\Omega _{3}.
\end{equation}%
Indeed, a first integration of $\Omega _{3}$ along the curve $\mathcal{C}%
_{\sigma =0}$ of $\sigma $-plane containing the origin $\sigma =0$ gives the
reduced integral
\begin{equation}
\int_{D}\Omega _{2}
\end{equation}%
of the holomorphic 2-form
\begin{equation}
\Omega _{2}=dU^{+\alpha }dV^{-\beta }\varepsilon _{\alpha \beta }
\end{equation}%
over the 2-cycle D. Thus we have,%
\begin{equation}
\mu =\int_{D\times \mathcal{C}_{\sigma =0}}\Omega _{3}=\int_{D}\Omega _{2}.
\end{equation}%
By Stokes theorem, one can bring the 2D integral to a 1D integration on the
non zero boundary $\partial D$.%
\begin{equation}
\int_{\partial D}\Omega _{1}=\int_{\partial D}U^{+\alpha }dV^{-\beta
}\varepsilon _{\alpha \beta }.
\end{equation}%
Next consider the curve $\mathcal{C}_{a^{+},b^{-}}$ with the following
features: \newline
(\textbf{i}) $\mathcal{C}_{a^{+},b^{-}}$ contains the particular conifold
point
\begin{equation}
\left( U^{+\alpha },V_{\beta }^{-}\right) =\left( a^{+\alpha },b_{\beta
}^{-}\right)
\end{equation}%
satisfying obviously the constraint eq
\begin{equation}
a^{+\alpha }b_{\alpha }^{-}=\mu
\end{equation}%
and surrounding the pole singularities
\begin{equation}
\frac{a^{+\alpha }}{a^{+}.U^{+}}\quad ,\qquad \frac{b^{-\alpha }}{b^{-\alpha
}V_{\alpha }^{-}}\quad ,
\end{equation}%
with residue $\left( a^{+\alpha },b_{\beta }^{-}\right) $. \newline
(\textbf{ii}) solves locally the equation $U^{+\alpha }V_{\alpha }^{-}=\mu $
inside of $\mathcal{C}_{a^{+},b^{-}}$; i.e in the vicinity of $\left(
a^{+\alpha },b_{\beta }^{-}\right) $, by expressing locally $U^{+\alpha }$
in terms of the inverse of $V_{\alpha }^{-}$. We have
\begin{equation}
U^{+\alpha }=\mu \frac{b^{-\alpha }}{b^{-}.V^{-}},
\end{equation}%
where we have set $b^{-}.V^{-}=b^{-\alpha }V_{\alpha }^{-}$. A similar
treatment may be done also for $V^{-\alpha }$ in terms of the inverse of $%
U^{+\alpha }$. We have
\begin{equation}
V^{-\alpha }=\mu \frac{a^{+\alpha }}{a^{+}.U^{+}}
\end{equation}%
with $a^{+}.U^{+}=a^{+\alpha }U_{\alpha }^{+}$. Now, if we take the boundary
$\partial D$ as given by the curve $\mathcal{C}_{a^{+},b^{-}}$, we get%
\begin{equation}
\int_{\mathcal{C}_{a^{+},b^{-}}}U^{+\alpha }dV^{-\beta }\varepsilon _{\alpha
\beta }=\mu \int_{\mathcal{C}_{a^{+},b^{-}}}\frac{b^{-\alpha }dV^{-\beta }}{%
b^{-}.V^{-}}\varepsilon _{\alpha \beta }.
\end{equation}%
Since $a^{+\alpha }$ and $b_{\beta }^{-}$ are fixed moduli ($da^{+\alpha
}=db_{\beta }^{-}=0$), the 1-form $b^{-\alpha }dV^{-\beta }$ read also as $%
d\upsilon $ with $\upsilon =\left( b^{-}.V^{-}\right) $ and so
\begin{equation}
\int_{\mathcal{C}_{a^{+},b^{-}}}\frac{b^{-\alpha }dV^{-\beta }}{b^{-}.V^{-}}%
\varepsilon _{\alpha \beta }=\int_{\mathcal{C}_{a^{+},b^{-}}}\frac{d\upsilon
}{\upsilon }=2i\pi .
\end{equation}%
We get at the end
\begin{equation}
\int_{A}\Omega _{3}=2i\pi \mu .
\end{equation}%
Note that the same result can be obtained by using the 1-form $V^{-\beta
}dU^{+\alpha }\varepsilon _{\alpha \beta }$. In this case the corresponding
one dimensional integral reads as
\begin{equation}
\int_{\mathcal{C}_{a^{+},b^{-}}}\frac{a^{+\alpha }dU^{+\beta }}{a^{+}.U^{+}}%
\varepsilon _{\alpha \beta }=\int_{\mathcal{C}_{a^{+},b^{-}}}\frac{d\left(
a^{+}.U^{+}\right) }{a^{+}.U^{+}}
\end{equation}%
and has a pole singularity at $a^{+}=U^{+}$.

\subsubsection{Fluctuations}

In quantum cosmology model on $S^{3}$, one is interested in the fluctuations
of the real volume of the round 3-sphere which becomes then
\begin{equation}
\Omega =e^{\Phi }\Omega _{0}.
\end{equation}%
In this relation, the local scaling factor $e^{\Phi }$ is just the absolute
value of the Jacobian $J=\left\vert \frac{\partial U^{\pm \prime }}{\partial
U^{\pm }}\right\vert $ of the general coordinate transformations
\begin{equation}
U^{\pm }\rightarrow U^{\pm \prime }=U^{\pm \prime }\left( U^{\pm }\right)
\end{equation}%
mapping $\Omega _{0}$ into $\Omega $. These fluctuations, which on $c=1$ non
critical string side describe momenta and winding corrections, are generated
by local perturbations of the parameter $p$ deforming the shape of the
3-sphere. To implement such deformations, the previous round sphere analysis
relying on $U^{+\gamma }U_{\gamma }^{-}=p$ should be now extended to the
hypersurface
\begin{equation}
U^{^{\prime }+\gamma }U_{\gamma }^{\prime -}=F\left( p\right)
\end{equation}%
with
\begin{equation}
p\rightarrow p^{\prime }=F\left( p\right) =p-\xi \left( p,u^{\pm }\right) ,
\end{equation}%
where $\xi \left( p,u^{\pm }\right) $ is a real harmonic function on $S^{3}$
and $\left( p,u^{\pm }\right) $ are the local coordinates of $R^{+}\times
SU\left( 2\right) \sim R^{4}\sim C_{u}^{2}$.\ Note that since there is a
zero mode $\xi _{0}$ inside $\xi \left( p,u^{\pm }\right) $ and seen that it
is already singled out ($p=-\xi _{0}$), the right way to write the
deformation $\xi \left( p,u^{\pm }\right) $ would be,%
\begin{equation}
\xi \left( p,u^{\pm }\right) =D^{++}\overline{\xi }^{--}\left( p,u^{\pm
}\right) +D^{--}\xi ^{++}\left( p,u^{\pm }\right) ,
\end{equation}%
where the function $\overline{\xi }^{--}\left( p,u^{\pm }\right) $ is the
complex conjugate of $\xi ^{++}\left( p,u^{\pm }\right) $. Note also that
thinking about $U^{^{\prime }\pm \gamma }$ as $\sqrt{p^{\prime }}u^{\pm
\gamma }$, that is
\begin{equation}
U^{^{\prime }\pm \gamma }=u^{\pm \gamma }\sqrt{F\left( p\right) }
\end{equation}%
and $u^{^{\prime }\pm \gamma }=u^{\pm \gamma }$, one sees that the local
deformations of the shape of $S^{3}$ comes from the variation $dp$ and read
as,
\begin{eqnarray}
dp\qquad  &\rightarrow &\qquad dp^{\prime } \\
dp^{\prime } &=&\left( 1-\frac{\partial \xi }{\partial p}\right) dp+\left(
D^{++}\xi \right) d\tau ^{--}+\left( D^{--}\xi \right) d\tau ^{++}+\left(
D^{0}\xi \right) d\tau ^{0}\quad ,  \notag
\end{eqnarray}%
where $D^{++},$ $D^{--}$ and $D^{0}$ are as in eqs(4.54) and
\begin{eqnarray}
d\tau ^{++} &=&u^{+}du^{+}\quad ,  \notag \\
d\tau ^{--} &=&u^{-}du^{-}\quad , \\
d\tau ^{0} &=&\frac{1}{2}\left( u^{+}du^{-}-u^{-}du^{+}\right) \quad .
\notag
\end{eqnarray}%
Like for the conifold analysis, one may use the fibration $S^{1}\times S^{2}$
to rewrite the local deformation parameter $\xi \left( p,u^{\pm }\right) $
as given by the projective function $\mathrm{\varphi }\left( p;\sigma
,u^{+},u^{-}\right) $ with $\sigma =e^{i\theta }$ and $u^{+\gamma }u_{\gamma
}^{-}=1$ together with the identification
\begin{equation}
u^{+\gamma }\equiv e^{i\phi }u^{+\gamma },\qquad u_{\gamma }^{-}\equiv
e^{-i\phi }u_{\gamma }^{-},\qquad \theta \equiv \theta -\phi \quad ,
\end{equation}%
and
\begin{eqnarray}
\mathrm{\varphi }\left( p;\sigma ,u^{+},u^{-}\right)  &=&\mathrm{\varphi }%
\left( p;e^{-i\phi }\sigma ,e^{i\phi }u^{+},e^{-i\phi }u^{-}\right) \quad ,
\notag \\
\frac{\partial }{i\partial \theta } &\equiv &D^{0},\qquad d\theta \equiv
d\tau ^{0}\quad .
\end{eqnarray}%
In the projective coordinates $\left( p;\sigma ,u^{+},u^{-}\right) $, the
projective function $\mathrm{\varphi }\left( p;\sigma ,u^{+},u^{-}\right) $
may be expanded twice: First in a Fourrier series as follows,%
\begin{eqnarray}
\mathrm{\varphi }\left( p;\sigma ,u^{+},u^{-}\right)  &=&\mathrm{\phi }%
^{0}\left( p;u^{+},u^{-}\right) +\sum_{n>0}e^{in\theta }\text{ }\mathrm{\phi
}^{+n}\left( p;u^{+},u^{-}\right)   \notag \\
&&+\sum_{n>0}e^{-in\theta }\text{ }\mathrm{\phi }^{-n}\left(
p;u^{+},u^{-}\right) \quad ,
\end{eqnarray}%
with
\begin{equation}
D^{0}\text{ }\mathrm{\phi }^{\pm n}\left( p;u^{+},u^{-}\right) =\pm n\text{ }%
\mathrm{\phi }^{\pm n}\left( p;u^{+},u^{-}\right) \quad .
\end{equation}%
Second in a harmonic series on the 2-sphere. In perturbation approach, the
projective function $\mathrm{\varphi }\left( p;\sigma ,u^{+},u^{-}\right) $
and its Laurent modes $\mathrm{\phi }^{\pm n}\left( p;u^{+},u^{-}\right) $
are treated as
\begin{eqnarray}
\mathrm{\varphi }\left( p;\sigma ,u^{\pm }\right)  &=&\mathrm{\phi }\left(
p;\sigma ,u^{+}\right) +\overline{\mathrm{\phi }}\left( p;\sigma
,u^{-}\right) +\mathcal{O}\left( \mathrm{\phi }\overline{\mathrm{\phi }}%
\right) \quad ,  \notag \\
\mathrm{\phi }^{+n}\left( p;u^{\pm }\right)  &=&p^{\frac{n}{2}}u^{\left(
+n\right) }\mathrm{\zeta }_{n}+\delta \eta ^{+n}\left( p;u^{\pm }\right)
\quad ,  \notag \\
\mathrm{\phi }^{-n}\left( p;u^{\pm }\right)  &=&p^{\frac{n}{2}}u^{\left(
-n\right) }\overline{\mathrm{\zeta }}_{-n}+\delta \eta ^{-n}\left( p;u^{\pm
}\right) \quad ,  \label{spl} \\
\delta \eta ^{\pm n}\left( p;u^{\pm }\right)  &=&\delta \eta ^{\pm n}\left(
\mathrm{\zeta },\overline{\mathrm{\zeta }}\right) \quad ,  \notag
\end{eqnarray}%
where we have set,
\begin{eqnarray}
u^{(+n)} &\equiv &u^{+(\alpha _{1}}...u^{+\alpha _{n})},\qquad
u^{(-n)}\equiv u_{(\beta _{1}}^{-}...u_{\beta _{n})}^{-}\quad ,  \notag \\
\mathrm{\zeta }_{n} &\equiv &\mathrm{\zeta }_{\left( n,0\right) }\equiv
\mathrm{\zeta }_{\left( \alpha _{1}...\alpha _{n}\right) },\qquad \overline{%
\mathrm{\zeta }}_{-n}\equiv \mathrm{\zeta }^{\left( \beta _{1}...\beta
_{n}\right) }\equiv \mathrm{\zeta }_{\left( 0,n\right) }\quad .
\end{eqnarray}%
Furthermore, referring to $S^{3}$ cosmology analysis of previous subsection,
one can write down the the harmonic volume 3-form $\Omega $ of the deformed
3-sphere. Using the standard relation
\begin{equation}
\Omega =\frac{\Omega _{0}}{\partial F/\partial p},
\end{equation}%
we get%
\begin{equation}
\Omega =\frac{\Omega _{0}}{1-\left( \frac{\partial \mathrm{\varphi }}{%
\partial p}\right) }\quad .
\end{equation}%
Comparing with $\exp \left( \Phi \right) =\Omega /\Omega _{0}$, we obtain
the relation between the conformal factor $\Phi \left( p,u^{\pm }\right) $
and the local complex deformation function $\mathrm{\varphi }\left( p,u^{\pm
}\right) $,
\begin{equation}
\Phi \left( p,u^{\pm }\right) =-\ln \left( 1-\frac{\partial \mathrm{\varphi }%
}{\partial p}\right) =-\ln \left( 1-\frac{\partial \mathrm{\phi }}{\partial p%
}-\frac{\partial \overline{\mathrm{\phi }}}{\partial p}-\mathcal{O}\left(
\mathrm{\phi }\overline{\mathrm{\phi }}\right) \right) \quad .
\end{equation}%
Upon splitting $\Phi \left( p,u^{\pm }\right) $ as the sum over the field
variable $\mathrm{\Phi }\left( p,u^{+}\right) $ and its complex conjugate $%
\overline{\mathrm{\Phi }}\left( p,u^{-}\right) $, we get at the first order
in $\zeta $,%
\begin{equation}
\mathrm{\Phi }\left( p,u^{+}\right) \simeq \frac{\partial \mathrm{\phi }}{%
\partial p},\qquad \overline{\mathrm{\Phi }}\left( p,u^{-}\right) \simeq
\frac{\partial \overline{\mathrm{\phi }}}{\partial p}\quad .  \label{flu}
\end{equation}%
For later use, it is interesting to note the two following useful relations.
First, observe that the quadratic functional field quantity,%
\begin{equation}
\frac{1}{g_{s}^{2}}S_{2}=\frac{2}{g_{s}^{2}}\sum_{n>0}\frac{p^{n}}{n}\left(
\int_{S^{2}}\overline{\mathrm{\zeta }}^{-n}\left( u^{-}\right) \text{ }%
\mathrm{\zeta }^{+n}\left( u^{+}\right) \right) \quad ,  \label{s0}
\end{equation}%
may be usually expressed as a harmonic integral over the real 3-sphere. One
way to do it is to use the relation
\begin{equation}
p\frac{\partial }{\partial p}\left( p^{\frac{n}{2}}\mathrm{\zeta }%
^{+n}\left( u^{+}\right) \right) =\frac{n}{2}\left( p^{\frac{n}{2}}\mathrm{%
\zeta }^{+n}\left( u^{+}\right) \right)
\end{equation}%
to put above integral into the form
\begin{equation}
\frac{1}{g_{s}^{2}}\int_{S^{3}}\overline{\mathrm{\zeta }}\left(
p,u^{-}\right) \frac{1}{p\frac{\partial }{\partial p}}\mathrm{\zeta }\left(
p,u^{+}\right) .  \label{ex}
\end{equation}%
Note that we have the operator $p\partial /\partial p$ counting the number
of $p$'s reads in terms of the harmonic coordinates as follows,%
\begin{equation}
p\frac{\partial }{\partial p}=D^{++}D^{--}+D^{--}D^{++}\equiv \left\{
D^{++},D^{--}\right\} .
\end{equation}%
Using the obvious identities,%
\begin{equation}
D^{++}\mathrm{\zeta }\left( p,u^{+}\right) =0\quad ,\qquad D^{--}\overline{%
\mathrm{\zeta }}\left( p,u^{-}\right) =0,
\end{equation}%
expressing respectively holomorphy in $u^{+}$ and $u^{-}$, and standard
relations on pseudo-differential operator analysis, in particular
\begin{equation}
\partial ^{-k}f=\sum_{s\geq 0}a_{s}\partial ^{s}f\partial ^{-s-k}
\end{equation}%
with\ somes numbers $a_{s}$, one can rewrite eq(\ref{ex}) as,%
\begin{equation}
\frac{1}{g_{s}^{2}}\left( \int_{S^{3}}\mathrm{\zeta }\left( p,u^{+}\right)
\frac{1}{D^{--}D^{++}}\overline{\mathrm{\zeta }}\left( p,u^{-}\right)
\right) \quad .
\end{equation}%
Then by integration by part, we can bring it to,%
\begin{equation}
\frac{1}{g_{s}^{2}}S_{2}=\frac{-1}{g_{s}^{2}}\left[ \int_{S^{3}}\left( \frac{%
1}{D^{--}}\mathrm{\zeta }\left( p,u^{+}\right) \right) \left( \frac{1}{D^{++}%
}\overline{\mathrm{\zeta }}\left( p,u^{-}\right) \right) \right] \quad .
\label{s1}
\end{equation}%
In this relation the measure $\int_{S^{3}}$ refers to the normalized
integral over the three sphere
\begin{equation}
\int_{S^{3}}1=1
\end{equation}%
which factorises in terms of normalized measures on $S^{1}$ and $S^{2}$ as
\begin{equation}
\int_{S^{2}}\left( \int_{S^{1}}1\right) =1.
\end{equation}%
Similarly, we can express the dimensionless cubic functional field quantity,%
\begin{equation}
\frac{1}{g_{s}^{2}}S_{3}=-\frac{1}{6g_{s}^{2}}\sum_{n_{1}+n_{2}=n_{3}}p^{%
\frac{\left\vert n_{1}\right\vert +\left\vert n_{2}\right\vert +\left\vert
n_{3}\right\vert }{2}-1}\left( \int_{S^{2}}\mathrm{\zeta }^{+n_{1}}\mathrm{%
\zeta }^{+n_{2}}\overline{\mathrm{\zeta }}^{-n_{3}}+\overline{\mathrm{\zeta }%
}^{-n_{1}}\overline{\mathrm{\zeta }}^{-n_{2}}\mathrm{\zeta }^{+n_{3}}\right)
\quad ,  \label{s10}
\end{equation}%
like a harmonic integral over the three sphere as shown below,%
\begin{equation}
\frac{1}{g_{s}^{2}}S_{3}=\frac{1}{6g_{s}^{2}}\left( \frac{1}{p}\int_{S^{3}}%
\mathrm{\zeta }^{2}\left( p,u^{+}\right) \overline{\mathrm{\zeta }}\left(
p,u^{-}\right) +\mathrm{\zeta }\left( p,u^{+}\right) \overline{\mathrm{\zeta
}}^{2}\left( p,u^{-}\right) \right) \quad .  \label{s2}
\end{equation}%
Note that these $S_{2}$ and $S_{3}$ terms, which will be used later, are in
fact just the leading two terms of a series
\begin{equation}
S=\sum_{n\geq 2}S_{n}  \label{ss}
\end{equation}%
describing the perturbative expansion of genus zero free energy $\mathcal{F}%
_{0}$ of the partition function of the topological string B model on
conifold. To prove the statements (\ref{s1},\ref{s2}), one needs the
harmonic expansions (5.44) and use the\ following result on the harmonic
integrals%
\begin{equation}
I_{k,l}\left( x,\mathrm{\zeta },\overline{\mathrm{\zeta }}\right)
=\int_{S^{3}}\left( \mathrm{\zeta }\right) ^{k}\left( \overline{\mathrm{%
\zeta }}\right) ^{l}
\end{equation}%
with $k$ and $l$\ two positive integers,%
\begin{equation}
I_{k,l}\left( x,\mathrm{\zeta ,}\overline{\mathrm{\zeta }}\right)
=\sum_{n_{1},...,m_{l}\neq 0}x^{\frac{\left\vert n_{1}\right\vert
+...+\left\vert n_{k}\right\vert +\left\vert m_{1}\right\vert
+...+\left\vert m_{l}\right\vert }{2}}\delta
_{n_{1}+...+n_{k},m_{1}...m_{l}}\int_{S^{2}}\left( \dprod\limits_{j=1}^{k}%
\mathrm{\zeta }^{+n_{j}}\dprod\limits_{s=1}^{l}\overline{\mathrm{\zeta }}%
^{-m_{s}}\right) .
\end{equation}%
For $k=l=0$, the integral $I_{0,0}\left( x,\mathrm{\zeta },\overline{\mathrm{%
\zeta }}\right) $ is just the normalized volume of $S^{3}$ and for $k=1,$ $%
l=0$, the integral $I_{1,0}\left( x,\mathrm{\zeta },\overline{\mathrm{\zeta }%
}\right) $ vanishes identically as there is no $SU\left( 2,C\right) $
singlet within $\mathrm{\zeta }$. For $\left( k,l\right) =\left( 1,1\right) $
and $\left( 2,1\right) $, we have the following
\begin{eqnarray}
I_{1,1}\left( x,\mathrm{\zeta },\overline{\mathrm{\zeta }}\right)
&=&\sum_{n\neq 0}x^{\left\vert n\right\vert }\left( \int_{S^{2}}\mathrm{%
\zeta }^{+n}\overline{\mathrm{\zeta }}^{-n}\right) , \\
I_{2,1}\left( x,\mathrm{\zeta },\overline{\mathrm{\zeta }}\right)
&=&\sum_{n_{1}+n_{2}\neq 0}x^{\frac{\left\vert n_{1}\right\vert +\left\vert
n_{2}\right\vert +\left\vert n_{1}+n_{2}\right\vert }{2}}\left( \int_{S^{2}}%
\mathrm{\zeta }^{+n_{1}}\mathrm{\zeta }^{+n_{2}}\overline{\mathrm{\zeta }}%
^{-n_{1}-n_{2}}\right) .
\end{eqnarray}%
Putting altogether these relations, one gets the desired results.

\subsection{Correlation functions}

In the harmonic frame work of the GSV model of $S^{3}$ quantum cosmology, we
can compute the $SU\left( 2,C\right) $ manifestly covariant $N$ points Green
functions $G_{N}=G\left( U_{1}^{\pm },...,U_{N}^{\pm }\right) $,
\begin{equation}
G_{N}=<\Phi \left( U_{1}^{\pm }\right) ...\Phi \left( U_{N}^{\pm }\right)
>\quad .  \label{gr}
\end{equation}%
These functions describe the correlations between the conformal fields $\Phi
_{i}=\Phi \left( U_{i}^{\pm }\right) $ at the points $U_{i}^{\pm }$ with
\begin{equation}
U_{i}^{+\alpha }U_{\alpha i}^{-}=p
\end{equation}%
and their evaluation uses Hartle-Hawking probability density $\varrho \left(
p,\mathrm{\zeta },\overline{\mathrm{\zeta }}\right) =\left\vert \Psi \left(
p,\mathrm{\zeta },\overline{\mathrm{\zeta }}\right) \right\vert ^{2}$.

\subsubsection{Hartlee-Hawking probability density}

The probability density $\varrho \left( p,\mathrm{\zeta },\overline{\mathrm{%
\zeta }}\right) $ involves the Hartle-Hawking universe wave function
\begin{equation}
\Psi =\Psi \left( p,\mathrm{\zeta },\overline{\mathrm{\zeta }}\right)
\end{equation}%
whose expression has been shown to coincide with the topological string
partition function
\begin{equation}
\left\vert \mathcal{Z}_{top}\left( p,\mathrm{\zeta },\overline{\mathrm{\zeta
}}\right) \right\vert ^{2}=\exp 2\func{Re}\left( \mathcal{F}\left( p,\mathrm{%
\zeta },\overline{\mathrm{\zeta }}\right) \right) \quad .
\end{equation}%
With this result in mind and following $\cite{30}$ as well as using the
analysis of previous section and standard techniques from quantum field
theory, the generic N points Green function $G_{N}$ (\ref{gr}) may be also
formulated as follows,%
\begin{equation}
G_{N}=\mathcal{N}\int D\mathrm{\zeta }D\overline{\mathrm{\zeta }}\left(
\dprod\limits_{i=1}^{N}\Phi _{i}\left( U_{i}^{\pm }\right) \right) \exp
\left( -\frac{1}{g_{s}^{2}}\mathcal{S}\left[ p,\mathrm{\zeta },\overline{%
\mathrm{\zeta }}\right] \right) \quad ,  \label{no}
\end{equation}%
where
\begin{equation}
D\mathrm{\zeta }D\overline{\mathrm{\zeta }}=\dprod\limits_{n\geq 0}\left( d%
\mathrm{\zeta }^{+n}d\overline{\mathrm{\zeta }}^{-n}\right) \quad ,
\end{equation}%
the hermitian field $\Phi _{i}$ standing for the infinitesimal complex field
variable $\mathrm{\Phi }\left( p_{i},u_{i}^{+}\right) $ together with its
complex conjugate $\overline{\mathrm{\Phi }}\left( p_{i},u_{i}^{-}\right) $
and where we have set,%
\begin{equation}
\varrho \left( p,\mathrm{\zeta },\overline{\mathrm{\zeta }}\right) =\mathcal{%
N}\exp \left( -\frac{1}{g_{s}^{2}}\mathcal{S}\left[ p,\mathrm{\zeta },%
\overline{\mathrm{\zeta }}\right] \right) .
\end{equation}%
The normalization factor $\mathcal{N}$ is fixed as usual by the unitary
condition
\begin{equation}
\mathcal{N}\int D\mathrm{\zeta }D\overline{\mathrm{\zeta }}\exp \left( -%
\frac{1}{g_{s}^{2}}\mathcal{S}\left[ p,\mathrm{\zeta },\overline{\mathrm{%
\zeta }}\right] \right) =1
\end{equation}%
and, in genus zero approximation of free energy, the factor $\mathcal{S}%
\left[ p,\mathrm{\zeta },\overline{\mathrm{\zeta }}\right] $ is defined as,%
\begin{eqnarray}
\mathcal{S}\left[ p,\mathrm{\zeta },\overline{\mathrm{\zeta }}\right]
&=&\sum_{n>0}\frac{2}{n}p^{n}\left( \int_{S^{2}}\mathrm{\zeta }^{+n}%
\overline{\mathrm{\zeta }}^{-n}\right)  \notag \\
&&-\frac{1}{3}\sum_{n_{1}+n_{2}=n_{3}}p^{\frac{\left\vert n_{1}\right\vert
+\left\vert n_{2}\right\vert +\left\vert n_{3}\right\vert -2}{2}}\left(
\int_{S^{2}}\mathrm{\zeta }^{+n_{1}}\mathrm{\zeta }^{+n_{2}}\overline{%
\mathrm{\zeta }}^{-n_{3}}\right) +O\left( \left[ \mathrm{\zeta }^{4}\right]
\right) .
\end{eqnarray}%
Observe that under the change%
\begin{equation}
p\rightarrow \lambda p,\qquad \zeta \rightarrow \lambda \zeta ,\qquad
g_{s}\rightarrow \lambda g_{s},\qquad \zeta ^{\pm n}\rightarrow \lambda ^{1-%
\frac{\left\vert n\right\vert }{2}}\zeta ^{\pm n},
\end{equation}%
$\mathcal{S}\left[ p,\mathrm{\zeta },\overline{\mathrm{\zeta }}\right] $
scales as,%
\begin{equation}
\mathcal{S}\left[ \lambda p,\lambda \mathrm{\zeta },\lambda \overline{%
\mathrm{\zeta }}\right] =\lambda ^{2}\mathcal{S}\left[ p,\mathrm{\zeta },%
\overline{\mathrm{\zeta }}\right] ,
\end{equation}%
and so the ratio $\frac{1}{g_{s}^{2}}\mathcal{S}\left[ p,\mathrm{\zeta },%
\overline{\mathrm{\zeta }}\right] $ appearing in the exponential of the
Hartlee-Hawking probability density is invariant.

\subsubsection{$\frac{1}{p}$ expansion}

Using the identities (\ref{s0}-\ref{s2}), eq(\ref{ss}) may be also
formulated as follows,%
\begin{equation}
\mathcal{S}\left[ p,\mathrm{\zeta },\overline{\mathrm{\zeta }}\right]
=\int_{S^{3}}\mathrm{\zeta }\frac{1}{\left\{ D^{--},D^{++}\right\} }%
\overline{\mathrm{\zeta }}-\frac{1}{6p}\int_{S^{3}}\left( \mathrm{\zeta }^{2}%
\overline{\mathrm{\zeta }}+\mathrm{\zeta }\overline{\mathrm{\zeta }}%
^{2}\right) +\mathcal{O}\left( \frac{1}{p^{2}}\left( \mathrm{\zeta }%
\overline{\mathrm{\zeta }}\right) ^{2},...\right) \quad .  \label{sr}
\end{equation}%
This expression recalls the usual field theory action; but with a non local
operator for the quadratic $\zeta $\ term. It may be imagined as a field
theory action with a tower interacting terms $\mathrm{\zeta }^{n}$ and
coupling constant $\lambda _{n}$ proportional to
\begin{equation*}
\frac{1}{p^{n-2}}.
\end{equation*}%
In the limit of a 3-sphere with large volume, one may treat these
interacting terms as perturbations around the Gaussian like factor,
\begin{equation}
\mathcal{S}_{0}\left[ p,\mathrm{\zeta },\overline{\mathrm{\zeta }}\right]
=\int_{S^{3}}\mathrm{\zeta }\frac{1}{\left\{ D^{--},D^{++}\right\} }%
\overline{\mathrm{\zeta }},
\end{equation}%
which we discuss below.\newline
Thinking about the normalization condition of the probability density $\rho %
\left[ p,\mathrm{\zeta },\overline{\mathrm{\zeta }}\right] $ as just the
value $\mathcal{Z}\left[ p,J=0\right] $ of some partition function $\mathcal{%
Z}\left[ p,\mathrm{J}\right] $ defined as,
\begin{equation}
\mathcal{Z}\left[ p,\mathrm{J},\overline{\mathrm{J}}\right] =\mathcal{N}\int
D\mathrm{\zeta }D\overline{\mathrm{\zeta }}\exp \left( -\frac{1}{g_{s}^{2}}%
\mathcal{S}\left[ p,\mathrm{\zeta },\overline{\mathrm{\zeta }}\right]
+\int_{S^{3}}\left( \mathrm{J\zeta }+\overline{\mathrm{J}}\overline{\mathrm{%
\zeta }}\right) \right) ,
\end{equation}%
where $\mathrm{J}$ and its complex conjugate $\overline{\mathrm{J}}$ are
local external source fields living on the 3-sphere, one can compute Green
functions type
\begin{equation}
<\mathrm{\zeta }\left( p_{1},u_{1}\right) ...\mathrm{\zeta }\left(
p_{N_{1}},u_{N_{1}}\right) \overline{\mathrm{\zeta }}\left(
p_{N_{1}+1},u_{N_{1}+1}\right) ...\overline{\mathrm{\zeta }}\left(
p_{N_{1}+N_{2}},u_{N_{1}+N_{2}}\right) >
\end{equation}%
and more generally
\begin{equation}
<\left( \dprod\limits_{i=1}^{N_{1}}\frac{\partial ^{n_{i}}\mathrm{\zeta }%
\left( p_{i},u_{i}^{\pm }\right) }{\partial p_{i}^{n_{i}}}\right) \left(
\dprod\limits_{j=1}^{N_{2}}\frac{\partial ^{m_{j}}\mathrm{\zeta }\left(
p_{j},u_{j}^{\pm }\right) }{\partial p_{j}^{m_{j}}}\right) >,\qquad
p_{i}=p_{j}=p,
\end{equation}%
describing the correlations between the field variable $\mathrm{\zeta }%
\left( x,u\right) $, $\overline{\mathrm{\zeta }}\left( x,u\right) $ and
their derivatives. For $<\mathrm{\zeta }\left( p_{1},u_{1}^{+}\right) ...%
\overline{\mathrm{\zeta }}\left( p_{N},u_{N}^{-}\right) >$, we have,%
\begin{equation}
<\mathrm{\zeta }\left( p_{1},u_{1}^{+}\right) ...\overline{\mathrm{\zeta }}%
\left( p_{N},u_{N}^{-}\right) >=\frac{\delta ^{N}\mathcal{Z}\left[ p,\mathrm{%
J},\overline{\mathrm{J}}\right] }{\delta \mathrm{J}\left(
p_{1},u_{1}^{+}\right) ...\delta \overline{\mathrm{J}}\left(
p_{N},u_{N}^{-}\right) }|_{\mathrm{J}=\overline{\mathrm{J}}=0}.
\end{equation}%
To compute these correlation functions, it is interesting to bring the above
generating functional $\mathcal{Z}\left[ p,\mathrm{J},\overline{\mathrm{J}}%
\right] $ into a manageable form. With the idea of a perturbation theory in $%
\left( 1/p\right) $ in mind and using the usual trick
\begin{equation}
\mathrm{\zeta }\left( p,u^{\pm }\right) =\frac{\delta }{\delta \mathrm{J}%
\left( p,u^{\pm }\right) },
\end{equation}%
the above relation can be also expressed as,
\begin{equation}
\mathcal{Z}\left[ p,\mathrm{J},\overline{\mathrm{J}}\right] =\mathcal{N}%
\left[ \exp \left( \frac{1}{6p}\int_{S^{3}}\left( \frac{\delta ^{3}}{\delta
\mathrm{J}^{2}\left( p,u^{+}\right) \delta \overline{\mathrm{J}}\left(
p,u^{-}\right) }+cc\right) +O\left( 4\right) \right) \right] \mathcal{Z}_{0}%
\left[ p,\mathrm{J},\overline{\mathrm{J}}\right] ,
\end{equation}%
where,
\begin{equation}
\mathcal{Z}_{0}\left[ p,\mathrm{J},\overline{\mathrm{J}}\right] =\mathcal{N}%
\int D\mathrm{\zeta }D\overline{\mathrm{\zeta }}\exp \left( -\frac{1}{%
g_{s}^{2}}\mathcal{S}_{0}\left[ p,\mathrm{\zeta },\overline{\zeta }\right]
+\int_{S^{3}}\left( \mathrm{J\zeta }+cc\right) \right) .
\end{equation}%
Up on integrating with respect to the field variable $\mathrm{\zeta }$, $%
\mathcal{Z}_{0}\left[ p,\mathrm{J},\overline{\mathrm{J}}\right] $ reads also
as,%
\begin{equation}
\mathcal{Z}_{0}\left[ p,\mathrm{J},\overline{\mathrm{J}}\right] =\exp \left[
-\frac{g_{s}^{2}}{2}\int_{S_{i}^{3}\times S_{j}^{3}}\overline{\mathrm{J}}%
\left( p_{i},u_{i}^{-}\right) \left[ \delta \left( u_{i}^{\pm },u_{j}^{\pm
}\right) \left\{ D_{j}^{++},D_{j}^{--}\right\} \right] \mathrm{J}\left(
p_{i},u_{j}^{+}\right) \right] ,  \label{z0}
\end{equation}%
where $\delta \left( u_{i}^{\pm },u_{j}^{\pm }\right) $ is a harmonic
distribution defined as,%
\begin{equation}
\int_{S_{j}^{3}}\delta \left( u_{i}^{\pm },u_{j}^{\pm }\right) F\left(
p_{j},u_{j}^{\pm }\right) =F\left( p_{i},u_{i}^{\pm }\right) .
\end{equation}%
The free field propagators
\begin{equation}
<\mathrm{\zeta }\left( p_{i},u_{i}^{+}\right) \overline{\mathrm{\zeta }}%
\left( p_{j},u_{j}^{-}\right) >_{0}
\end{equation}%
and the corresponding correlation function
\begin{equation}
<\mathrm{\Phi }\left( p_{i},u_{i}^{\pm }\right) \overline{\mathrm{\Phi }}%
\left( p_{j},u_{j}^{\mp }\right) >_{0}
\end{equation}%
as well as higher order points Green functions are easily obtained from eq(%
\ref{z0}). For example, we have
\begin{equation}
<\zeta \left( p_{i},u_{i}^{+}\right) \zeta \left( p_{j},u_{j}^{+}\right) >=0.
\end{equation}%
The same result is valid for any correlation function$\ $involving one
handed complex deformations as shown below,%
\begin{eqnarray}
&<&\dprod\limits_{i=1}^{N_{1}}\frac{\partial ^{n_{i}}\mathrm{\zeta }\left(
p_{i},u_{i}^{\pm }\right) }{\partial p_{i}^{n_{i}}}>=0,  \notag \\
&<&\dprod\limits_{j=1}^{N_{2}}\frac{\partial ^{m_{j}}\mathrm{\zeta }\left(
p_{j},u_{j}^{\pm }\right) }{\partial p_{j}^{m_{j}}}>=0.
\end{eqnarray}%
The simplest non trivial results are given by the propagator,%
\begin{equation}
<\zeta \left( p_{i},u_{i}^{+}\right) \overline{\zeta }\left(
p_{j},u_{j}^{-}\right) >_{0}=g_{s}^{2}\left\{ D_{j}^{++},D_{j}^{--}\right\}
\delta \left( u_{i}^{\pm },u_{j}^{\pm }\right) ,  \label{p1}
\end{equation}%
We also have,
\begin{equation}
<\mathrm{\Phi }\left( p_{i},u_{i}^{+}\right) \overline{\mathrm{\Phi }}\left(
p_{j},u_{j}^{-}\right) >_{0}=\frac{g_{s}^{2}}{p^{2}}\left( \left\{
D_{j}^{++},D_{j}^{--}\right\} \right) ^{3}\delta \left( u_{i}^{\pm
},u_{j}^{\pm }\right) .  \label{p2}
\end{equation}%
Using these relations, one can compute all correlation functions. All of
them preserve manifestly the $SU\left( 2,C\right) $ isometry of the 3-sphere.

\section{Conclusion}

In this paper we have studied the partition function $\mathcal{Z}_{top}$ of
the B-model topological string on conifold and the correlations functions of
the scale operator field of Gukov-Saraikin-Vafa quantum cosmology model on $%
S^{3}$. To that purpose, we have first developed harmonic analysis for
conifold and shown that this formalism gives in fact a unified description
of $T^{\ast }S^{3}$ and its sub-varieties $T^{\ast }S^{2}$, $S^{3}$ and $%
S^{2}$. Recall that the idea of using harmonic space has been considered in
the past by Galperin \textit{et al} for the case of $S^{2}$ for the
construction of an off shell superfield formulation of extended
supersymmetic gauge theories with $SU_{R}\left( 2\right) $\ symmetry. The
harmonic space method developed in this paper goes beyond and has the
following basic properties: \newline
(\textbf{1}) It preserves manifestly conifold $SL\left( 2,C\right) $
isometry and allows to complete partial results for the computation of the
partition function of topological strings on conifold. Recall that in
standard complex analysis, computation of $\mathcal{Z}_{top}\left( T^{\ast
}S^{3}\right) $ deals with the subset of conifold local complex deformations
restricted to its subspace $T^{\ast }S^{1}$. Harmonic space method for
conifold covers local complex deformation over all conifold points. Moreover
the harmonic variables $U_{\alpha }^{+}$ and $V_{\beta }^{-}$ get a
remarkable interpretation in 2D $c=1$ non critical string theory. There $%
U_{1}^{+}$ and $U_{2}^{+}$ are respectively associated with positive unit
momentum and winding modes. $V_{1}^{-}$ and $V_{2}^{-}$ are associated with
negative unit mode momentum and winding. \newline
(\textbf{2}) The harmonic method applies as well for the study of
restrictions down to $T^{\ast }S^{2}\sim SL\left( 2\right) /C^{\ast }$ and
to the real slices $S^{3}\sim SU\left( 2\right) $ and $S^{2}\sim SU\left(
2\right) /U\left( 1\right) $. The harmonic formalism for $T^{\ast }S^{2}$ is
recovered naturally from that of $T^{\ast }S^{3}$ by fixing the $C^{\ast }$
symmetry of $T^{\ast }S^{1}$ fiber. This harmonic space method may be
applied as well for studying local complex deformation moduli space of
string on $K3$ with deformed A$_{1}$ singularity.\newline
(\textbf{3}) Harmonic space analysis on $T^{\ast }S^{3}$ has a remarkable 1
to 1 correspondence Laurent analysis on $T^{\ast }S^{1}$. This property,
which is also valid for harmonic analysis $S^{3}$ and Fourier analysis on $%
S^{1}$, allowed us to :

(\textbf{a}) construct the dictionary (\ref{di}) giving the correspondence
between $T^{\ast }S^{3}$ and $T^{\ast }S^{1}$ and similarly between their
real compact slices.

(\textbf{b}) use the dictionary (\ref{di}) to derive the explicit form of
the B-model topological string partition function on conifold. This obtained
partition function preserves manifestly conifold $SL\left( 2\right) $
isometry.

With this machinery at hand, and guided by known results on the ground ring
of 2D $c=1$ string theory $\cite{16}$, we have reconsidered the study the
local complex deformations of $T^{\ast }S^{3}$ keeping track with manifest $%
SL\left( 2,C\right) $ isometry of the conifold. We have also derived the
explicit expression of the manifestly $SL\left( 2,C\right) $ invariant form
of the partition function $\mathcal{Z}_{top}$ of B-model topological string
on conifold; see also theorem of section 7. The harmonic space method has
been also used to study stringy quantum cosmology of Gukov Saraikin and
Vafa; in particular to compute the manifestly $SU\left( 2,C\right) $
covariant correlation functions $<\Phi \left( S^{3}\right) ...\Phi \left(
S^{3}\right) >$ of the GSV quantum cosmology model. Actually this analysis
completes the results of $\cite{30}$

\begin{acknowledgement}
\qquad We thank Protars III program, D12/25/CNRST- Brane physics, Rabat. We
also thank ICTP, the International Centre for Theoretical Physics, Trieste
Italy, for kind hospitality and the Network 626 OEA for support.
\end{acknowledgement}

\section{Appendix: Harmonic analysis on conifold and subvarieties}

\qquad In this section, we give some useful details on the harmonic analysis
for conifold $T^{\ast }S^{3}$, the cotangeant bundle of complex one
dimension projective space $T^{\ast }P^{1}$ and their corresponding real
slices namely the spheres $S^{3}$ and $S^{2}$. We give the main lines on
harmonic expansions on these manifolds, the harmonic differential calculus,
forms, harmonic integration and the various types of harmonic distributions.

\subsection{General set up}

\qquad We begin by recalling that in harmonic frame work, the conifold $%
T^{\ast }S^{3}$ is defined by
\begin{equation}
u^{+\alpha }v_{\alpha }^{-}=1,\qquad u^{+\alpha }u_{\alpha }^{+}=0,\qquad
v^{-\alpha }v_{\alpha }^{-}=0
\end{equation}%
with $\left( u^{+1},u^{+2},v_{1}^{-},v_{2}^{-}\right) $ four complex
coordinates in $C^{4}$ and where, for simplicity, we have set $\mu =1$. Its
two complex dimension coset
\begin{equation}
T^{\ast }P^{1}\sim T^{\ast }S^{3}/C^{\ast }
\end{equation}%
is defined by similar relations,%
\begin{equation}
u^{+\alpha }v_{\alpha }^{-}=1,\qquad u^{+\alpha }u_{\alpha }^{+}=0,\qquad
v^{-\alpha }v_{\alpha }^{-}=0,
\end{equation}%
but now $u^{+\alpha }$ and $v_{\alpha }^{-}$ belong to to the weighted
projective space $WP^{3}\left( 1,1,-1,-1\right) $, i.e with projective
transformations,%
\begin{equation}
u^{+\alpha }\rightarrow \lambda u^{+\alpha },\qquad v_{\alpha
}^{-}\rightarrow \frac{1}{\lambda }v_{\alpha }^{-},\qquad \lambda \in
C^{\ast }.  \label{ch}
\end{equation}%
For later use, note that the variable $v_{\alpha }^{-}$ in the global
defined equation $u^{+\alpha }v_{\alpha }^{-}=1$ may be locally solved in
the vicinity of $u^{+\alpha }=\eta ^{+\alpha }$ and $v_{\alpha }^{-}=\eta
_{\alpha }^{-}$ \ as follows
\begin{equation}
v_{\alpha }^{-}=\frac{1}{u^{+\alpha }\eta _{\alpha }^{+}}\eta _{\alpha
}^{+},\qquad \eta ^{+\alpha }\eta _{\alpha }^{-}=1\qquad \eta ^{\pm \alpha
}\eta _{\alpha }^{\pm }=0.  \label{in}
\end{equation}%
Harmonic functions $\mathcal{F}$ on conifold are defined by help of harmonic
functions $F^{q}=F^{q}(u^{+},v^{-})$ living on $T^{\ast }P^{1}$. These are
homogeneous holomorphic functions
\begin{equation}
F^{q}(\lambda u^{+},\frac{1}{\lambda }v^{-})=\lambda ^{q}F^{q}(u^{+},v^{-})
\end{equation}%
depending on the harmonic variables $u^{+}$ and $v^{-}$; but no dependence
on $u^{-}$ and $v^{+}$; i.e,%
\begin{equation}
\frac{\partial F^{q}}{\partial u^{-}}=0,\qquad \frac{\partial F^{q}}{%
\partial v^{+}}=0.
\end{equation}%
To get the expansion of $\mathcal{F}$, one thinks about conifold as embedded
in the projective space $WP^{4}\left( -1,1,1,-1,-1\right) $ with the
projective coordinates
\begin{equation}
\sigma \equiv \frac{1}{\lambda }\sigma ,\qquad u^{+\alpha }\equiv \lambda
u^{+\alpha },\qquad v_{\alpha }^{-}\equiv \frac{1}{\lambda }v_{\alpha }^{-}.
\end{equation}%
with $\lambda $\ a $C^{\ast }$ gauge parameter. Function $\mathcal{F}\left(
\sigma ,u^{+},v^{-}\right) $ on conifold are projective function
\begin{equation}
\mathcal{F}\left( \frac{1}{\lambda }\sigma ,\lambda u^{+},\frac{1}{\lambda }%
v^{-}\right) =\mathcal{F}\left( \sigma ,u^{+},v^{-}\right)
\end{equation}%
and have then the following Laurent expansion,
\begin{equation}
F\left( \sigma ,u^{+},v^{-}\right) =\sum_{q\in Z}\sigma
^{q}F^{q}(u^{+},v^{-})
\end{equation}%
where $F^{q}(u^{+},v^{-})$ are as before. Laurent modes
\begin{equation}
F^{q}(u^{+},v^{-})=\doint \frac{d\sigma }{2i\pi }\sigma ^{-q-1}\mathcal{F}%
\left( \sigma ,u^{+},v^{-}\right)
\end{equation}%
carry a well defined $C^{\ast }$ integer charge $q$ as shown below,
\begin{equation}
\nabla ^{0}F^{q}=\left( u^{+\alpha }\frac{\partial }{\partial u^{+\alpha }}%
-v^{-\alpha }\frac{\partial }{\partial v^{-\alpha }}\right) F^{q}=qF^{q},
\end{equation}%
and have the following typical harmonic expansion%
\begin{equation}
F^{q}(u^{+},v^{-})=\sum_{n\geq 0}F^{\left( \alpha _{1}...\alpha _{q+n}\beta
_{1}...\beta _{n}\right) }\text{ }u_{(\alpha _{1}}^{+}...u_{\alpha
_{q+n}}^{+}v_{\beta _{1}}^{-}...v_{\beta _{n})}^{-}.  \label{an}
\end{equation}%
In this relation, the harmonic monomial quantity
\begin{equation}
u_{(\alpha _{1}}^{+}...u_{\alpha _{n}}^{+}v_{\beta _{1}}^{-}...v_{\beta
_{m})}^{-}
\end{equation}%
stands for a complete symmetrization of the harmonic variables,
\begin{equation}
u_{(\alpha _{1}}^{+}...u_{\alpha _{n}}^{+}v_{\beta _{1}}^{-}...v_{\beta
_{m})}^{-}=\frac{1}{\left( n+m\right) !}\sum_{\sigma \in \mathcal{S}%
_{n+m}}u_{(\alpha _{\sigma \left( 1\right) }}^{+}...u_{\alpha _{\sigma
\left( n\right) }}^{+}v_{\beta _{\sigma \left( 1\right) }}^{-}...v_{\beta
_{\sigma \left( m\right) })}^{-},
\end{equation}%
and $F^{\left( \alpha _{1}...\alpha _{q+n}\beta _{1}...\beta _{n}\right) }$
stands for the harmonic modes obtained by inverting the relation. This is
achieved by help of harmonic integration rules on $T^{\ast }P^{1}$ which we
describe below. Before, note that the identification between the operator $%
\nabla ^{0}$ and $\sigma \frac{\partial }{\partial \sigma }$ follows from
the splitting $SL\left( 2\right) =C^{\ast }\times SL\left( 2\right) /C^{\ast
}$.

\subsubsection{Harmonic differential forms}

Differential forms on the conifold, in particular harmonic 1-forms,
preserving manifestly the $SL\left( 2,C\right) $ isometry group covariance
may be also obtained by starting from the holomorphic 1-forms
\begin{equation}
du^{+\beta }
\end{equation}%
and the anti-holomorphic ones
\begin{equation}
dv_{\beta }^{-}
\end{equation}%
on the ambient complex space $C^{4}$ and build the appropriate isotriplet of
1-forms where $SL\left( 2,C\right) $ indices are contracted. One way to do
it is to note that instead of $du^{+\beta }$ and $dv_{\beta }^{-}$ one may
use the equivalent expressions
\begin{eqnarray}
&&u^{+\alpha }du_{\alpha }^{+},\qquad v_{\alpha }^{-}dv^{-\alpha },\qquad
\frac{1}{2}\left( v_{\alpha }^{-}du^{+\alpha }-v_{\alpha }^{-}du^{+\alpha
}\right) , \\
&&\frac{1}{2}\left( u^{+\alpha }dv_{\alpha }^{-}+v_{\alpha }^{-}du^{+\alpha
}\right) .
\end{eqnarray}%
The three first 1-forms constitute altogether an $SL\left( 2,C\right) $
isotriplet; while the fourth one, which reads also as
\begin{equation}
\frac{1}{2}d\left( u^{+\alpha }v_{\alpha }^{-}\right) ,
\end{equation}%
is a $SL\left( 2,C\right) $ isosinglet. On the conifold and $T^{\ast }P^{1}$
where $u^{+\alpha }v_{\alpha }^{-}$ is a constant, the differential $d\left(
u^{+\alpha }v_{\alpha }^{-}\right) $ vanishes identically and one ends with
the following harmonic one forms,%
\begin{eqnarray}
d\tau ^{\left( ++,0\right) } &=&\varepsilon _{\alpha \beta }u^{+\alpha
}du^{+\beta },  \notag \\
d\tau ^{\left( +,-\right) } &=&\frac{1}{2}\left( u^{+\alpha }dv_{\alpha
}^{-}-v_{\alpha }^{-}du^{+\alpha }\right) , \\
d\tau ^{\left( 0,--\right) } &=&\varepsilon ^{\alpha \beta }v_{\alpha
}^{-}dv_{\beta }^{-},  \notag
\end{eqnarray}%
where the convention notation $\left( p,q\right) $ refers to p charges $%
u^{+} $ and q charges $v^{-}$. Note that because of the constraint eq $%
u^{+\alpha }v_{\alpha }^{-}=1$, the 1-form $d\tau ^{\left( +,-\right) }$
reads also as $u^{+\alpha }dv_{\alpha }^{-}$ or equivalently $v_{\alpha
}^{-}du^{+\alpha }$.

In the coordinate frame $\left( \sigma ,u^{+\alpha },v_{\alpha }^{-}\right) $
of $WP^{4}\left( -1,1,1,-1,-1\right) $, the one forms $d\tau ^{\left(
++,0\right) }$ and $d\tau ^{\left( 0,--\right) }$ map respectively as $%
\lambda ^{2}d\tau ^{\left( ++,0\right) }$ and $\lambda ^{-2}d\tau ^{\left(
0,--\right) }$ and the non covariant $d\tau ^{\left( +,-\right) }$ gets
identified locally with
\begin{equation}
\frac{d\sigma }{\sigma }.
\end{equation}%
Using eq(\ref{in}), one may express the local identification as $\sigma
=u^{+\alpha }\eta _{\alpha }^{+}$ and so we have,%
\begin{equation}
\doint v_{\alpha }^{-}du^{+\alpha }=\doint \frac{d\left( u^{+\alpha }\eta
_{\alpha }^{+}\right) }{u^{+\alpha }\eta _{\alpha }^{+}}=\doint \frac{%
d\sigma }{\sigma }\sim 1.
\end{equation}%
In the frame $\left( \sigma ,u^{+\alpha },v_{\alpha }^{-}\right) $ frame the
projective one forms $d\tau ^{\left( ++,0\right) }$ and $d\tau ^{\left(
0,--\right) }$ are the holomorphic 1-forms on $T^{\ast }P^{1}$. The
holomorphic volume 2-form of $T^{\ast }P^{1}$ is
\begin{equation}
d\tau ^{\left( ++,0\right) }\wedge d\tau ^{\left( 0,--\right) }
\end{equation}%
and the holomorphic three form $\Omega $ on the conifold reads then as
\begin{equation}
\frac{d\sigma }{\sigma }\wedge d\tau ^{\left( 0,--\right) }\wedge d\tau
^{\left( ++,0\right) }.
\end{equation}%
The results on the real slices $S^{3}$ and $S^{2}$\ are recovered by setting
$v^{-}=u^{-}$ and underlying constraints. In what follows, we consider some
aspect on this real truncation.

\subsubsection{3-sphere}

\qquad Let us start by recalling that in the harmonic frame work one of the
remarkable ways to define the volume form $\omega _{0}$ of a \textit{unit}
three sphere is as follows,%
\begin{equation}
\omega _{0}=A\text{ }d\tau ^{0}\wedge d\tau ^{--}\wedge d\tau ^{++},\qquad
\int_{S^{3}}\omega _{0}=1,  \label{c1}
\end{equation}%
where one recognizes
\begin{equation}
d\tau ^{++}\wedge d\tau ^{--}
\end{equation}%
as the real volume 2-form on the real 2-sphere. The advantage of this
definition is that all harmonic variables are treated on equal footing.
Moreover $\omega _{0}$ is manifestly $SU\left( 2,C\right) $ invariant since
it is gauge invariant under projective transformation
\begin{equation}
u^{\pm \alpha \prime }=e^{\pm i\theta }u^{\pm \alpha },
\end{equation}%
a property captured by the sum of Cartan Weyl charge which add exactly to
zero. To have more insight in this way of doing, note that for a unit sphere
parameterized by small harmonics
\begin{equation}
u^{+\alpha }u_{\alpha }^{-}=1,\qquad u^{\pm \alpha }u_{\alpha }^{\pm }=0,
\end{equation}%
the harmonic 1-forms $d\tau ^{\left( q\right) }$, $q=0,\pm 2$ read as
\begin{eqnarray}
d\tau ^{++} &=&2u^{+\alpha }du_{\alpha }^{+}=2\varepsilon _{\alpha \beta
}u^{+\alpha }du^{+\beta },  \notag \\
d\tau ^{--} &=&2u^{-\alpha }du_{\alpha }^{-}=2\varepsilon ^{\alpha \beta
}u_{\beta }^{-}du_{\alpha }^{-},  \label{sh} \\
d\tau ^{0} &=&\left( u^{+\alpha }du_{\alpha }^{-}-u_{\alpha }^{-}du^{+\alpha
}\right) .  \notag
\end{eqnarray}%
They form altogether an $SU\left( 2,C\right) $ vector as one sees by help of
eqs(\ref{13}) and the highest weight state condition%
\begin{equation}
\left[ D^{0},d\tau ^{++}\right] =2d\tau ^{++},\qquad \left[ D^{++},d\tau
^{++}\right] =0,
\end{equation}%
together with the following relations,%
\begin{eqnarray}
\left[ D^{--},d\tau ^{++}\right] &=&d\tau ^{0}  \notag \\
\left[ D^{--},d\tau ^{0}\right] &=&d\tau ^{--}, \\
\left[ D^{--},d\tau ^{--}\right] &=&0.  \notag
\end{eqnarray}%
The fourth remaining symmetric combination namely
\begin{equation}
u^{+\alpha }du_{\alpha }^{-}+u_{\alpha }^{-}du^{+\alpha }
\end{equation}%
vanishes identically due to the constraint eq $u^{+\alpha }u_{\alpha }^{-}=1$
and because of the property,
\begin{equation}
\left( u^{+\alpha }du_{\alpha }^{-}+u_{\alpha }^{-}du^{+\alpha }\right)
=d\left( u^{+\alpha }u_{\alpha }^{-}\right) =0  \label{d4}
\end{equation}%
With help of the above relations, one can easily check the $SU\left(
2,C\right) $ invariance of the three form $\omega _{0}$,
\begin{equation}
\left[ D^{0},\omega _{0}\right] =\left[ D^{++},\omega _{0}\right] =\left[
D^{--},\omega _{0}\right] =0.  \label{cn}
\end{equation}%
To establish eq(\ref{c1}), one starts as usual from the real volume form $%
v_{4}$ of the complex space $C^{2}$ and implement the constraint relations
of the real hypersurface. In harmonic language, this corresponds to taking
\begin{equation}
v_{4}=A\text{ }du^{+1}\wedge du^{+2}\wedge du_{1}^{-}\wedge du_{2}^{-}
\end{equation}%
where A is a normalization factor to be fixed later; then implement the
constraint eqs $u^{+\alpha }u_{\alpha }^{-}=1$ and $u^{\pm \alpha }u_{\alpha
}^{\pm }=0$. These harmonic eqs may be handled in different manners; for
instance by singling out a harmonic variable say $u_{2}^{-}$ and solves it
by help of above constraint eqs as
\begin{equation}
u_{2}^{-}+\frac{u^{+1}u_{1}^{+}}{u^{-2}}=0,\qquad u_{2}^{-}+\frac{%
u^{+1}u_{1}^{-}-1}{u^{+2}}.
\end{equation}%
This method breaks however the $SU\left( 2,C\right) $ isometry we have been
preserving so far. An other way is to use $SU\left( 2,C\right) $ invariance
of $v_{4}$ to rewrite like
\begin{equation}
v_{4}=\frac{-A}{4}\varepsilon _{\alpha \beta }\varepsilon ^{\gamma \delta
}du^{+\alpha }\wedge du^{+\beta }\wedge du_{\gamma }^{-}\wedge du_{\delta
}^{-}
\end{equation}%
and implement the relation $u^{+\alpha }u_{\alpha }^{-}=1$ to first reduces
it to
\begin{equation}
\omega _{0}=\frac{-A}{8}d\tau ^{--}\wedge du^{+\alpha }\wedge du^{+\beta
}\varepsilon _{\alpha \beta }
\end{equation}%
and then substitute the expression of the antisymmetric tensor
\begin{equation}
\varepsilon _{\alpha \beta }=\left( u_{\alpha }^{-}u_{\beta }^{+}-u_{\beta
}^{-}u_{\alpha }^{+}\right)
\end{equation}%
to put $\left( du^{+\alpha }\wedge du^{+\beta }\right) \varepsilon _{\alpha
\beta }$ like
\begin{equation}
\frac{1}{2}d\tau ^{0}\wedge d\tau ^{++}
\end{equation}%
which should be compared with eq(\ref{c1}). The factor is fixed by the
condition
\begin{equation}
\int_{S^{3}}\omega _{0}=1.
\end{equation}%
Note that under the change $u^{\pm }\rightarrow e^{\pm i\theta }u^{\pm }$,
the harmonic differentials $d\tau ^{++}$ and $d\tau ^{--}$ transform
covariantly as
\begin{eqnarray}
d\tau ^{++} &\rightarrow &e^{2i\theta }d\tau ^{++},  \notag \\
d\tau ^{--} &\rightarrow &e^{-2i\theta }d\tau ^{--}
\end{eqnarray}%
while the non covariant $d\tau ^{0}$ gets identified with one form on circle
$S^{1}$. The above description extends naturally to the case of real
3-spheres $U^{+\gamma }U_{\gamma }^{-}=p$ with generic radii.

\subsection{Harmonic integration and distributions}

\subsubsection{ Integration rules}

\qquad Because of the constraint eqs $u^{+\alpha }v_{\alpha }^{-}=1$ and $%
u^{+\alpha }u_{\alpha }^{+}=0$, $v^{-\alpha }v_{\alpha }^{-}$, one can
reduces monomials in harmonic variables, belonging to tensor product
representations of isospinors, as a sum of terms transforming in $SL\left(
2,C\right) $ irreducible representations. These reductions are useful in
harmonic integration analysis involving traces on irreducible
representations of $SL\left( 2,C\right) $. There are some typical relations
which are particularly interesting in performing integration calculus. A set
of these relations corresponds those given by the two following standard
reduction formulas,%
\begin{equation}
u_{\alpha }^{+}u_{(\beta _{1}}^{+}..u_{\beta _{n}}^{+}v_{\gamma
_{1}}^{-}..v_{\gamma _{m)}}^{-}=u_{(\alpha }^{+}u_{\beta
_{1}}^{+}..v_{\gamma _{m)}}^{-}+\frac{m}{m+n+1}\varepsilon _{\alpha (\gamma
_{1}}u_{\beta _{1}}^{+}..u_{\beta _{n}}^{+}v_{\gamma _{2}}^{-}..v_{\gamma
_{m)}}^{-},
\end{equation}%
and,%
\begin{equation}
v_{\gamma }^{-}u_{(\beta _{1}}^{+}..u_{\beta _{n}}^{+}v_{\gamma
_{1}}^{-}..v_{\gamma _{m)}}^{-}=v_{(\gamma }^{-}u_{\beta
_{1}}^{+}..v_{\gamma _{m)}}^{-}-\frac{n}{m+n+1}\varepsilon _{\gamma (\beta
_{1}}u_{\beta _{2}}^{+}..u_{\beta _{n}}^{+}v_{\gamma _{1}}^{-}..v_{\gamma
_{m)}}^{-}.
\end{equation}%
As an illustration, we give the following leading examples for $n,m=0,1$.
For $n=1$ and $m=0$, we have
\begin{equation}
u_{\alpha }^{+}u_{\beta }^{+}=u_{(\alpha }^{+}u_{\beta )}^{+}
\end{equation}%
and
\begin{equation}
v_{\alpha }^{-}u_{\beta }^{+}=v_{(\alpha }^{-}u_{\beta )}^{+}-\frac{1}{2}%
\varepsilon _{\alpha \beta }
\end{equation}%
while for $n=0$ and $m=1$, we have
\begin{equation}
v_{\alpha }^{-}v_{\beta }^{-}=v_{(\alpha }^{-}v_{\beta )}^{-}
\end{equation}
and
\begin{equation}
u_{\alpha }^{+}v_{\beta }^{-}=u_{(\alpha }^{+}v_{\beta )}^{-}+\frac{1}{2}%
\varepsilon _{\alpha \beta }.
\end{equation}
For $n=m=1$, we have the following reduction,
\begin{eqnarray}
u_{\alpha }^{+}u_{(\beta }^{+}v_{\gamma )}^{-} &=&u_{(\alpha }^{+}u_{\beta
}^{+}v_{\gamma )}^{-}+\frac{1}{3}\varepsilon _{\alpha (\gamma }u_{\beta
)}^{+},  \notag \\
v_{\alpha }^{-}u_{(\beta }^{+}v_{\gamma )}^{-} &=&v_{(\alpha }^{-}u_{\beta
}^{+}v_{\gamma )}^{-}-\frac{1}{3}\varepsilon _{\alpha (\beta }v_{\gamma
)}^{-}.
\end{eqnarray}%
These relations are important for integration on harmonic variables. Since
the key point in harmonic integration is mainly taking traces retaining $%
SL\left( 2,C\right) $ singlets, we have then the following integration rules,%
\begin{equation}
\int_{T^{\ast }P^{1}}(u^{+})^{(m}(v^{-})^{n)}(u^{+})_{(k}(v^{-})_{l)}=\frac{%
(-1)^{n}m!n!}{(m+n+1)!}\delta _{(\beta _{1}}^{(\alpha _{1}}\cdots \delta
_{\beta _{k+l)}}^{\alpha _{m+n)}}  \label{tp}
\end{equation}%
which is non zero only if $m=l$\ {and}${\ }n=k$. Here the harmonic monomials
$(u^{+})^{(m}(v^{-})^{n)}$ stand for the short of the following upper index
quantity,
\begin{equation}
(u^{+})^{(m}(v^{-})^{n)}\equiv u^{+(\alpha _{1}}...u^{+\alpha _{m}}v^{-\beta
_{1}}...v^{-\beta _{n)}},
\end{equation}%
and a similar relation for lower indices. As far as eq(\ref{tp}) it vanishes
identically whenever
\begin{equation}
m+n\neq k+l.
\end{equation}%
In the case when this condition is fulfilled, we have non zero results. For
instance, we have for $n=m=0$, the normalized holomorphic volume of $T^{\ast
}P^{1}$, that is
\begin{equation}
\int_{T^{\ast }P^{1}}=1,
\end{equation}%
and for the case $n=k=0$, the above harmonic integral reduces to%
\begin{equation}
\int_{T^{\ast }P^{1}}(u^{+})^{m}(v^{-})_{l}=\frac{1}{(m+1)}\delta _{(\beta
_{1}}^{(\alpha _{1}}\cdots \delta _{\beta _{l)}}^{\alpha _{m)}},
\end{equation}%
whose non trivial leading term is
\begin{equation}
\int_{T^{\ast }P^{1}}u^{+\alpha }v_{\beta }^{-}=\frac{1}{2}\delta _{\beta
}^{\alpha }.
\end{equation}%
On the real 2-sphere obtained from previous analysis by setting
\begin{equation}
v^{-}=u^{-},
\end{equation}
the above rules reduce to
\begin{eqnarray}
u_{\alpha }^{+}u_{(\beta _{1}}^{+}\cdots u_{\beta _{n}}^{+}u_{\gamma
_{1}}^{-}\cdots u_{\gamma _{m)}}^{-} &=&u_{(\alpha }^{+}u_{\beta
_{1}}^{+}\cdots u_{\gamma _{m)}}^{-}+\frac{m}{m+n+1}\varepsilon _{\alpha
(\gamma _{1}}u_{\beta _{1}}^{+}\cdots u_{\beta _{n}}^{+}u_{\gamma
_{2}}^{-}\cdots u_{\gamma _{m)}}^{-},  \notag \\
u_{\gamma }^{-}u_{(\beta _{1}}^{+}..u_{\beta _{n}}^{+}u_{\gamma
_{1}}^{-}..u_{\gamma _{m)}}^{-} &=&u_{(\gamma }^{-}u_{\beta
_{1}}^{+}..u_{\gamma _{m)}}^{-}-\frac{n}{m+n+1}\varepsilon _{\gamma (\beta
_{1}}u_{\beta _{2}}^{+}..u_{\beta _{n}}^{+}u_{\gamma _{1}}^{-}..u_{\gamma
_{m)}}^{-}.
\end{eqnarray}%
We also have the following harmonic integration rule on the unit 2-sphere,%
\begin{equation}
\int_{S^{2}}du(u^{+})^{(m}(u^{-})^{n)}(u^{+})_{(k}(u^{-})_{l)}=\frac{%
(-1)^{n}m!n!}{(m+n+1)!}\delta _{(\beta _{1}}^{(\alpha _{1}}\cdots \delta
_{\beta _{k+l)}}^{\alpha _{m+n)}}.
\end{equation}

\subsubsection{Delta function and harmonic distributions}

\qquad An important tool in the harmonic integration on the real 2-sphere $%
S^{2}$, on which we focus our attention now on, is given by harmonic $\delta
$ functions
\begin{equation}
\delta ^{(q,-q)}(u_{1},u_{2})
\end{equation}%
extending usual one dimensional Dirac distribution. Like the usual ordinary $%
\delta $ function,%
\begin{equation}
\delta {(x_{1}-{x_{2}})}=\frac{1}{2\pi }\int_{-\infty }^{+\infty
}exp[ip(x_{1}-{x_{2}})]dp
\end{equation}%
diverging $x_{1}={x_{2}}$, this is a distribution with quite same features.
It is defined as follows,%
\begin{equation}
\int_{S^{2}}d^{2}u_{2}\text{ }\delta ^{(q,-q)}(u_{1}^{\pm },u_{2}^{\pm
})F^{p}(u_{2}^{\pm })=F^{q}(u_{1}^{\pm })\delta ^{pq},\qquad q\in Z.
\end{equation}%
Viewed as a function on the 2-sphere, the $\delta ^{(q,-q)}(u_{1},u_{2})$
functions have the generic harmonic expansion,%
\begin{equation}
\delta ^{(q,-q)}(u_{1},u_{2})=\sum_{n=0}^{^{\infty }}(-1)^{n+q}\frac{%
(2n+q+1)!}{(n)!(n+q)!}%
(u_{1}^{+})_{(n+q}(u_{1}^{-})_{n)}(u_{2}^{+})^{(n}(u_{2}^{-})^{n+q)}
\end{equation}%
which is clearly divergent for
\begin{equation}
u_{1}^{\pm }=u_{2}^{\pm }.
\end{equation}%
The previous integration rule allows us to determine the coefficients $%
F^{\left( \alpha _{1}...\alpha _{q+n}\beta _{1}...\beta _{n}\right) }$ of
the harmonic expansion (\ref{an}). We have
\begin{equation}
F^{\left( \alpha _{1}...\alpha _{q+n}\text{ }\beta _{1}...\beta _{n}\right)
}=\frac{(-1)^{n+q}(2n+q+1)!}{(n+q)!n!}\int_{S^{2}}d^{2}u\text{ }{%
(u^{+})^{(n}(u^{-})^{n+q)}F^{(q)}(u)}
\end{equation}%
Along with the generalized Dirac $\delta ^{\left( q,-q\right) }$\
distribution, there are others harmonic distributions which are particularly
relevant in quantum field theory. A class of these harmonic distributions
correspond to
\begin{equation}
\frac{1}{(u_{1}^{+}u_{2}^{+})^{n}}
\end{equation}%
and
\begin{equation}
\frac{1}{(v_{1}^{-}v_{2}^{-})^{n}}
\end{equation}%
with singularity at $u_{1}^{+}=u_{2}^{+}$ and $v_{1}^{-}=v_{2}^{-}$. For $%
\frac{1}{(u_{1}^{+}u_{2}^{+})^{n}}$, we have,%
\begin{equation}
\frac{1}{(u_{1}^{+}u_{2}^{+})^{n}}=\frac{1}{n!}\sum_{k=0}^{\infty }(-1)^{n+k}%
\frac{(2k+n+1)!}{k!(k+1)!}\frac{n}{n+k}%
(u_{1}^{+})_{(k}(v_{1}^{-})_{k+n)}(u_{2}^{+})^{(k}(v_{2}^{-})^{k+n)},
\end{equation}%
with $n>0$. A similar harmonic distributions may be also written down for $%
(v_{1}^{-}v_{2}^{-})^{-n}$. These distributions appear in the computation of
Green functions. For recent explicit applications see for instance $\cite{58}
$-$\cite{60}$ and references therein.

\end{document}